\documentclass[twocolumn,twocolappendix]{aastex6}
\usepackage{amssymb,amsmath,latexsym,graphics, graphicx,epsfig,multirow,comment,hyperref,appendix,feyn,slashed,xcolor,afterpage, makecell} 
\usepackage{booktabs}
\usepackage{tabularx}

\newcommand{\abs}[1]{\left\lvert #1 \right\rvert}
\newcommand{\overbar}[1]{\mkern 2mu\overline{\mkern-2mu#1\mkern-2mu}\mkern 2mu}

\newcommand {\be} {\begin {equation}}
\newcommand {\ee} {\end {equation}} 

\newcommand {\bes} {\begin {equation*}}
\newcommand {\ees} {\end {equation*}}

\newcommand{\es}[2] {\begin{equation} \label{#1} \begin{split} #2 \end{split} \end{equation}}

\usepackage{csvsimple}

\newcommand\Tstrut{\rule{0pt}{2.6ex}}         
\newcommand\Bstrut{\rule[-0.9ex]{0pt}{0pt}}   

\newcommand{\beq}{\begin{equation}}
\newcommand{\eeq}{\end{equation}}

\begin{document}

\title{
Deciphering Contributions to the Extragalactic Gamma-Ray Background \\ from  2 GeV to 2 TeV
}
\preprint{MIT/CTP-4808
}

\author{Mariangela Lisanti and Siddharth Mishra-Sharma}
\affil{Department of Physics, Princeton University, Princeton, NJ 08544}

\author{Lina Necib and Benjamin R. Safdi}
\affil{Center for Theoretical Physics, Massachusetts Institute of Technology, Cambridge, MA 02139}

\begin{abstract}
Astrophysical sources outside the Milky Way, such as active galactic nuclei and star-forming galaxies, leave their imprint on the gamma-ray sky as nearly isotropic emission referred to as the Extragalactic Gamma-Ray Background (EGB).  While the brightest of these sources may be individually resolved, their fainter counterparts contribute diffusely.  In this work, we use a recently-developed analysis method, called the Non-Poissonian Template Fit, on up to 93 months of publicly-available data from the \emph{Fermi} Large Area Telescope to determine the properties of the point sources that comprise the EGB.  This analysis takes advantage of photon-count statistics to probe the aggregate properties of these source populations below the sensitivity threshold of published catalogs.  We measure the source-count distributions and point-source intensities, as a function of energy, from $\sim$2 GeV to 2~TeV. 
We find that the EGB is dominated by point sources, likely blazars, in all seven energy sub-bins considered.  These results have implications for the interpretation of IceCube's PeV neutrinos, which may originate from sources that contribute to the non-blazar component of the EGB.  
Additionally, we comment on implications for future TeV observatories such as the Cherenkov Telescope Array.  We provide sky maps showing locations most likely to contain these new sources at both low ($\lesssim 50$ GeV) and high ($\gtrsim 50$ GeV) energies for use in future observations and cross-correlation studies.
\end{abstract}
\maketitle

\pagebreak
\section{Introduction}

The Extragalactic Gamma-Ray Background (EGB) is the nearly isotropic all-sky emission that arises from sources outside of the Milky Way.  The OSO-3~\citep{1968ApJ...153L.203C, 1972ApJ...177..341K} and SAS-2 satellites~\citep{1975ApJ...198..163F,1978ApJ...222..833F} were the first to see hints of the EGB and have since been followed by EGRET~\citep{1998ApJ...494..523S,2004ApJ...613..956S} and, most recently, the \emph{Fermi} Large Area Telescope\footnote{\url{http://fermi.gsfc.nasa.gov/}}~\citep{Ackermann:2014usa,TheFermi-LAT:2015ykq}.  The origin of the EGB remains an open question.  The dominant contributions are likely due to blazars~\citep{Stecker:1993ni, Stecker:1996ma, Muecke:1998cs, Narumoto:2006qg,Dermer:2007fg, Pavlidou:2007dv, Ajello:2009ip, Collaboration:2010gqa, Abazajian:2010pc,  Stecker:2010di,Singal:2011yi,  Ajello:2011zi, Ajello:2013lka,DiMauro:2013zfa, Ajello:2015mfa, Ackermann:2015uya}, star-forming galaxies (SFGs)~\citep{Soltan:1998jg,Pavlidou:2002va, Bhattacharya:2009yv, Ando:2009nk, Fields:2010bw, Makiya:2010zt, Ackermann:2012vca, Chakraborty:2012sh, Lacki:2012si, Tamborra:2014xia}, and misaligned active galactic nuclei (mAGN)~\citep{Stawarz:2005tq, Inoue:2011bm, Massaro:2011ww, DiMauro:2013xta, Hooper:2016gjy}.  Understanding the relative contributions of these source components to the EGB has taken on a new sense of importance in light of IceCube's observation of ultra-high-energy extragalactic neutrinos~\citep{Aartsen:2013bka,Aartsen:2013jdh,Aartsen:2015knd,Aartsen:2015rwa}, the origin of which still remains a mystery.  For instance, the same sources that dominate the extragalactic neutrino background at $\sim$PeV energies may also contribute significantly to the EGB from $\sim$GeV--TeV energies~\citep{Murase:2013rfa, Tamborra:2014xia,Hooper:2016gjy}.  In addition, the EGB may harbor the imprints of more exotic physics such as dark-matter annihilation or decay~\citep{Bengtsson:1990xf,Bergstrom:2001jj,Ullio:2002pj,Bottino:2004qi,Bertone:2004pz,Bringmann:2012ez,Ajello:2015mfa, DiMauro:2015tfa, Ackermann:2015tah}, as well as contributions from truly diffuse processes such as propagating ultra-high-energy cosmic rays~\citep{Loeb:2000aa,Kalashev:2007sn,Ahlers:2011sd,Murase:2012df,Taylor:2015rla} and structure formation shocks in clusters of galaxies~\citep{Murase:2008yt,Zandanel:2014pva}.
Given the potential wealth of information that can be extracted from the EGB, deciphering its constituents remains a high priority.     

Most recently, \emph{Fermi} presented a measurement of the EGB intensity from 100~MeV to 820~GeV~\citep{Ackermann:2014usa}. The total EGB intensity is the \textit{sum of all resolved point sources (PSs) and smooth isotropic emission}. The smooth emission, referred to as the Isotropic Gamma-Ray Background (IGRB), arises from \textit{PSs that are too faint to be resolved individually as well as other truly diffuse processes.}    It is also important to note that both the EGB and IGRB may be contaminated by cosmic rays that are mis-identified as gamma rays; this emission is expected to be smoothly distributed across the sky.
Of the known gamma-ray emitting PSs at high latitudes, which are captured by Fermi's 3FGL~\citep{Acero:2015hja} catalog from 0.1--300 GeV and the more recent 2FHL~\citep{Ackermann:2015uya} catalog from 50--2000 GeV, the dominant source class is blazars.

In this work, we use a novel analysis method, called Non-Poissonian Template Fitting (NPTF), to study the source populations that contribute to the EGB in a data-driven manner.  The method relies on photon-count statistics to illuminate the aggregate properties of a source population, even when its constituents are not individually resolvable~\citep{Malyshev:2011zi,Lee:2014mza, Lee:2015fea}.  This allows us to constrain the contribution of PSs to the EGB whose flux is too dim to be detected individually.  While at very low fluxes the NPTF also loses the ability to distinguish PSs from smooth emission, the threshold for PS detection is lower for the NPTF than it is for other techniques that rely on finding individually-significant sources.  This is because the NPTF only measures the aggregate properties of a PS population. 

Using the NPTF, we are able to recover, for the first time, the source-count distribution (\emph{e.g.}, flux distribution) for isotropically distributed PSs at high Galactic latitudes, as a function of energy from 1.89 GeV to 2~TeV.  This builds on previous studies that use related methods to obtain the source-count distributions in single energy bins from $\sim$2--12~GeV~\citep{Zechlin:2015wdz,Lee:2015fea} and from 50--2000~GeV~\citep{TheFermi-LAT:2015ykq}. 

The source-count distribution for a given astrophysical population convolves information about its cosmological evolution.  For a flat, non-expanding universe, a uniformly distributed population of galaxies has a differential source-count distribution $dN/dF \propto F^{-5/2}$, where $F$ is the source flux at Earth and $dN$ is the differential number of sources~\citep{Sandage}.  This is the well-known Euclidean limit.  However, the power-law index changes when one takes the standard $\Lambda$CDM cosmology and more realistic assumptions for the redshift evolution of source-dependent observables such as luminosity.  Therefore, the features of the source-count distribution---especially, its power-law indices and/or flux breaks---encode information about the number of source classes contributing to the EGB as well as their cosmological evolution.

These source-count distributions provide the keys for interpreting the GeV--TeV sky.  For example, they enable us to obtain the intensity spectrum for PSs, down to a certain flux threshold, as a function of energy.  We find that while the EGB is dominated by PSs, likely blazars, in the entire energy range from 1.89--2000~GeV, there is also room for other source classes, which contribute flux more diffusely, to produce a sizable fraction of the EGB.    
Our findings may therefore leave open the possibility that IceCube's PeV neutrinos~\citep{Aartsen:2013bka,Aartsen:2013jdh,Aartsen:2015knd,Aartsen:2015rwa} can be explained by $pp$ hadronic interactions in \emph{e.g.}, SFGs~\citep{Murase:2013rfa,Tamborra:2014xia,Ando:2015bva} or mAGN~\citep{Hooper:2016jls}, which---as we show in Sec.~\ref{sec:simulations}---show up as smooth isotropic emission under the NPTF.  Additionally, the high-energy source-count distributions allow us to make predictions for the number of blazars, which dominate the high-energy data, that will be resolved by upcoming TeV observatories such as the Cherenkov Telescope Array (CTA)~\citep{2011arXiv1111.2183C, Dubus:2012hm}.  
While our analysis does not let us conclusively identify the locations of these sources, we provide maps showing the locations on the sky where, statistically, there are most likely to be PSs.

This paper is organized as follows.  We begin in Sec.~\ref{sec:methodology} by reviewing the analysis methods.  Sec.~\ref{sec:simulations} then applies these methods to simulated sky maps.  We cannot stress the importance of these simulated data studies enough; they are crucial for proving the stability of the analysis methods and laying the foundation for the data results that follow.  Our data study is divided into two separate analyses for low (1.89--94.9~GeV) and high (50--2000~TeV) energies, described in Sec.~\ref{sec:lowenergy} and \ref{sec:highenergy}, respectively. The global fits to the full energy range, as well as their implications, are discussed in Sec.~\ref{sec:conclusions}.  Further details on the creation of the simulated data maps and supplementary analysis plots are provided in the Appendix.  The main results of this work are summarized in a few key figures.  In particular, the source-count distributions for the low and high-energy analyses are shown in Figs. \ref{fig:dndsdata}, \ref{fig:dndsdata_HE}, and \ref{fig:dnds_0_10}, respectively, while Fig. \ref{fig:global} presents a spectral fit to the PS intensity from 2 GeV to 2 TeV.

\section{Methodology} 
\label{sec:methodology}

In this paper, we make use of both Poissonian and non-Poissonian template-fitting techniques.  
Poissonian template fitting is a standard tool in astrophysics for decomposing a sky map into component ``templates" with different spatial morphologies.  The NPTF builds upon this technique by allowing for the addition of templates whose spatial morphology traces the distribution of a PS population, even if the exact position of the sources that make up that population are not known.  More precisely, in both template-fitting procedures one starts with a data set $d$ that consists of counts $n_p$ in each pixel $p$.\footnote{We will only work with a single energy bin at a time for simplicity, though in principle model parameters may be shared between energy bins.  In this case, the likelihood function over the full energy range may be written as the product of the likelihood functions in the energy sub-bins.}  One then fits a model ${\cal M}$ with parameters $\theta$ to the data by calculating the likelihood function 
\be
p(d |\theta, \mathcal{M}) = \prod_p p_{n_p}^{(p)}(\theta) \,,
\label{eq:likelihood}
\ee
where $p_{n_p}^{(p)}(\theta)$ denotes the probability of observing $n_p$ photons in pixel $p$ with model parameters $\theta$.

In Poissonian template fits, the probabilities $p_{n_p}^{(p)}(\theta)$ are Poisson distributions, with the model parameters $\theta$ only determining the means of the distributions.  That is, the mean expected number of photon counts at each pixel $p$ may be written as 
\es{}{
\mu_p(\theta) = \sum_\ell \mu_{p, \ell} (\theta) \,,
}
where the sum is over template components and $\mu_{p, \ell} (\theta)$ denotes the mean of the $\ell^\text{th}$ component for model parameters $\theta$.  The $\theta$ may parameterize, for example, the overall normalization of the templates or the shapes of the templates.  Then, the probability $p_{n_p}^{(p)}(\theta)$ is simply given by the Poisson distribution with mean $\mu_p$.  

In the NPTF, the situation is more complicated because we do not know where the PSs are.  As a result, if we want to calculate the probability of observing $n_p$ photons in a given pixel $p$, we must first calculate the probability that a PS (or a collections of PSs) exists in the vicinity of the pixel $p$, with a given flux (or set of fluxes).  Then, for that PS population, we calculate the probability of $n_p$ photons being produced in pixel $p$.  Convolving these two calculations together leads to distinctly non-Poissonian probabilities.  In particular, the probability distributions in the presence of unresolved PSs tend to be broader than Poisson distributions, if both distributions have the same mean expected number of photon counts.  The intuition behind this fact is that relative to a diffuse source, a collection of PSs leads to more ``hot'' pixels with many photons (where there are PSs) and more ``cold'' pixels with very few photons (where there are no PSs).  

\subsection{The Templates}
We include three Poissonian templates for (1) diffuse gamma-ray  emission in the Milky Way, assuming the \emph{Fermi} \texttt{p8r2} (\emph{gll\_iem\_v06.fits}) foreground model, (2) uniform emission from the \emph{Fermi} bubbles~\citep{Su:2010qj}, and (3) smooth  isotropic emission.  Each of these templates is associated with a single model parameter describing its overall normalization.  Variations to the choice of foreground model and bubbles template will be discussed in Sec.~\ref{sec:systematictests}.  

The model parameters specific to the isotropic-PS population enter into the source-count distribution $dN/dF$, which we characterize as a triply-broken power law:
\begin{widetext}
\be
\frac{dN}{dF} = A^\text{PS}_\text{iso} \,\begin{cases} 
 \left(  \frac{F}{F_{b,3}}\right)^{-n_4} &  F < F_{b,3}\\ 
\left( \frac{F}{F_{b,3}}\right)^{-n_3}  & F_{b,3} \leq F< F_{b,2}  \\ 
\left( \frac{F_{b,2}}{F_{b,3}}\right)^{-n_3} \left( \frac{F}{F_{b,2}}\right)^{-n_2}   & F_{b,2} \leq F< F_{b,1}  \\ 
\left( \frac{F_{b,2}}{F_{b,3}}\right)^{-n_3} \left( \frac{F_{b,1}}{F_{b,2}}\right)^{-n_2}  \left(  \frac{F}{F_{b,1}}\right)^{-n_1} & F_{b,1} \leq F \end{cases} \,.
\label{eq:sourcecount3break}
\ee
\end{widetext}
In particular, there are three breaks, $F_{b,1...3}$, along with four indices, $n_{1..4}$, and the overall normalization, $A^\text{PS}_\text{iso}$.\footnote{Note that the NPTF can also handle PS templates with non-trivial spatial distribution, as was done in the Inner Galaxy analyses in~\cite{Lee:2015fea,Linden:2016rcf}, though in this work we will only consider the isotropic-PS template.}
  The justification for a triply-broken power law is that $F_{b,1}$ designates the high-flux loss of sensitivity, beyond which $dN/dF$ cannot be probed because no sources exist with such high flux.  The break $F_{b,3}$ designates the low-flux sensitivity, below which PS emission cannot be distinguished from smooth emission.  This leaves $F_{b,2}$ to probe any physical break in the source-count distribution in the flux region where the NPTF can constrain it.  We have verified, however, that the results do not change significantly if the source-count distribution is fit by a doubly broken power law. 

It is important to stress that the photon-count probabilities are non-Poissonian in the presence of unresolved PSs because their locations are unknown.  Once we know where a PS is, we can fix its location and describe it through a Poissonian template with a free parameter for the overall normalization of the source.  However, even  resolved sources with known locations may be characterized by the non-Poissonian template if we do not also put down Poissonian templates at their locations.  This is the approach that we take throughout this work; that is, we model both the resolved (in the 3FGL and 2FHL catalogs) and unresolved PS populations through a single $dN/dF$ distribution, without individually specifying the locations of any sources.    

The point-spread function (PSF) must be properly accounted for in the template-fitting procedure.  The diffuse models are smoothed according to the PSF using the {\it Fermi} Science Tools routine \texttt{gtsrcmaps}.  The bubbles template is smoothed with a Gaussian approximation to the PSF, with width set to give the correct 68\% containment radius in each energy bin.  We follow the prescription developed in~\cite{Malyshev:2011zi} to account for the PSF in the calculation of the non-Poissonian photon-count probabilities; for this, we use the King function parameterization of the PSF provided with the instrument response function for the given data set.  In Sec.~\ref{sec:systematictests}, however, we show that consistent results are obtained when using a Gaussian approximation to the PSF instead.    
 
\subsection{Bayesian Fitting Procedure} 
 
The formalism developed in~\cite{Malyshev:2011zi,Lee:2014mza,Lee:2015fea} (see also ~\cite{Zechlin:2015wdz} and ~\cite{Linden:2016rcf}) is used to calculate the photon-count probability distributions in each pixel as a function of the Poissonian and non-Poissonian model parameters $\theta$.  Then, Bayesian techniques are used to construct a posterior distribution $p(\theta | d, {\cal M})$ for the parameters $\theta$ and the likelihood function in~\eqref{eq:likelihood}.  We construct the posterior distribution numerically using the \texttt{MultiNest} package~\citep{Feroz:2008xx,Buchner:2014nha} with 700 live points, importance nested sampling and constant efficiency mode disabled, and sampling efficiency set for model-evidence evaluation.  
 
 All prior distributions are taken to be flat except for $A_\text{iso}^\text{PS}$, which is taken to be log-flat.  The prior ranges for the model parameters are shown in Tab.~\ref{tab:priors}.  
 \begin{table*}[t]
\renewcommand{\arraystretch}{1.4}
\setlength{\tabcolsep}{5.2pt}
\begin{center}
\begin{tabular}{  c | c ||| c | c ||c |c}
Parameter	 & Prior Range &Parameter	& Prior Range & Parameter	& Prior Range   \Tstrut\Bstrut	\\   
\hline 
$A_\text{diff}$  & $[0, 2]$  & $\log_{10}A_\text{iso}^{\text{PS}}$  & [-10, 20] & $n_1 $ & $[2.05, 5]$\\
$A_\text{bub}$ & $[0, 2]$  & $S_{b,3}$ & $[0.1, 1]$ ph  &$n_2$ & $[1.0, 3.5]$ \Tstrut\Bstrut \\ 
$A_\text{iso}$  & $[0, 2]$  &$S_{b,2}$ & $[1, 30]$ ph &$n_3 $ & $[1.0, 3.5]$\Tstrut\Bstrut \\ 
& &$S_{b,1}$ & $[30, 2 \times S_{b,\text{max}}]$ ph &$n_4$ & $[-1.99, 1.99]$ \Tstrut\Bstrut \\
\end{tabular}
\end{center}
\caption{
Parameters and associated prior ranges for the templates used in the NPTF.   The priors on the breaks $S_{b,1...3}$ are given in terms of counts, defined relative to the mean exposure $\langle \overbar{ \mathcal{E}}^{(p)} \rangle$ in the ROI.  $S_\text{b,max}$ is the maximum number of photons in the 3FGL~\citep{Acero:2015hja} (2FHL~\citep{Ackermann:2015uya}) catalog in the energy bin of interest for the low (high)-energy analysis.  Note that all prior distributions are linear-flat, except for that of $A_\text{iso}^\text{PS}$, which is log-flat.  The baseline normalizations of the $A_\ell$ are described in the text.  }
\label{tab:priors}
\end{table*}  
These prior ranges successfully reconstruct the source-count distributions of simulated data sets, as discussed in Sec.~\ref{sec:simulations}.  Variations to the prior ranges in Tab.~\ref{tab:priors} are considered in Sec.~\ref{sec:systematictests}.

In Tab.~\ref{tab:priors}, the parameter $A_\ell$ denotes the normalization of the $\ell^\text{th}$ template, which is defined in terms of a baseline value.    The baseline value is obtained by first performing a Poissonian template fit over 17 (10) log-spaced energy sub-bins between $1.89$ and $94.9$ GeV (50 and 2000 GeV) for the low (high)-energy analysis.  When this procedure is applied to the low-energy analysis where the known PSs are very bright, we mask the 300 brightest and most variable 3FGL sources, at 95\% containment.  At both high and low energies, we include a PS model constructed from the 3FGL catalog.\footnote{Importantly, we do not include the PS model or mask any PSs in the NPTF analyses.}  The fitting procedure then allows us to recover the normalizations for the diffuse background, bubbles, and isotropic templates in each energy sub-bin.

The actual energy bins used for the NPTF studies presented in this work are larger than the sub-bins described above.  Therefore, the baseline normalizations used to define the NPTF priors in the energy range $[E_\text{min}, E_\text{max}]$ are found by applying the best-fit Poissonian normalizations from the individual sub-bins to the corresponding templates, which are then combined.\footnote{In practice, however, this prescription for combining the templates between energy sub-bins does not significantly affect our results.  
}
  Therefore, $A_\ell = 1$ in the NPTF analysis implies that the normalization of the $\ell^\text{th}$ template is the same as that computed from the Poissonian scans.  The benefit of this approach is that it allows one to keep track of how the individual Poissonian templates react to the addition of non-Poissonian ones.  For example, the normalization of the diffuse-background template should remain consistent between a standard template analysis, where PSs are accounted for by the 3FGL model, and the NPTF analysis, where PSs are accounted for by the non-Poissonian template; indeed, we find that is the case in all of the analyses we perform. 
  
\subsection{Exposure Correction}  
  
While the source-count distribution $dN/dF$ is defined in terms of flux, $F$, with units of $\text{ph}/\text{cm}^2/\text{s}$, the priors for the breaks in Tab.~\ref{tab:priors} are written in terms of counts, $S_{b,1...3}$.  To convert from flux to counts, we multiply by the exposure of the instrument, with units of cm$^2$ s.  However, the relation between flux and counts is complicated by the fact that the exposure of the instrument varies both with energy and position in the sky. Below, we describe how we deal with both complications, starting first with the energy dependence.  
  
The exposure map in the $i^\text{th}$ energy sub-bin is given by $\mathcal{E}^{(p)}_i$.  To construct the exposure map $\overbar{ \mathcal{E}}^{(p)}$ in the larger energy range from $[E_\text{min}, E_\text{max}]$, which contains multiple energy sub-bins, we average over the $\mathcal{E}^{(p)}_i$ of the individual sub-bins, weighted by a power-law spectrum $dN/dE \sim E^{-2.2}$, as this is generally consistent with the isotropic spectrum over most of our energy range.   
This procedure introduces a source of systematic uncertainty in going from counts to flux, as not all source components have an energy spectrum consistent with this spectrum.  However, we have checked that variations to this procedure---such as weighting the exposures in the sub-bins by power laws of the form $E^{-n}$, with $n$ varying between $1$ and $3$---do not significantly change the results.\footnote{We have also checked that weighting the exposures in the sub-bins by the intensities  computed from the Poissonian template scans gives consistent results.}   The weighting procedure is most important at very high energies, on the order of hundreds of GeV, where the exposure map varies strongly across the energy sub-bins.   
 
The breaks $S_{b,1...3}$ in Tab.~\ref{tab:priors}, with units of counts, are defined relative to the mean exposure $\langle \overbar{ \mathcal{E}}^{(p)} \rangle$, averaged over all pixels in the region of interest (ROI).  Because the NPTF is performed at the level of counts and not flux, we must also convert the source-count distribution $dN/dF$ to a distribution $dN^{(p)}/dS$, which is unique to each pixel $p$:
 \es{dNdFdS}{
 {dN^{(p)} \over dS} (S) = {1 \over \overbar{\mathcal{E}}^{(p)}} \left. {dN \over dF} \right\vert_{F = S / \overbar{\mathcal{E}}^{(p)}} \,.
 }   
 Then, the photon-count probability distribution must be computed uniquely at each pixel.  In practice, however, it is numerically expensive to perform this procedure for every pixel in the ROI.  Instead, we follow~\cite{Zechlin:2015wdz} and break the ROI up into $N_\text{exp}$ regions by exposure.  Within each region, we assume that all pixels have the same exposure, which is taken to be the mean over all pixels in the sub-region.  The likelihood function is then computed uniquely in each exposure region, and the total likelihood function for the ROI is the product of the likelihoods across exposure regions.  In practice, we find that our results are convergent for $N_\text{exp} \geq 10$.  We will take $N_\text{exp} = 15$ throughout this work, though we have checked that our main results are consistent with those found using $N_\text{exp} = 25$.
 
 \subsection{Data Samples}
 \label{sec:data}
 
 We run the NPTF analysis, as described above, on \emph{Fermi} data, considering low (1.89--94.9~GeV) and high (50--2000~GeV) energies separately.  The former is discussed in Sec.~\ref{sec:lowenergy}, while the latter is the focus of Sec.~\ref{sec:highenergy}.  The primary difference between the data sets used in these studies is the data-quality cuts; moving to higher energies requires loosening these criteria to avoid being limited by statistics.  The overlap in energy between the two studies allows us to compare the consistency of the results when transitioning between analyses.  

The low-energy study uses the Pass 8 \emph{Fermi} data from $\sim$August 4, 2008 to June 3, 2015.  The primary studies use the top quartile of the \emph{ultracleanveto} event class (PSF3) as ranked by angular resolution, although the top-three quartiles (PSF1--3) are also studied separately.\footnote{The PSF quartiles indicate the quality of the reconstructed photon direction, with `PSF3' being the best and `PSF0' being the worst.}  As a systematic check, we also consider the top-three quartiles of {\it source} data.  The \emph{ultracleanveto} event class is the cleanest event class released with the Pass 8 data and is recommended for studies of the EGB.  However, the {\it source} event class has an enhanced exposure and thus may be advantageous at high energies where statistics become limited.  On the other hand, we expect the {\it source} data to have additional cosmic-ray contamination relative to the \emph{ultracleanveto} data.    

The recommended\footnote{\url{http://fermi.gsfc.nasa.gov/ssc/data/analysis/documentation/Cicerone/Cicerone_Data_Exploration/Data_preparation.html}} event quality cuts are applied, requiring that all photons have a zenith angle less than $90^\circ$ and satisfy ``\texttt{DATA\_QUAL==1 \&\& LAT\_CONFIG==1 \&\& ABS(ROCK\_ANGLE)$ < 52$}.''   A HEALPix~\citep{Gorski:2004by} pixelation is used with \emph{nside}=128, which corresponds to pixels roughly $0.5^\circ$ to a side.   We consider four separate energy bins:  $[1.89, 4.75]$, $[4.75, 11.9]$, $[11.9, 30]$, and $[30, 94.9]$~GeV. 

In the low-energy analysis with {\it ultracleanveto} PSF3 data, the means of the weighted exposure maps in the four increasing energy bins are $[5.78 \times 10^{10}, 5.40 \times 10^{10}, 5.18 \times 10^{10}, 5.38 \times 10^{10}]$ cm$^2$ s over the region of interest with $|b| \geq 30^\circ$.  The 68\% containment radii for the PSF, averaged over the isotropic spectra in the energy sub-bins, are $[0.20, 0.11, 0.06, 0.04]$ degrees.  Going to PSF1--3 data, the exposures increase to $[1.69 \times 10^{11}, 1.66 \times 10^{11}, 1.63 \times 10^{11}, 1.67 \times 10^{11}]$ cm$^2$ s, while the 68\% containment radii of the PSF degrade to $[0.32, 0.16, 0.10, 0.08]$ degrees.  Going to {\it source} data with PSF1--3, the exposures ($[2.10 \times 10^{11}, 2.07 \times 10^{11}, 2.07 \times 10^{11},2.15 \times 10^{11}]$ cm$^2$ s) increase further, while the 68\% containment radii ($[0.32,0.16,0.10,0.08]$ degrees) are essentially the same as in the {\it ultracleanveto} case.

The high-energy analysis uses the Pass 8 \emph{Fermi} data from $\sim$August 4, 2008 to May 2, 2016 and all PSF quartiles of either the \emph{ultracleanveto} or \emph{source} event class.  The ROI is also extended to $|b| >10^\circ$.  We include more data in the high-energy analysis as there are far fewer photons than at lower energies.  We employ the recommended event-quality cuts as in the low-energy analysis and also choose \emph{nside}=128 HEALPix pixelation.  Results are presented for the three energy bins $[50, 151], [151, 457]$, and $[457, 2000]$~GeV.  With {\it ultracleanveto} data, the weighted exposures in the energy bins are $[2.48 \times 10^{11}, 2.31 \times 10^{11}, 1.69 \times 10^{11}]$ cm$^2$ s, while with {\it source} data the exposures become $[3.23 \times 10^{11}, 3.20 \times 10^{11}, 2.87 \times 10^{11}]$ cm$^2$ s.  For both data sets, the 68\% containment radii are approximately $[0.14, 0.12, 0.11]$ degrees.  We will also discuss results of analyses performed over a single wide-energy bin from $[50, 2000]$~GeV.

\section{Simulated Data Studies}
\label{sec:simulations}

To study the behavior of the NPTF, we apply it to simulated data sets of the gamma-ray sky.  These results are crucial both for understanding systematics associated with the NPTF as well as for interpreting the results of the NPTF in terms of evidence for or against the existence of these source populations. 

A simulated data map can be created starting from a particular source population that contributes to the EGB.  Using a theory model for the energy spectrum and luminosity function, the source-count distribution for that population can be derived in a specified energy range---see Appendix~\ref{app:sims} for further details on this procedure.  The appropriate number of sources is then drawn from this function and randomly distributed across the sky, with counts chosen to follow the intensity spectrum. Sources are then smeared with the appropriate Gaussian PSF to mimic the desired {\it Fermi} data set
bin-wise in energy, and Poisson counts are drawn to obtain the simulated map for the population.  This is then combined with the simulated contribution of the \texttt{p8r2} foreground model and the \emph{Fermi} bubbles, whose normalizations are determined from the Poissonian template fits to the real data, as described in Sec.~\ref{sec:methodology}.  

For most of this section, we simulate data corresponding to the PSF3 event type (best PSF quartile) of the \emph{ultracleanveto} event class and focus on the following four energy bins: [1.89, 4.75], [4.75, 11.9], [11.9, 30], and [30, 94.9]~GeV.\footnote{Potential systematic issues arising from going to lower energies are discussed in Appendix~\ref{app:verylowenergies}.}  However, we also simulate data corresponding to the PSF1--3 (top 3 PSF quartiles) instrument response function to illustrate potential advantages in going to the more inclusive data set, albeit with a slightly worse PSF.   Once the simulated data maps are created, we run them through the NPTF analysis pipeline.  First, we analyze the case where either blazars or SFGs fully account for the EGB, and then we analyze a perhaps more realistic scenario where both populations contribute significantly to the flux.  The particular blazar and SFG models used here are merely meant for illustration.  They are chosen as examples that span the range of possibilities between smooth and PS isotropic contributions.  As mAGN are fainter and more numerous then blazars, they likely act similarly to SFGs in the context of the NPTF and so we do not consider them separately here.  A detailed analysis of how the NPTF responds to the broader class of theoretical models for these source classes is beyond the scope of this work.

\subsection{Blazars}

Active galactic nuclei (AGN) are the highly luminous central regions of galaxies where emission is dominated by accretion onto a supermassive black hole~\citep{Urry:1995mg}.  If the black hole is spinning, then relativistic jets may also form.  Blazars are a subclass of AGN in which the jet is oriented within $14^\circ$ of the line-of-sight~\citep{Angel:1980}.  The spectral energy distribution of these objects is bimodal with a peak in the ultraviolet due to synchrotron radiation of electrons in the jet, and another peak in the gamma band from inverse Compton scattering of the same electrons \citep{Fossati:1998zn,Ghisellini:1998it,Ghisellini:2009fj,Boettcher:2013wxa}.  There is also the possibility of a hadronic contribution to blazar gamma-ray spectra, although this is likely to be sub-dominant~\citep{Tavecchio:2013fwa,Cerruti:2014iwa,Zdziarski:2015rsa}. Blazars may be further classified as either BL Lacertae (BL~Lacs) or Flat Spectrum Radio Quasars (FSRQs), which are characterized by the absence or presence of broad optical/ultraviolet emission lines, respectively.  

Before \emph{Fermi}, few blazars had been identified in gamma rays, and to estimate the size of this population, one had to extrapolate based on those observed at lower frequencies.  However, \emph{Fermi} brought the discovery of many more blazars in the gamma-ray band, making it possible to study their properties directly~\citep{Collaboration:2010gqa, Ajello:2011zi,Ajello:2013lka, DiMauro:2013zfa, Giommi:2015ela, Padovani:2014cha}.  Most recently, 403 blazars (with $|b| > 15^\circ$) from the First LAT AGN Catalog~\citep{Abdo:2010ge} were studied~\citep{Ajello:2015mfa}.  FSRQs and BL~Lacs were considered together in the same sample to improve statistics.  We use the best-fit luminosity and spectral energy distributions given in~\cite{Ajello:2015mfa} (specifically, the luminosity-dependent density evolution, or LDDE, scenario) to model the blazar component in our simulated data and refer to it as the ``Blazar--1'' model.  Alternatively, we also consider BL~Lacs and FSRQs separately, adding up their respective contributions using the LDDE1 model from~\cite{Ajello:2013lka} and the LDDE model from~\cite{Ajello:2011zi}, which we refer to as the ``Blazar--2'' model. This model predicts a much flatter source-count distribution below the \emph{Fermi} detection threshold, with more low-flux sources. The two source-count models approximately bracket the current theoretical uncertainty in the faint-end slope of blazars, and we use them to study the response of different blazar models to the NPTF, although this is meant to be purely illustrative and by no means exhaustive.

\begin{figure*}[tbhp] 
   \centering
   \includegraphics[width=0.9\textwidth]{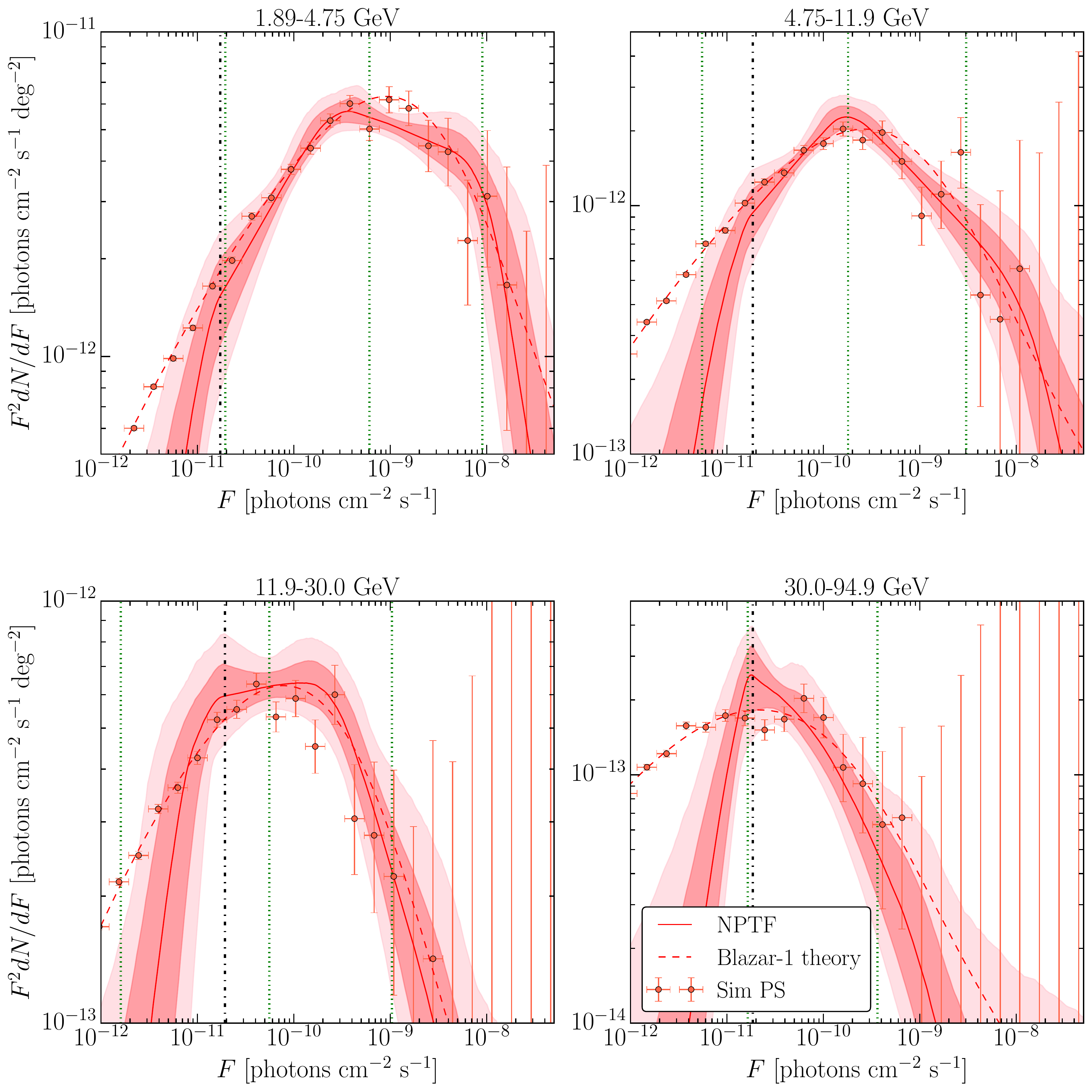} 
   \caption{The source-count distribution of the isotropic-PS population obtained by running the NPTF on simulated data in which the EGB arises from the Blazar--1 model~\citep{Ajello:2015mfa}.  Results are presented for the four energy bins considered.  The source-count distribution of the input blazar model (dashed red) matches the posterior for the isotropic PSs (68 and 95\% credible intervals, constructed pointwise, shaded in red) well at fluxes corresponding to counts above $\sim$1 photon (vertical, dot-dashed black).  The vertical dotted green lines indicate the fluxes at which 90\%, 50\%, and 10\% of the flux is accounted for, on average, by sources with larger flux (from left to right, respectively).  The red points show the histogram of the simulated PSs, with 68\% Poisson error bars (vertical).  Note that the NPTF loses sensitivity to sources contributing less than $\sim$1 photon; as a result, the NPTF result does not match the simulated data well below the dot-dashed black line.  }
   \label{fig:bl1dnds}
\end{figure*}

Figure~\ref{fig:bl1dnds} shows the best-fit source-count distributions recovered when the NPTF analysis is run on the Blazar--1 simulated data map, assuming the PSF3 instrument response function.  In each panel, the dark (light) red band is the 68\% (95\%) credible interval for the isotropic-PS source-count distribution as recovered from the posterior and  constructed pointwise in flux. The red line shows the median source-count distribution, constructed in the same way. The dashed red curve, on the other hand, indicates the source-count distribution of the blazar model used to generate the simulated data.  A flux histogram of the simulated PSs for the particular realization shown here is given by the red points, with vertical error bars indicating the 68\% credible interval associated with Poisson counting statistics on the number of sources in that bin.  Notice that these error bars become large at high fluxes because there are very few sources per flux bin. 

In general, the reconstructed source-count distribution is in good agreement with the input source-count distribution at intermediate fluxes, with uncertainties becoming large at low and high fluxes.  At high flux, this is due to the fact that it is unlikely to draw a bright source from the underlying source-count distribution.  At low fluxes, it is difficult to distinguish PS emission from genuinely isotropic emission.  To illustrate this point, we also mark the flux that corresponds to a single photon on average (in the particular energy range, region-of-interest, and event class) with the vertical dot-dashed black line.  At fluxes corresponding to counts near or below $\sim$1 photon, it is difficult to distinguish PS emission from smooth emission with the NPTF, as evidenced by the growing uncertainties.  In this low-flux regime, we do not expect that the NPTF will be able to fully recover the properties of the input source-count distribution.  

The vertical dotted green lines in Fig.~\ref{fig:bl1dnds} correspond to the fluxes above which 90\%, 50\%, and 10\% (from left to right) of the photon counts are accounted for, on average, by sources with larger flux.  Note that in the lowest energy bin, 90\% of the flux arises from sources that contribute more than one photon.  Moving towards higher energies, a larger fraction of the flux arises from sources that contribute less than one photon.  In all energy bins, more than 50\% of the flux is accounted for by sources that contribute more than a single photon each.

\begin{figure*}[tb] 
   \centering
  $ \begin{array}{cc}
   \scalebox{0.43}{\includegraphics{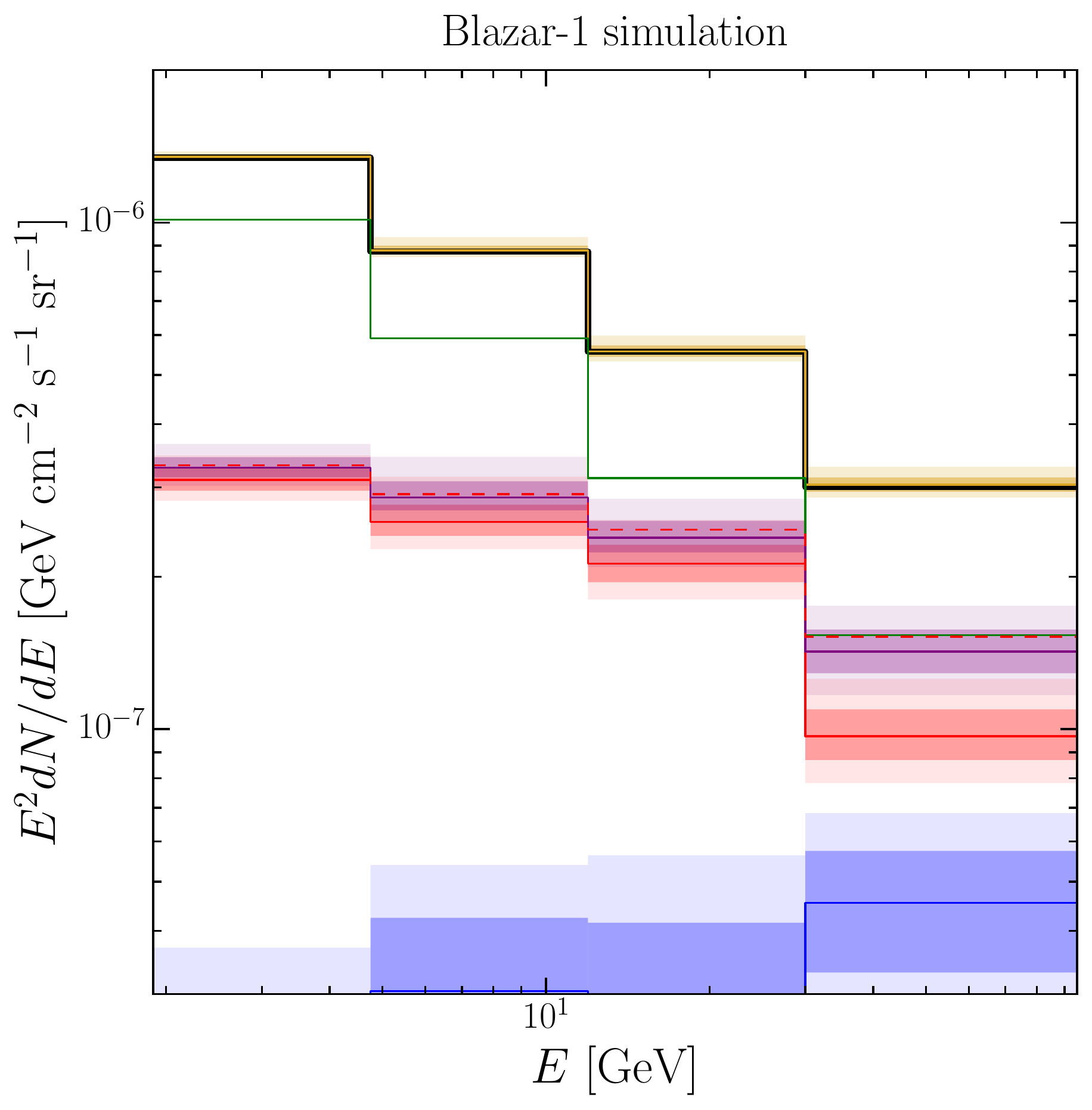}} &	\scalebox{0.43}{\includegraphics{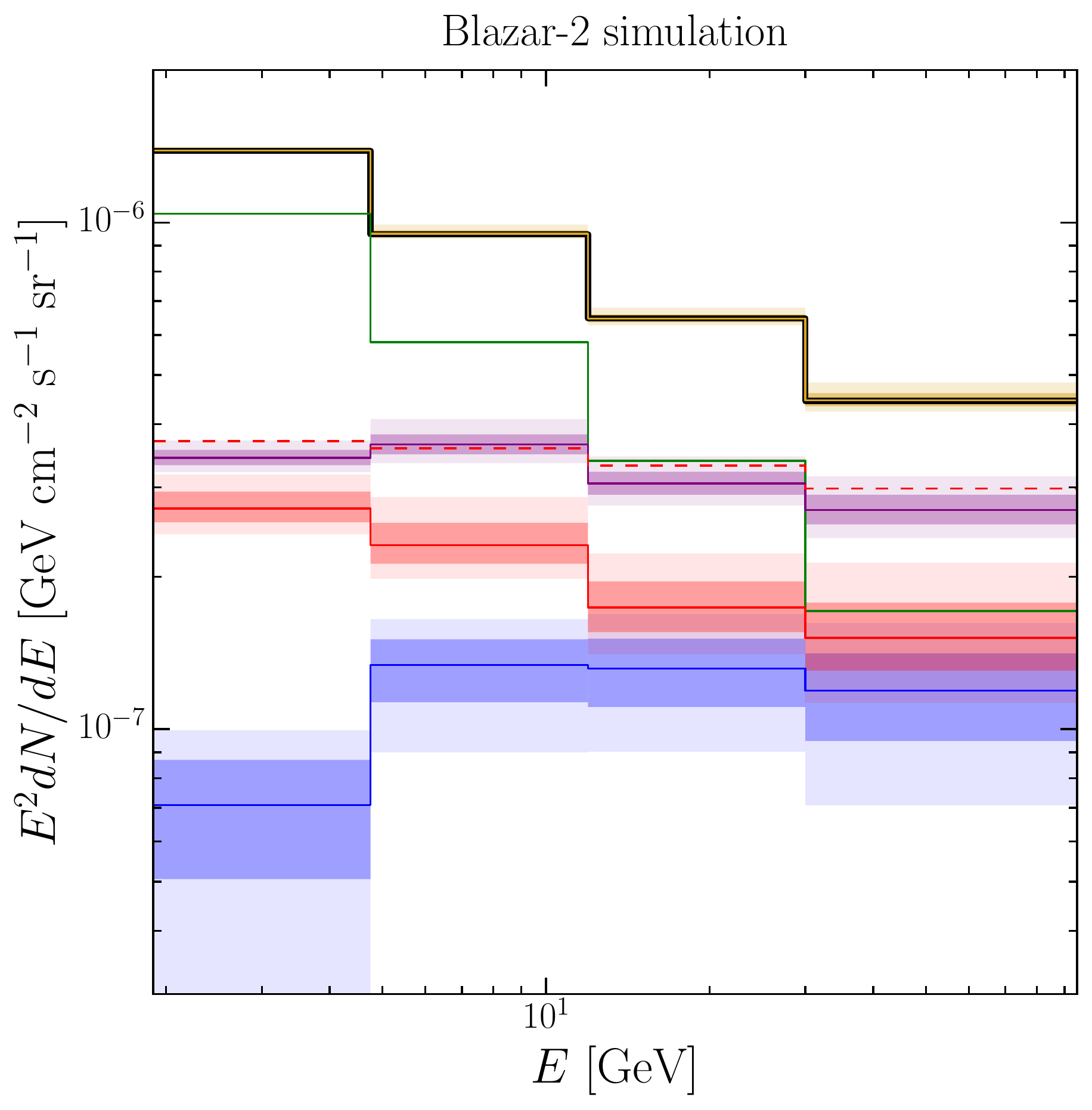}} \\
   \scalebox{0.43}{\includegraphics{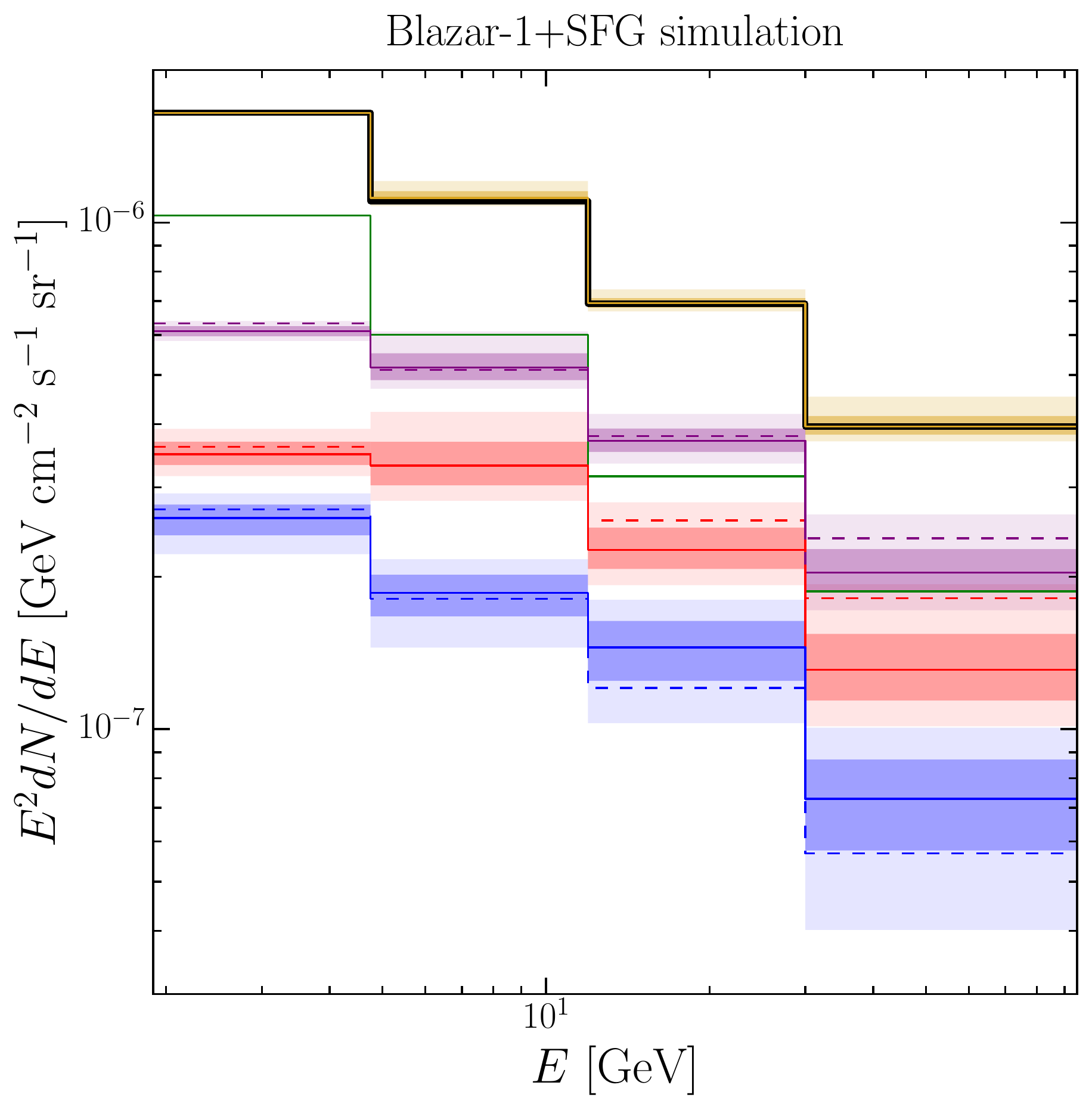}} &	\scalebox{0.43}{\includegraphics{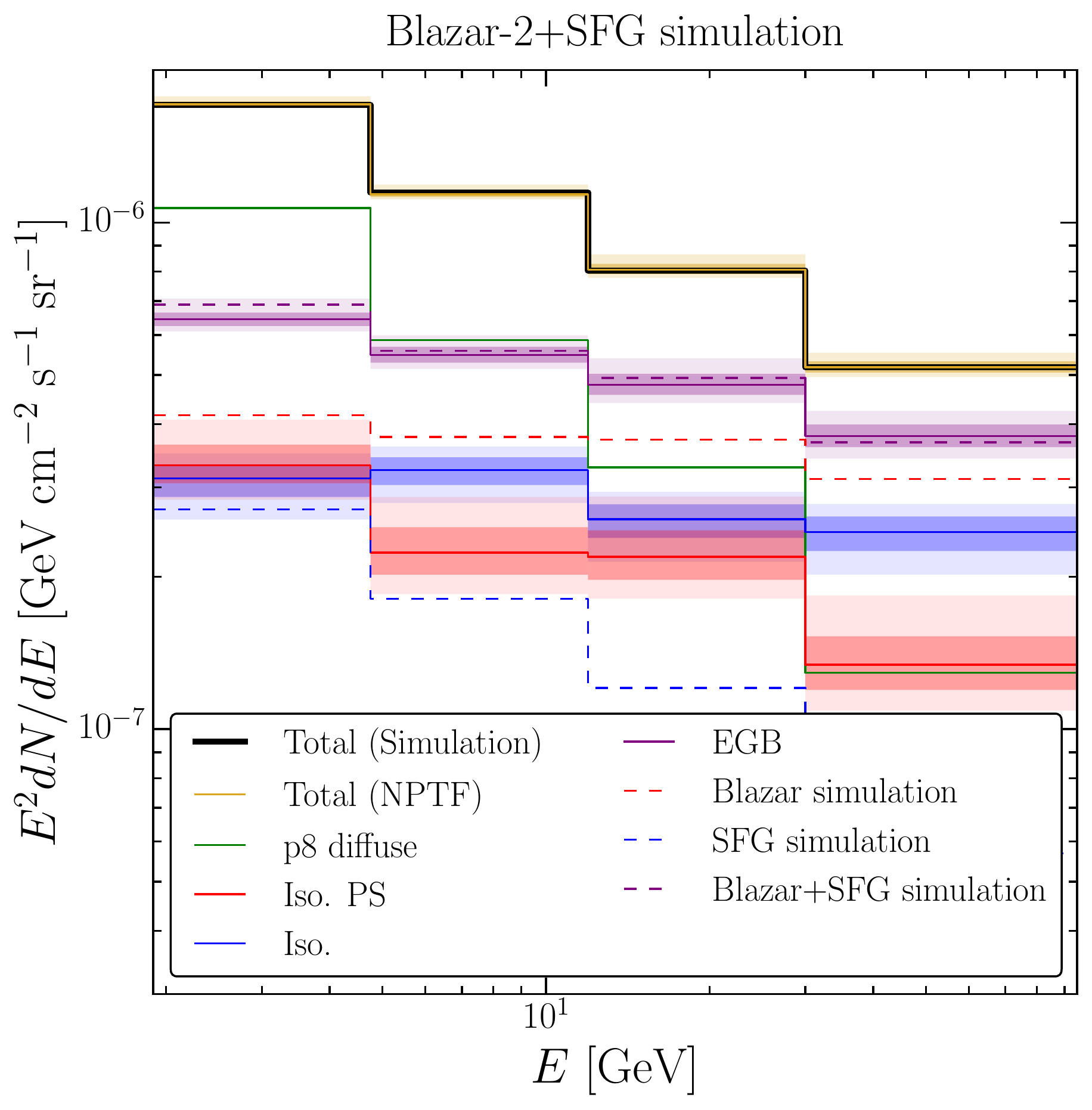}}
   \end{array}$
   \caption{The energy spectra for the isotropic and isotropic-PS templates in each energy bin considered; the 68 and 95\% credible intervals, constructed from the posterior distributions, are shown in blue and red, respectively.  The top row represents the results for simulated data, with {\it ultracleanveto} PSF3 instrument response function, in which the EGB consists of only Blazar--1 sources~\citep{Ajello:2015mfa} \emph{(top left)} or Blazar--2 sources~\citep{Ajello:2011zi, Ajello:2013lka} \emph{(top right)}.  The bottom row  shows the same results, except when SFGs~\citep{Tamborra:2014xia} are also included in the simulation.  The simulated spectrum for blazars (SFGs) is shown in dashed red (blue).  For the Blazar--1 model, the isotropic-PS template absorbs almost the entirety of the flux.   For the Blazar--2 model, both smooth and PS isotropic components absorb flux, but their sum (EGB, purple band) is consistent with the input.  When SFGs are also included, more emission is absorbed by the smooth isotropic template; however, the total emission absorbed by the smooth and PS isotropic templates is consistent with the expected total of SFG and blazar intensities.  The spectrum for Galactic diffuse emission is shown by the green line in each panel (median only).  The sum of all template emission (yellow band) agrees with the total spectrum of the simulated data.  Note that the energy spectrum of the bubbles template is not shown.}
   \label{fig:blESpec}
\end{figure*}

The corresponding energy spectra for the various templates are shown on the top left panel of Fig.~\ref{fig:blESpec}.  As is evident, these blazars show up as PSs under the NPTF; indeed, the smooth isotropic flux (blue) is sub-dominant in each energy bin.  
Overlaid in dashed red is the spectrum for the simulated Blazar--1 sources.  The sum of the smooth and PS isotropic components---which is simply the EGB intensity---is consistent with the simulated spectrum for the blazar model.  The green curve shows the median of the posterior for the galactic diffuse model spectrum.  The energy spectrum of the diffuse model is softer than that for blazars, so that the diffuse model dominates more at low energies than at high.  The sum of the components (yellow band) is consistent with the total flux in the simulated data (black lines) at 68\% confidence.

As a contrasting example, we also simulate the Blazar--2 model, which predicts more low-flux sources than the previous example we considered.  The best-fit source-count distributions for the Blazar--2 simulated maps are shown in Fig.~\ref{fig:bl2dnds}. Once again, we see good agreement between the input data and the recovered source-count distribution above the single-photon sensitivity threshold. In this case, however, the reference model predicts a larger fraction of flux coming from sources below this threshold.  For example, about 50\% of the flux comes from sub-single photon sources in the second energy bin, and this fraction only increases further at higher energies.  The corresponding energy spectrum is shown in the top right panel of Fig.~\ref{fig:blESpec}.  As expected, an increasing amount of flux is absorbed by the Poissonian isotropic template.  However, the EGB spectrum, shown by the purple band, is still consistent with the input spectrum for the Blazar--2 model.
\afterpage{
\begin{figure*}[htbp] 
   \centering
   \includegraphics[width=0.9\textwidth]{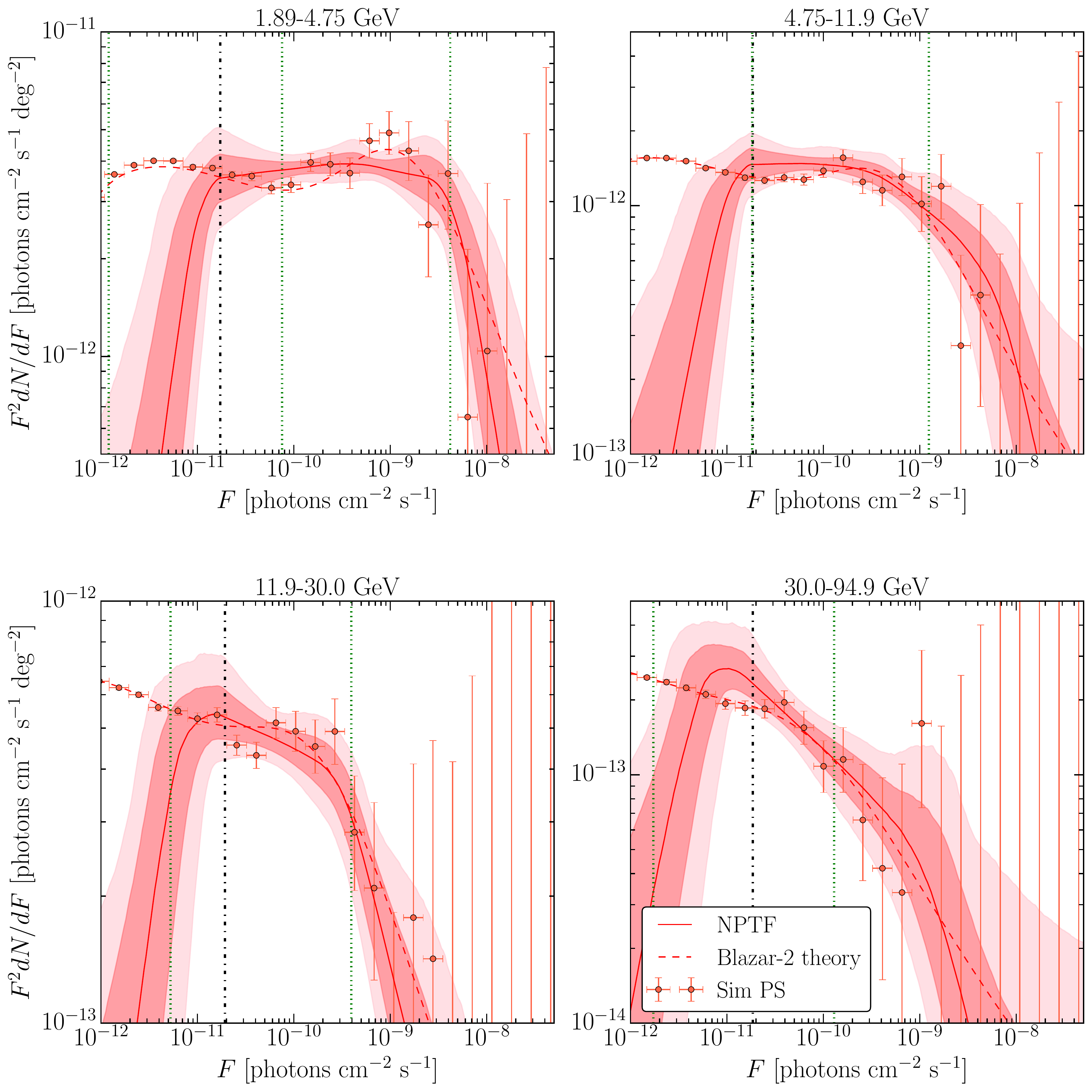} 
   \caption{Same as Fig.~\ref{fig:bl1dnds}, except for the Blazar--2 model~\citep{Ajello:2011zi, Ajello:2013lka}.}
   \label{fig:bl2dnds}
\end{figure*}
}

To further quantify the ability of the NPTF to reconstruct the blazar flux as PS emission, it is convenient to consider the ratios $I_\text{iso}^\text{PS} / I_\text{blazar-sim}$ in each energy bin, where $I_\text{iso}^\text{PS}$ is the PS intensity found by the NPTF and $ I_\text{blazar-sim}$ is the blazar intensity in the simulation.  For the Blazar--1 model, we find\footnote{Throughout this work, best-fit values indicate the 16$^\text{th}$, 50$^\text{th}$, and 84$^\text{th}$ percentiles of the appropriate posterior probability distributions. } 
\begin{equation}
\frac{I_\text{iso}^\text{PS}}{ I_\text{blazar-sim}} = [0.94_{-0.04}^{+0.05}, 0.88_{-0.05}^{+0.07}, 0.86_{-0.07}^{+0.08}, 0.64_{-0.07}^{+0.08}] \nonumber
\end{equation}
in each of the four respective energy bins, while for the Blazar--2 model, we find
\begin{equation}
\frac{I_\text{iso}^\text{PS}}{ I_\text{blazar-sim}} = [0.74_{-0.04}^{+0.06}, 0.64_{-0.05}^{+0.07}, 0.53_{-0.06}^{+0.07}, 0.51_{-0.07}^{+0.09}]\,, \nonumber
\end{equation}
 for the particular Monte Carlo realizations shown.\footnote{Different Monte Carlo realizations are found to induce variations consistent with the quoted statistical uncertainties, generally on the order of 5\%.}
For the Blazar--2 scenario, more flux goes into smooth isotropic emission, which is why the PS fractions are correspondingly smaller in each energy bin.  Note that, in both scenarios, the fraction of the blazar flux absorbed by the PS template decreases at higher energies, where the photon counts become less numerous and a higher fraction of the blazar flux is generated by sub-threshold sources.  As a result, the intensities $I_\text{iso}^\text{PS}$ should be interpreted as lower bounds on the blazar flux; this intuition is validated by the fact that the ratios  $I_\text{iso}^\text{PS} / I_\text{blazar-sim}$ tend to be less than unity.  

Next, we explore whether including more quartiles of the {\it ultracleanveto} data, as ranked by PSF, increases our ability to reconstruct the blazar flux as PSs under the NPTF.  When including more quartiles of data, there are two competing effects that determine our ability to constrain the PS flux: on the one hand,  we increase the effective area, but on the other hand, we worsen the angular resolution of the data set.  We investigate these effects by repeating the Monte Carlo tests described above using the PSF1--3 instrument response function.  The results of the PSF1--3 tests are described in Appendix~\ref{app:ucvsims}, and here we simply quote the fractions 
\begin{equation}
\frac{ I_\text{iso}^\text{PS}}{ I_\text{blazar-sim}} = [0.78_{-0.05}^{+0.06}, 0.81_{-0.06}^{+0.07}, 0.72_{-0.06}^{+0.06}, 0.57_{-0.05}^{+0.06}]\,  \nonumber
 \end{equation}
 for a generic realization of the Monte Carlo simulations for the Blazar--2 model.  The PSF1--3 event type increases our ability to distinguish between the blazar emission and smooth emission compared to the PSF3 event type.

\subsection{Star-Forming Galaxies}

Star-forming galaxies (SFGs) like our own Milky Way are individually fainter, though much more numerous, than blazars.  The modeling of SFGs in the gamma-ray band is highly uncertain, as \emph{Fermi} has only detected eight SFGs thus far~\citep{Fornasa:2015qua}.  However, SFGs could still contribute a sizable fraction of the total flux observed by \emph{Fermi}.  Even though SFGs are PSs, their flux is expected to be dominated by a large population of dim sources degenerate with smooth isotropic emission.  Under the NPTF, therefore, we expect that the majority of their  emission will be absorbed by the smooth isotropic template.  
To illustrate this point, we simulate SFGs using the luminosity function and energy spectrum from~\cite{Tamborra:2014xia}.  In that work, input from infrared wavelengths was used to construct a model for the infrared flux from SFGs.
Then, a scaling relation was used to convert from infrared to gamma-ray luminosities.  The contributions from quiescent and starburst SFGs were considered separately, along with SFGs that host an AGN.  Note, however, that other models predict less emission from SFGs than this particular case---see \emph{e.g.},~\cite{Makiya:2010zt,Inoue:2011bm,Ackermann:2012vca}.

The results of the SFG-only simulations are described in Appendix~\ref{app:sfgsims}.  We find that while the NPTF does detect a small PS component in the first few energy bins, as the result of a few SFGs above the sensitivity threshold of the NPTF in those energy bins, by far most of the SFG emission is detected as smooth isotropic emission, with the ratio $I_\text{iso}^\text{PS} / I_\text{iso}\lesssim 1/100$ in all energy bins, where $I_\text{iso}$ is the intensity of smooth isotropic emission.  Moreover, the intensity $I_\text{iso}$ is consistent with the simulated EGB (SFG flux) in all energy bins, at 68\% confidence.

\subsection{Blazar and SFG combination}
\begin{figure*}[t] 
   	\begin{center}$
	\begin{array}{cc}
	\scalebox{0.43}{\includegraphics{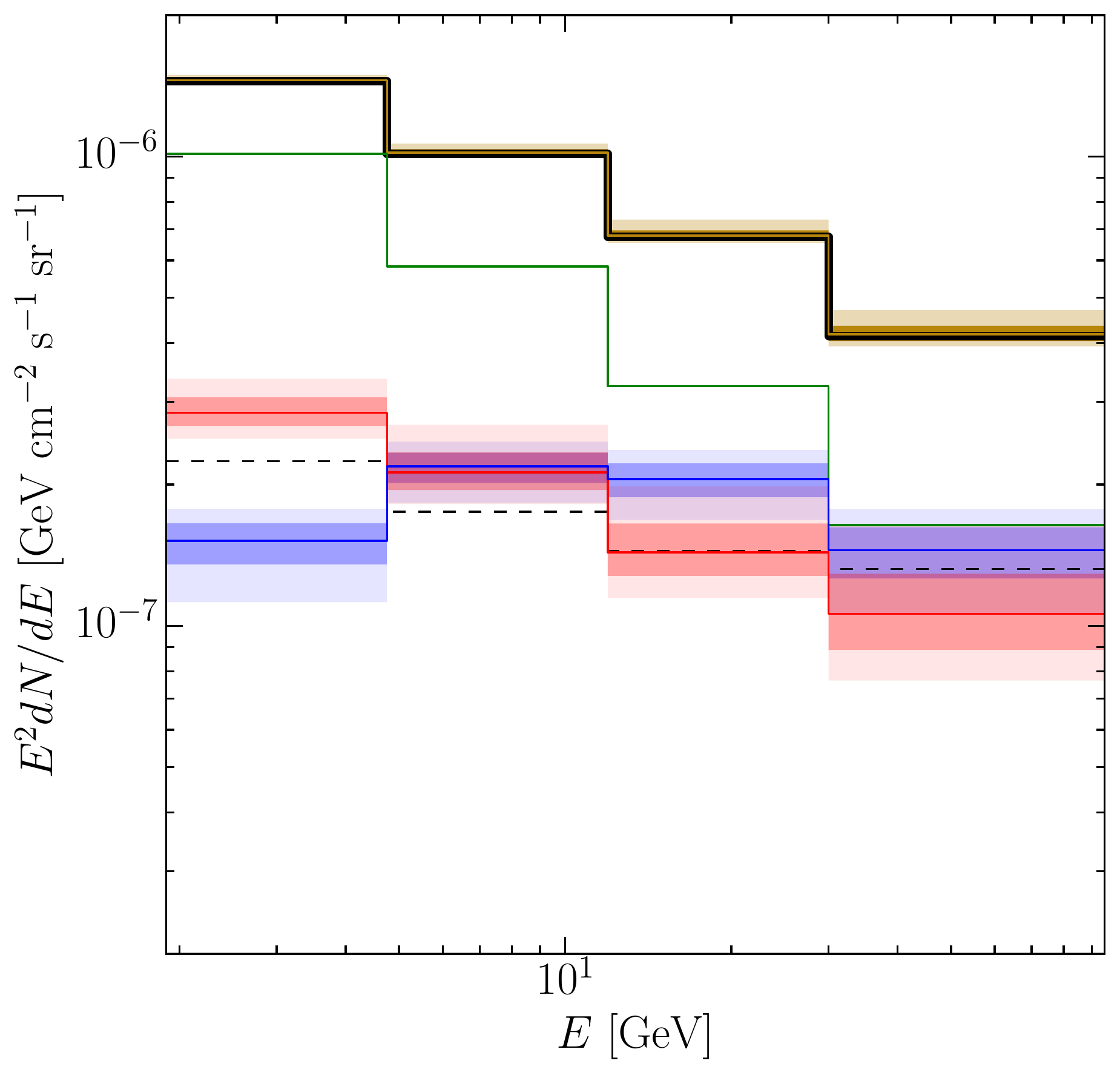}} &\scalebox{0.43}{\includegraphics{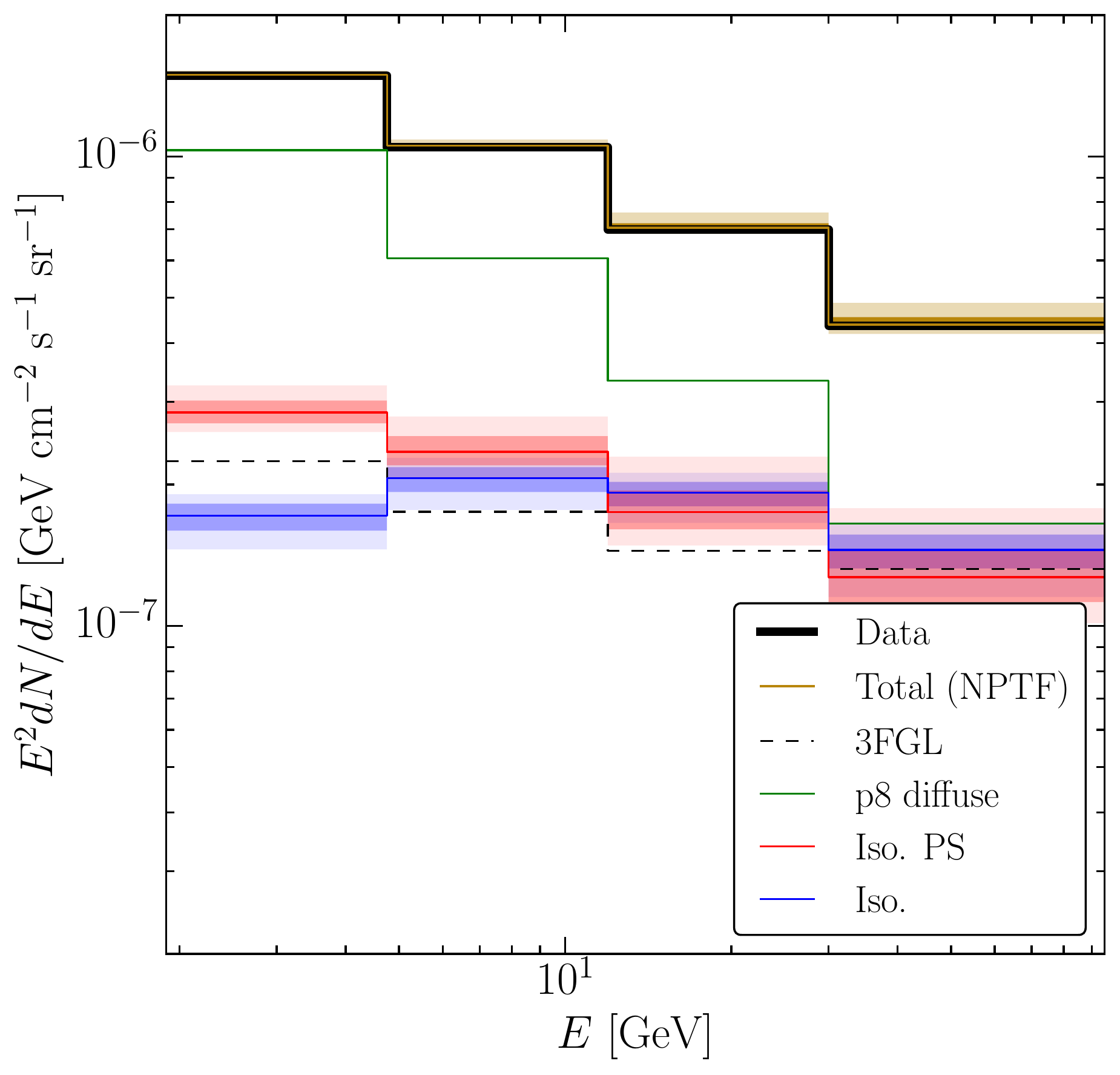}} 
	\end{array}$
	\end{center}
\caption{Best-fit energy spectra for the NPTF analysis using Pass~8 {\it ultracleanveto} data and the \texttt{p8r2} foreground model.  The left (right) panel shows the PSF3 (PSF1--3) results.  The 68 and 95\% credible intervals, constructed from the posterior distributions in each energy bin, are shown for the isotropic-PS and smooth isotropic templates in red and blue, respectively.  The median intensity for the foreground model is also shown (green).  The sum of all the components (yellow band) agrees with the total spectrum of the \emph{Fermi} data (black). The \emph{Fermi} bubbles contribution is subdominant (averaged over the full region of interest) and is thus not plotted.  For comparison, the spectrum of the 3FGL sources is shown in dashed black.  We caution the reader that, at higher energies, the 3FGL spectra are driven by extrapolations from low energies where the statistics are better.  The systematic uncertainties associated with this extrapolation are difficult to quantify and are not shown here.}
   \label{fig:nptfdata} 
\end{figure*}

A perhaps more realistic scenario for testing the NPTF is to consider a scenario where both SFGs and blazars contribute to the EGB.  Therefore, we create simulated maps that include both components and test them on the NPTF.  The recovered energy spectra for the SFG + Blazar--1 (Blazar--2) example is shown in the bottom left (right) panel of Fig.~\ref{fig:blESpec}.  In both cases, the PS spectrum is consistent with that found in the blazar-only simulations, which are shown in the top panels in that figure.  
The reconstructed source-count distributions for these examples are not shown, as they are consistent with those found in the blazar-only cases.

In the case of of the Blazar--1 model, the spectra of the smooth isotropic emission and the PS emission trace the spectra of the input SFG population and blazar population, respectively. In the case of the Blazar--2 model, the PS flux is further below the input blazar spectrum, as was found in the blazar-only simulations.  However, the smooth isotropic emission is further above the simulated SFG spectrum.  In both cases, the sum of the smooth isotropic emission and PS emission (EGB) is consistent with the simulated blazar plus SFG flux.  

There is, in fact, a subtle difference between the PS distribution recovered with and without the addition of a SFG population.  The difference becomes noticeable when comparing the fractions 
\begin{equation}
\frac{I_\text{iso}^\text{PS}}{I_\text{blazar-sim}} = [0.97_{-0.05}^{+0.06}, 1.00_{-0.09}^{+0.11}, 0.87_{-0.07}^{+0.09}, 0.72_{-0.09}^{+0.12}] \nonumber
\end{equation}
for SFG + Blazar--1 and 
\begin{equation}
\frac{I_\text{iso}^\text{PS}}{ I_\text{blazar-sim} }= [0.80_{-0.06}^{+0.08}, 0.59_{-0.06}^{+0.07}, 0.59_{-0.06}^{+0.08}, 0.43_{-0.05}^{+0.06}] \nonumber
\end{equation} 
for SFG + Blazar--2 to the corresponding values for the blazar-only simulations.  In the simulations with SFGs, the fractions  $I_\text{iso}^\text{PS} / I_\text{blazar-sim}$ are generally higher and have larger uncertainties.  The reason for this is that the SFG emission is degenerate with an enhanced sub-threshold component to the PS source-count distribution.

Simulating data with the PSF1--3 instrument response function, we find that the ratios $I_\text{iso}^\text{PS} / I_\text{blazar-sim}$ are somewhat closer to unity than in the PSF3 case.  In particular, for the SFG  +  Blazar--2 model simulations,
\begin{equation}
\frac{I_\text{iso}^\text{PS}}{ I_\text{blazar-sim}} = [1.03_{-0.13}^{+0.20}, 0.73_{-0.05}^{+0.06}, 0.66_{-0.06}^{+0.07}, 0.57_{-0.06}^{+0.07}] \,.  \nonumber
\end{equation}
The improved exposure allows the NPTF to probe lower fluxes and to therefore recover a larger fraction of the isotropic-PS emission.

\section{Low-Energy Analysis: 1.89--94.9 GeV} 
\label{sec:lowenergy}

The findings from the previous section illustrate that the NPTF procedure is able to set strong constraints on the PS (\emph{e.g.}, blazar) and smooth Poissonian (\emph{e.g.}, SFGs, mAGN) contributions to the EGB.
 In this section, we focus on the energy range from 1.89--94.9~GeV, and begin by presenting the results of our benchmark analysis on the real \emph{Fermi} data.  This is followed by a detailed discussion of potential systematic uncertainties and their effects on the conclusions.  

\subsection{ Pass~8~{\it ultracleanveto} Data}
\label{sec:benchmark}

\subsubsection{Top PSF Quartile}

We begin by analyzing the {\it ultracleanveto} PSF3 data for $\abs{b} \geq 30^\circ$, using the \texttt{p8r2} foreground model.  This is referred to as the ``benchmark analysis'' throughout the text.  Table~\ref{tab:bestfit} provides the best-fit intensities for each spectral component, as a function of energy, and the best-fit spectra are plotted in the left panel of Fig.~\ref{fig:nptfdata}.   The \texttt{p8r2} diffuse model is shown in green (median only), while the smooth isotropic and isotropic-PS posteriors are shown by the blue and red bands, respectively.  The best-fit spectrum for PSs with $|b| > 30^\circ$ in the 3FGL catalog~\citep{Acero:2015hja} is shown by the dashed black line in Fig.~\ref{fig:nptfdata}; the spectrum as plotted should be treated with care as systematic uncertainties are not properly accounted for.  In particular, the 3FGL catalog includes sources between 0.1--300~GeV.  At the high end of this range, the spectrum is driven to a large extent by extrapolations from lower energies, where the statistics are better.  The potential errors associated with such extrapolations are difficult to quantify and are not shown in Fig.~\ref{fig:nptfdata}.    
As a result, a direct comparison between the 3FGL spectrum and our results is difficult to make, especially in the highest energy bins.  For this reason, we have a dedicated NPTF study for energies greater than 50~GeV in Sec.~\ref{sec:highenergy}.  Those results are compared to the \emph{Fermi} 2FHL catalog~\citep{TheFermi-LAT:2015ykq}, which is explicitly constructed at higher energies and is likely a more faithful representation of above-threshold PSs in this regime.    

\begin{table*}[phtb]
\renewcommand{\arraystretch}{1.4}
\setlength{\tabcolsep}{5pt}
\begin{center}
\begin{tabular}{ c  | c  c  c c  c   }
 Energy & $I_\text{EGB}$&$I_\text{iso}^\text{PS}$ & $I_\text{iso}$ & $I_\text{diff}$ & $I_\text{bub}$   \\
$[\text{GeV}]$ &  \multicolumn{5}{c}{$\left[\text{cm}^{-2}\text{ s}^{-1}\text{ sr}^{-1}\right]$}    \\%
\hline
1.89--4.75  
&  $1.38_{-0.04}^{+0.05} \times 10^{-7}$ & $9.00_{-0.54}^{+0.66} \times 10^{-8}$ & $4.82_{-0.52}^{+0.43} \times 10^{-8}$ & $3.22_{-0.02}^{+0.02} \times 10^{-7}$ & $2.90_{-0.69}^{+0.67} \times 10^{-8}$\\
4.75--11.9 &    
$5.46_{-0.22}^{+0.24} \times 10^{-8}$ & $2.68_{-0.21}^{+0.26} \times 10^{-8}$ & $2.77_{-0.21}^{+0.18} \times 10^{-8}$ & $7.38_{-0.16}^{+0.15} \times 10^{-8}$ & $1.44_{-0.39}^{+0.39} \times 10^{-8}$   \\
11.9--30.0  & 
$1.76_{-0.09}^{+0.10} \times 10^{-8}$ & $7.17_{-0.76}^{+0.99} \times 10^{-9}$ & $1.04_{-0.08}^{+0.08} \times 10^{-8}$ & $1.63_{-0.07}^{+0.07} \times 10^{-8}$ & $5.18_{-2.23}^{+2.35} \times 10^{-9}$  \\  
30.0-94.9  &  
$5.74_{-0.41}^{+0.46} \times 10^{-9}$ & $2.40_{-0.38}^{+0.48} \times 10^{-9}$ & $3.30_{-0.42}^{+0.39} \times 10^{-9}$ & $3.73_{-0.33}^{+0.31} \times 10^{-9}$ & $1.46_{-0.92}^{+1.25} \times 10^{-9}$  \\
\end{tabular}
\end{center}
\caption{Best-fit intensities for all templates used in the NPTF analysis of Pass~8 {\it ultracleanveto} PSF3 data and the \texttt{p8r2} foreground model.  Note that the \emph{Fermi} bubbles template intensity is defined relative to the interior of the bubbles, while the intensities of the other templates are computed with respect to the region $\abs{b} \geq 30^\circ$.  The best-fit EGB intensity, which is the sum of the smooth and PS isotropic contributions, is also shown.  The posterior distributions for the template intensities are provided in Fig.~\ref{fig:p8triangle1}--\ref{fig:p8triangle4}.   }
\label{tab:bestfit}
\end{table*}
\begin{table*}[phtb]
\renewcommand{\arraystretch}{1.3}
\setlength{\tabcolsep}{3pt}
\begin{center}
\begin{tabular}{ c  | c  c  c c |  c c c   }
 Energy & $n_1$ & $n_2$ & $n_3$ & $n_4$ & $F_{b,3}$ & $F_{b,2}$ & $F_{b,1}$   \\
$[\text{GeV}]$ &  & & & & \multicolumn{3}{c}{$\left[\text{cm}^{-2}\text{ s}^{-1}\right]$}    \\\hline
1.89--4.75 &  
$3.96_{-0.80}^{+0.68}$ & $2.04_{-0.05}^{+0.05}$ & $1.74_{-0.37}^{+0.19}$ & $-0.40_{-1.05}^{+1.18}$ & $1.13_{-0.52}^{+0.39} \times 10^{-11}$ & $1.22_{-0.56}^{+2.00} \times 10^{-10}$ & $1.43_{-0.46}^{+0.51} \times 10^{-8}$
   \\
4.75--11.9 &    
$3.84_{-0.86}^{+0.78}$ & $2.13_{-0.13}^{+0.15}$ & $1.91_{-0.12}^{+0.09}$ & $-0.44_{-1.03}^{+1.21}$ & $1.16_{-0.51}^{+0.47} \times 10^{-11}$ & $2.95_{-1.79}^{+1.80} \times 10^{-10}$ & $5.52_{-2.06}^{+2.66} \times 10^{-9}$\\
11.9--30.0  & 
$3.54_{-0.91}^{+0.96}$ & $2.42_{-0.32}^{+0.41}$ & $1.97_{-0.13}^{+0.11}$ & $-0.14_{-1.15}^{+1.13}$ & $1.11_{-0.50}^{+0.52} \times 10^{-11}$ & $3.47_{-1.76}^{+1.56} \times 10^{-10}$ & $2.83_{-1.34}^{+1.34} \times 10^{-9}$    \\  
30.0-94.9  &  
$3.63_{-0.98}^{+0.89}$ & $1.83_{-0.47}^{+0.52}$ & $2.51_{-0.21}^{+0.29}$ & $-0.20_{-1.16}^{+1.15}$ & $1.02_{-0.46}^{+0.47} \times 10^{-11}$ & $2.48_{-1.36}^{+1.86} \times 10^{-10}$ & $1.68_{-0.65}^{+0.68} \times 10^{-9}$   \\
\end{tabular}
\end{center}
\caption{Best-fit parameters for the source-count distributions recovered for each energy bin; the flux breaks  $F_{b,i}$ and indices $n_i$ are labeled from highest to lowest ($F_{b,i} > F_{b,i+1}$).  These values correspond to the NPTF analysis of Pass~8 {\it ultracleanveto} PSF3 data with the \texttt{p8r2} foreground model. The median and 68\% credible  intervals are recovered from the posterior distributions, which are provided in Fig.~\ref{fig:p8triangle1}--\ref{fig:p8triangle4}.     }
\label{tab:bestfit_dndf}
\end{table*}

\begin{figure*}[tb] 
   \centering
   \includegraphics[width=\textwidth]{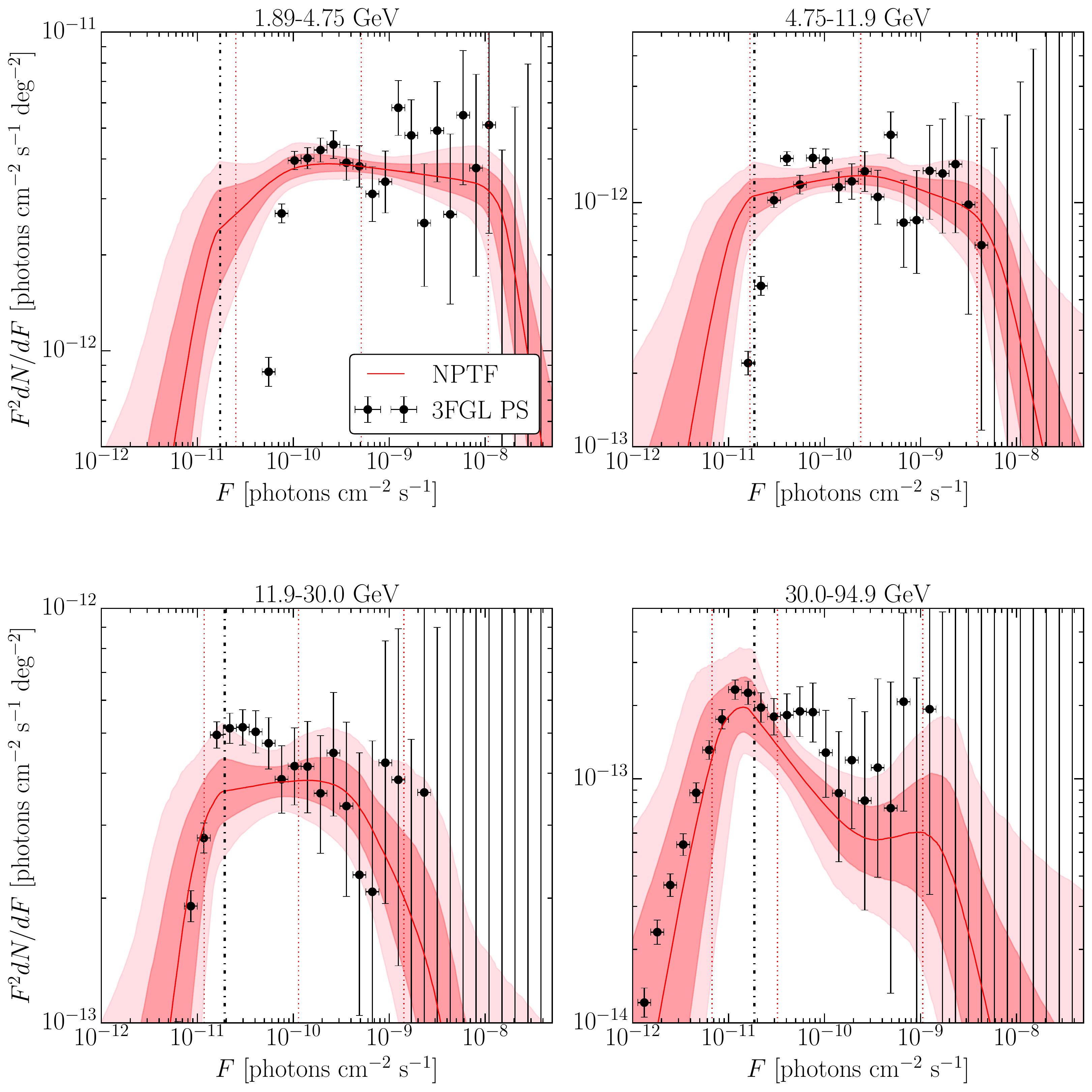} 
   \caption{The best-fit source-count distribution, as a function of energy, for the isotropic-PS population obtained by the NPTF analysis of Pass~8 {\it ultracleanveto} PSF3 data with the \texttt{p8r2} foreground model.  The median (red line) and 68 and 95\% credible intervals (shaded red bands) are shown.  The vertical dot-dashed black line denotes the $\sim$1 photon boundary, below which the NPTF begins to lose sensitivity.  The vertical dotted red lines indicate the fluxes at which 90\%, 50\%, and 10\% of the flux is accounted for, on average, by sources of larger flux (from left to right, respectively). The black points correspond to the \emph{Fermi} 3FGL sources, with 68\% statistical error bars (vertical).  The NPTF is expected to be sensitive down to the $\sim$1 photon limit, extending the reach to sources below the 3FGL detection threshold.  This is most apparent in the lowest energy bin, where the apparent 3FGL flux threshold is  $\sim$10 times higher than that for the NPTF.   
 We caution the reader that, at higher energies, the 3FGL spectra are driven by extrapolations from low energies where the statistics are better.  The systematic uncertainties associated with this extrapolation are difficult to quantify and are not included in the source counts shown here.}
   \label{fig:dndsdata}
\end{figure*}

The source-count distributions reconstructed from the NPTF are shown in Fig.~\ref{fig:dndsdata}, with best-fit parameters provided in Tab.~\ref{tab:bestfit_dndf}.  For comparison, the binned 3FGL source-count distributions are also plotted; the vertical error bars represent 68\% statistical uncertainties  and do not account for systematic uncertainties.  
A few trends are clearly visible.  First, each flux break tends to have large uncertainties.  This may be a reflection of the fact that the real source-count distribution is not a simple triply-broken power law, but rather a more complicated function, as in the blazar simulations of Sec.~\ref{sec:simulations}.  Therefore, the best-fit values for each of these parameters, when viewed independently, may be somewhat deceptive.  As is evident in Fig.~\ref{fig:dndsdata}, the posteriors for the breaks and indices are distributed in such a way as to describe a smooth concave function for $F^2 dN/dF$.

At very high and very low flux, the uncertainties on the indices ($n_1$ and $n_4$, respectively) become large.  At high flux, this is simply due to the fact that there are very few sources, so the source-count distribution falls off rapidly.  At low flux, the large uncertainties on $n_4$ arise from the difficulty in distinguishing the isotropic-PS contribution from its smooth counterpart.  Indeed, below the single-photon boundary (dot-dashed black line), the NPTF analysis starts to lose sensitivity.  The posterior distributions for the slopes above (below) the highest (lowest) break are highly dependent on the priors and so the quoted values in Tab.~\ref{tab:bestfit_dndf} should be treated with care.

The presence of any distinctive breaks encodes information about the number of source populations as well as their evolutionary properties.  In all energy bins, we see that the NPTF places the lowest break, $F_{b,3}$, close to the one-photon sensitivity threshold and the highest break, $F_{b,1}$, in the vicinity of the highest-flux 3FGL source (see Tab.~\ref{tab:bestfit_dndf} for the exact values).  The evidence for an additional break, $F_{b,2}$, at intermediate fluxes  varies depending on the energy bin.  From 1.89--4.75~GeV, there is strong indication for a break at fluxes $\sim$$10^{-10}$ ph cm$^{-2}$ s$^{-1}$, with the index $n_2 \approx 2.04$ above the break hardening to $n_3 \approx 1.74$ below the break.  In the two subsequent energy bins, up to $\sim$$30$ GeV, we also find evidence that the source-count distribution hardens as we move from high fluxes to below the second break, with the index $n_3$ below the second break $\sim$1.9-2.0 in both cases.
  In the last bin, the uncertainties are too large to determine if the source-count distribution changes slope at any flux above the lowest break $F_{b,1}$.

\subsubsection{Top Three PSF Quartiles}
\label{sec:benchmark_top3}
\begin{figure*}[phtb] 
   \centering
   \includegraphics[width=\textwidth]{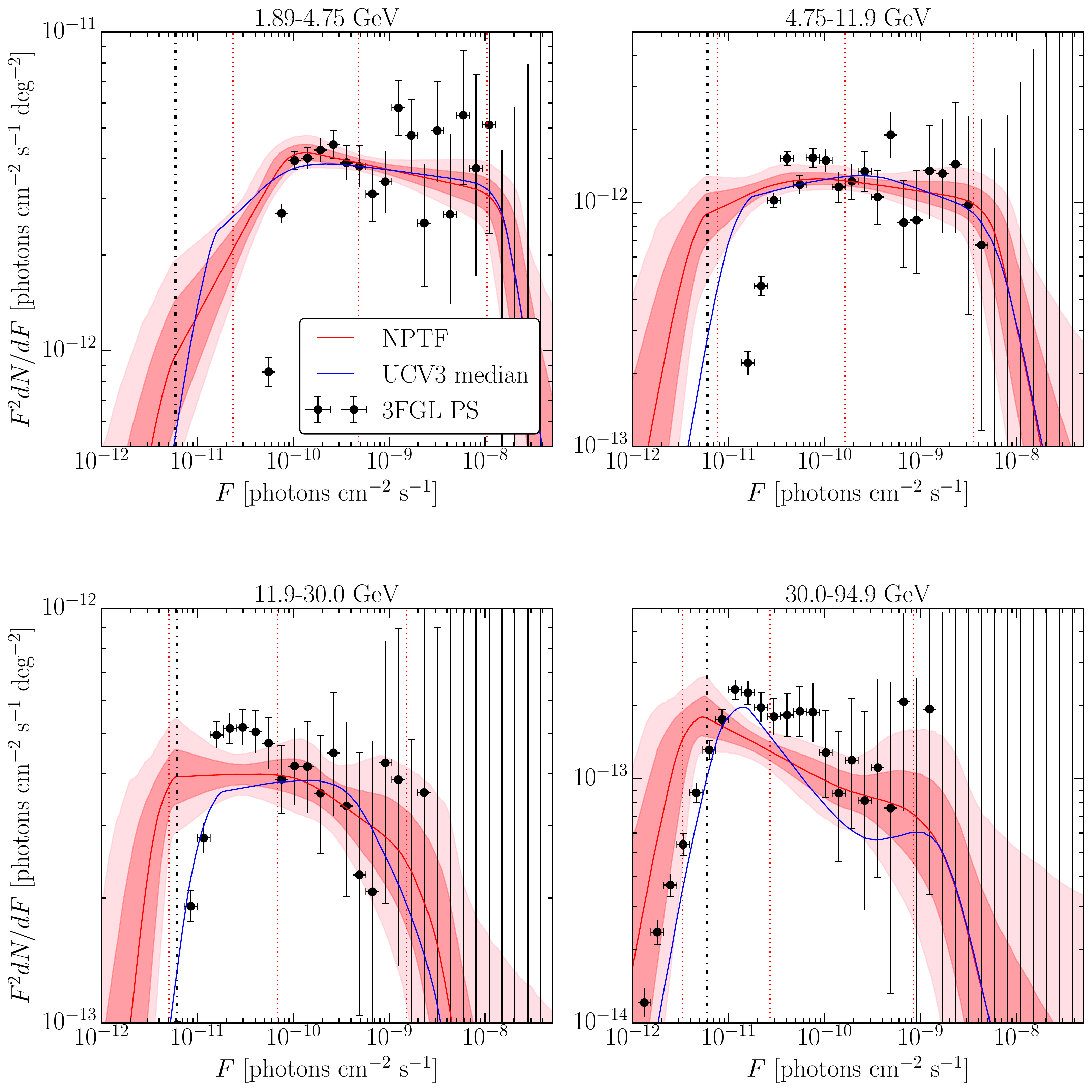} 
   \caption{The same as Fig.~\ref{fig:dndsdata}, except using the top three quartiles (PSF1--3) of the Pass 8 {\it ultracleanveto} data.  The median source-count distribution for the PSF3 analysis is shown in blue.  The best-fit values for the source-count distributions are provided in Tab.~\ref{tab:bestfit_dndf_lowQ3} of the Appendix. }
   \label{fig:dndsdata_top3}
\end{figure*}

The benchmark analysis described in the previous section used only the top quartile (PSF3) of the Pass 8 {\it ultracleanveto} data set.  This restriction selects events with the best angular resolution, but at the price of reducing the total photon count.  In Sec.~\ref{sec:simulations}, we showed that  including the top three quartiles of the Pass 8 {\it ultracleanveto} data may help constrain the source-count distribution at low fluxes.  With that in mind, we now investigate how the results of the benchmark analysis change when using the PSF1--3 {\it ultracleanveto} data set.

In general, the best-fit intensities for the individual spectral components are consistent within uncertainties with those obtained using only the top quartile of data.  (See Tab.~\ref{tab:bestfit_lowQ3} for specific values.)  The PS flux does increase slightly  in going from PSF3 to PSF1--3 in the upper energy bins due to the increased exposure.  More specifically, the ratios of the median PS intensities measured with {\it ultracleanveto} PSF1--3 data to those measured with PSF3 data are $[1.00,1.06,1.19,1.19]$ in the four increasing energy bins. 
This can also be seen in the associated spectral intensity plot (right panel of Fig.~\ref{fig:nptfdata}), where the red bands are further above the 3FGL line in the last energy bins than in the corresponding plot for the PSF3 analysis (left panel).   
The intensity of the EGB is seen to increase slightly, in all energy bins, when going from PSF3 to PSF1--3 data, potentially suggesting additional cosmic-ray contamination with the looser photon-quality cuts, though the increases in EGB intensities are within statistical uncertainties.     

The best-fit source-count distributions recovered by the NPTF with PSF1--3 data are shown in Fig.~\ref{fig:dndsdata_top3}.  For reference, the blue curve shows the best-fit for the PSF3--only analysis.  The most important difference between the PSF3 and PSF1--3 results is that the source-count distributions extend to lower flux with PSF1--3 data.  This is due to the fact that the exposure in each energy bin, averaged over the region of interest, is larger for the top three quartiles compared to the top quartile alone.  As a result, the flux corresponding to single-photon detection is lower (compare the vertical dot-dashed line in Fig.~\ref{fig:dndsdata_top3} with that in Fig.~\ref{fig:dndsdata}), which improves the NPTF reach.  Thus, the PSF1--3 analysis is sensitive to more sub-threshold sources.  
Note that the same trend was observed in the simulation tests in Sec.~\ref{sec:simulations} in going from PSF3 to PSF1--3 data sets.

Other than the location of the lowest break, which is lower due to the increased exposure, all other source-count distribution parameters are consistent, within uncertainties, between analyses.  At the lowest energy, the break at \mbox{$F_{b,2} \sim 10^{-10}$} photons~cm$^{-2}$~s$^{-1}$ is even more pronounced, with an index $n_2 \sim 2.10$ above the break and $n_3 \sim 1.75$ below the break.  In the highest energy bin, the structure observed in the source-count distribution for the benchmark analysis has smoothed out.

\subsection{Systematic Tests}
\label{sec:systematictests}

The previous subsection illustrated how the results of the NPTF change when additional {\it ultracleanveto} PSF quartiles are included in the analysis. We also tested the stability of our analysis to variations in the region of interest, \emph{Fermi} event class,  foreground modeling, \emph{Fermi} bubbles, PSF modeling, and choice of priors.  These systematic tests are described in detail in Appendix~\ref{app:systematics}.  

Figure~\ref{fig:systematicsplot} briefly summarizes the results.  The EGB intensity as measured by \emph{Fermi} is shown by the gray band.  To obtain this band, we use the best-fit power-law spectrum with exponential cut-off provided in~\cite{Ackermann:2014usa}; the width of the gray band is found by varying between best-fit values for the three foreground models considered in that paper (Models A/B/C) and does not include statistical uncertainties, which become increasingly important at high energies.  The smooth isotropic intensity, and thus the intensity of the EGB, is subject to large systematic uncertainties.   As expected, the variation in smooth isotropic intensity is most pronounced when using the {\it source} event class, which contains more cosmic-ray contamination.    However, the spectrum of emission from PSs as captured by the NPTF appears robust to all the systematic effects considered here.  This is the primary conclusion of this subsection. 

\afterpage{ 
\begin{figure*}[phtb] 
   \centering
   \includegraphics[width=\textwidth]{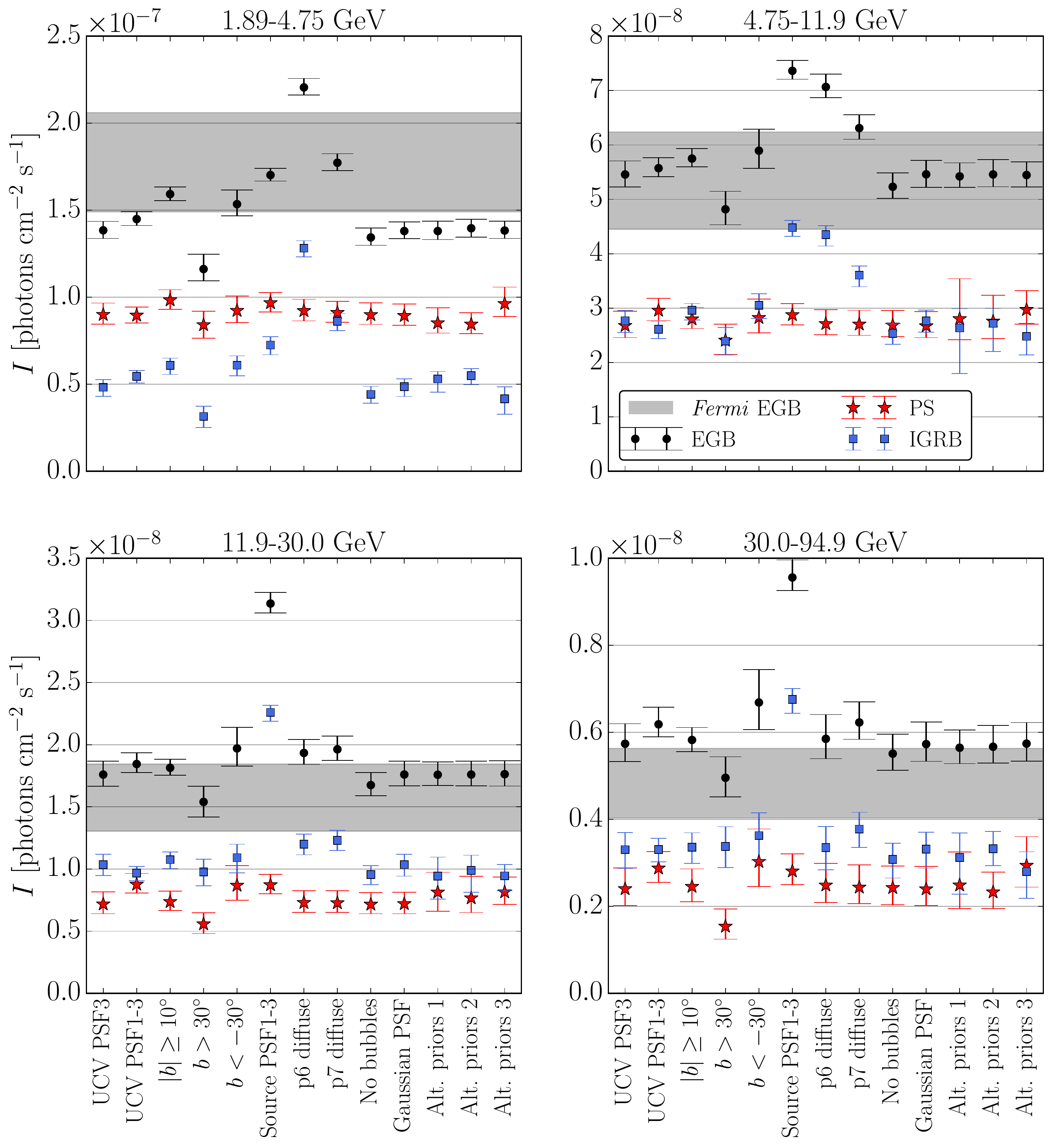} 
   \caption{Comparison of the EGB (black circles), IGRB (blue squares), and PS (red stars) intensities recovered by the NPTF for the various systematic tests described in Sec.~\ref{sec:systematictests} and Appendix~\ref{app:systematics}.  Note that `UCV' is shorthand for {\it ultracleanveto}.  The gray band is meant to indicate the systematic uncertainty associated with the measured \emph{Fermi} EGB~\citep{Ackermann:2014usa} (see text for more details).}
   \label{fig:systematicsplot}
\end{figure*}
}

\section{High-Energy Analysis: 50--2000 GeV}
\label{sec:highenergy}

We now consider the NPTF results at high energies from 50--2000~GeV.  The number of photons available decreases when moving to higher energies, so we loosen the restrictions on the PSF quartiles to maximize the sensitivity potential of the NPTF.  In this section, the majority of the analyses are done using all quartiles of the {\it ultracleanveto} data, though we also show results using all quartiles of {\it source} data.  For the same reason, we widen the ROI to $|b| > 10^\circ$ rather than $30^\circ$, although the results are not sensitive to this cut, as we will show.

\begin{figure*}[phtb] 
   \centering
   \includegraphics[width=\textwidth]{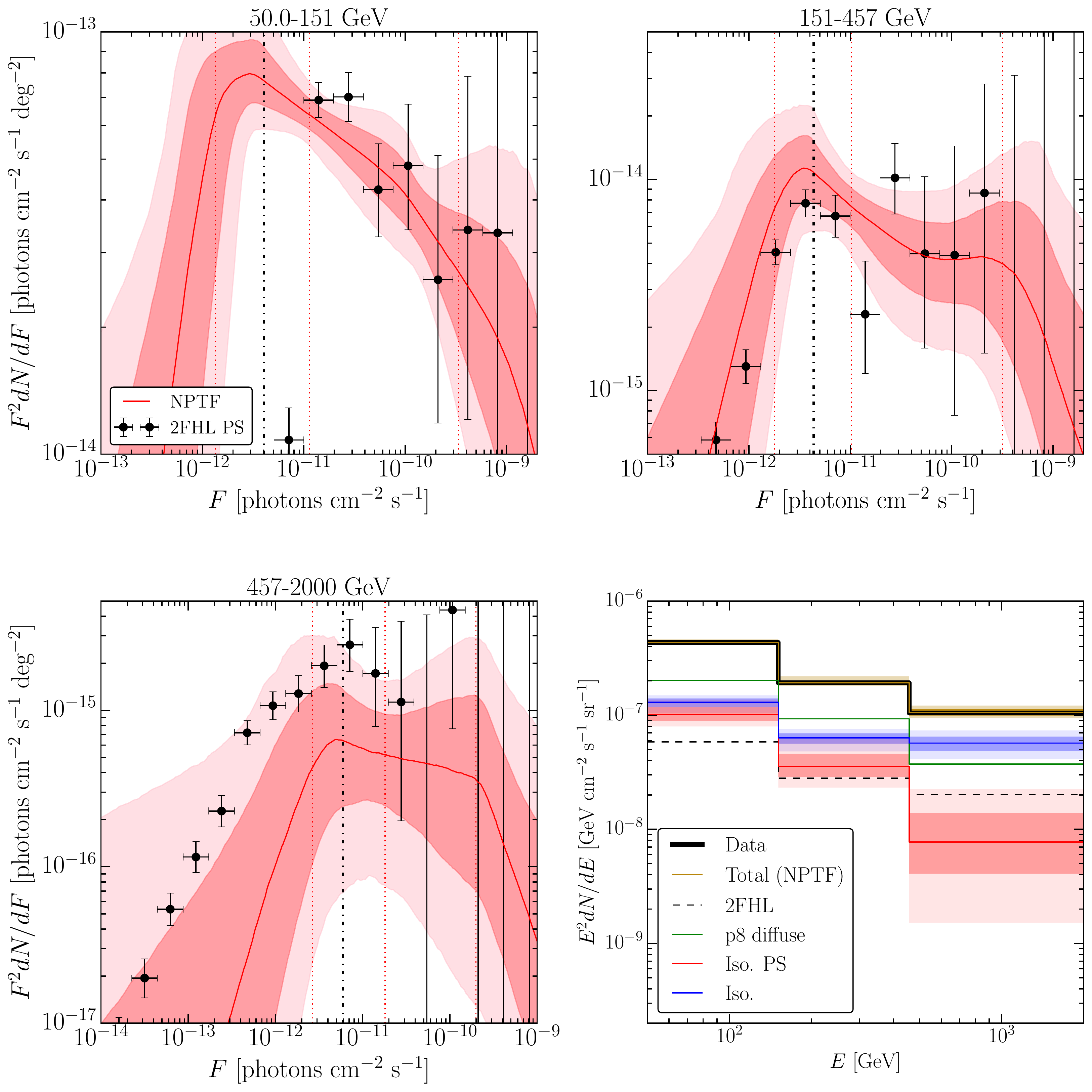} 
   \caption{NPTF results for the high-energy analysis of all quartiles of Pass8 {\it ultracleanveto} data. Top row and bottom left panel:  The best-fit source-count distribution for the isotropic-PS population, for each separate energy bin, is shown using the same format conventions as Fig.~\ref{fig:dndsdata}.  The black points correspond to the \emph{Fermi} 2FHL sources~\citep{Ackermann:2015uya}, with 68\% statistical error bars (vertical).  Bottom right panel: Best-fit energy spectrum.  The 68 and 95\% credible intervals are shown for the isotropic-PS and smooth isotropic templates in red and blue, respectively.  The median intensity for the foreground is also included (green).  The sum of all the components (yellow band) agrees with the total spectrum of the \emph{Fermi} data (black).  The spectrum of the 2FHL sources is provided in dashed black.  Best-fit values for the intensities and source-count distributions in each energy bin are provided in Tab.~\ref{tab:bestfit_highE_intensity} and \ref{tab:bestfit_highE_dnds} of the Appendix.  Note that, as for the 3FGL case, the spectra of 2FHL sources are driven at the high end by extrapolations from lower energies; the associated uncertainties are not shown here. }
   \label{fig:dndsdata_HE}
\end{figure*}

The best-fit energy spectra recovered by the NPTF analysis for the high-energy study of  {\it ultracleanveto} data is shown in the bottom right panel of Fig.~\ref{fig:dndsdata_HE}.\footnote{The intensities and best-fit values for the individual model parameters are given in Appendix~\ref{app:suppanalysis_high}.}  The fit results are compared with the best-fit  energy spectrum for sources in \emph{Fermi}'s 2FHL catalog~\citep{Ackermann:2015uya} (dashed black line).  This recently-published catalog is based on 80 months of data and focuses on hard sources in the range from 50--2000~GeV.  Statistical and systematic uncertainties are not accounted for in the determination of the 2FHL spectrum in Fig.~\ref{fig:dndsdata_HE}; these are likely non-negligible, especially at the highest energies.

The best-fit source-count distributions for the three energy bins are also shown in Fig.~\ref{fig:dndsdata_HE}, in the top row and bottom left panel.  The black points in those panels denote the 2FHL source-count distributions, with vertical error bars indicating 68\% Poisson errors.  The statistical errors on the 2FHL sources are large due to the fact that there are not many sources.  In all energy bins, the NPTF places the lowest break close to the single-photon sensitivity threshold (vertical dot-dashed line) and the highest break in the vicinity of the brightest 2FHL source, just as in the low-energy analysis.  Most notably in the 50--151~GeV bin, the NPTF probes unresolved sources with fluxes nearly an order-of-magnitude below the apparent 2FHL threshold.  We find no evidence for an additional intermediate-flux break in any of the energy bins, although it is difficult to make conclusive statements due to the large uncertainties in the individual source-count distributions. 

\begin{figure*}[phtb] 
   \centering
   \includegraphics[width=\textwidth]{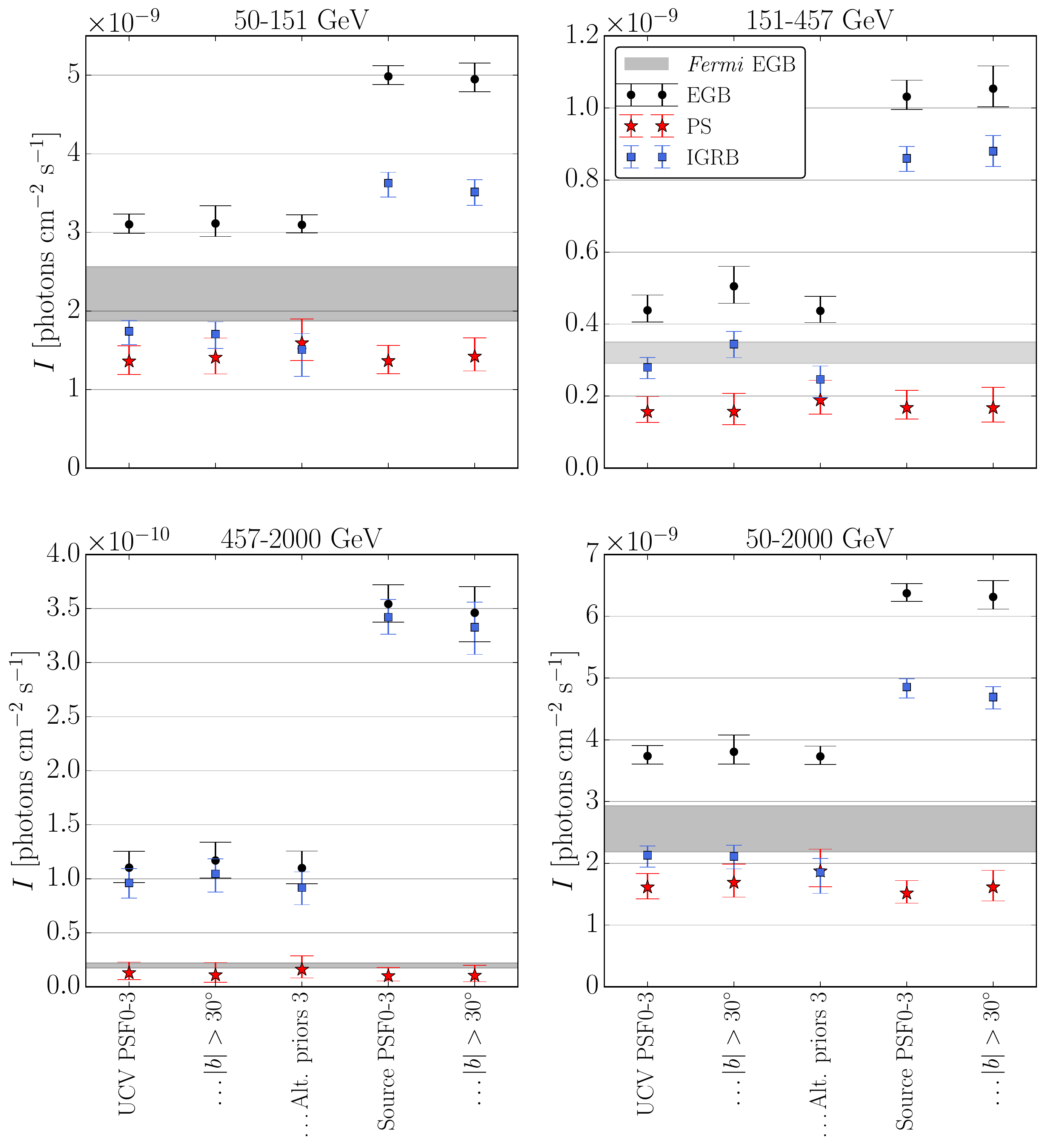} 
   \caption{Comparison of the EGB (black circles), IGRB (blue squares), and PS (red stars) intensities recovered by the NPTF for the various systematic tests specific to high energies.  The gray band indicates the systematic uncertainty associated with the measured \emph{Fermi} EGB~\citep{Ackermann:2014usa}.  The associated source-count distributions for these analyses are provided in Appendix~\ref{app:suppanalysis_high}. }
   \label{fig:systematics_HE}
\end{figure*}

We have completed a number of systematic tests of the high-energy analyses that include looking at all quartiles of the {\it source} data, requiring $|b| > 30^\circ$ for both event classes, and using the third alternate prior choice, with $n_4 > 1$.   The results are summarized in Fig.~\ref{fig:systematics_HE}, and the source-count distributions for each of the systematic tests are given in Appendix~\ref{app:suppanalysis_high}.  Importantly, the isotropic-PS intensity is consistent across all the tests.  However, the EGB intensities recovered by the NPTF are, in general, higher than those measured by \emph{Fermi}.  This discrepancy is likely due to increased cosmic-ray contamination above $\sim$100~GeV, as suggested by the high IGRB intensities recovered by the NPTF at these energies.  Indeed, the \emph{Fermi} EGB study on Pass~7 data~\citep{Ackermann:2014usa} used dedicated event classes with specific data cuts to minimize such contributions.  Such an analysis is beyond the scope of this work, as our primary focus is on the PS populations.  We simply caution the reader that the derived intensity for the smooth isotropic component in the high-energy analyses is subject to potentially large contamination.
\begin{figure*}[tb] 
   	\begin{center}$
	\begin{array}{cc}
	\scalebox{0.43}{\includegraphics{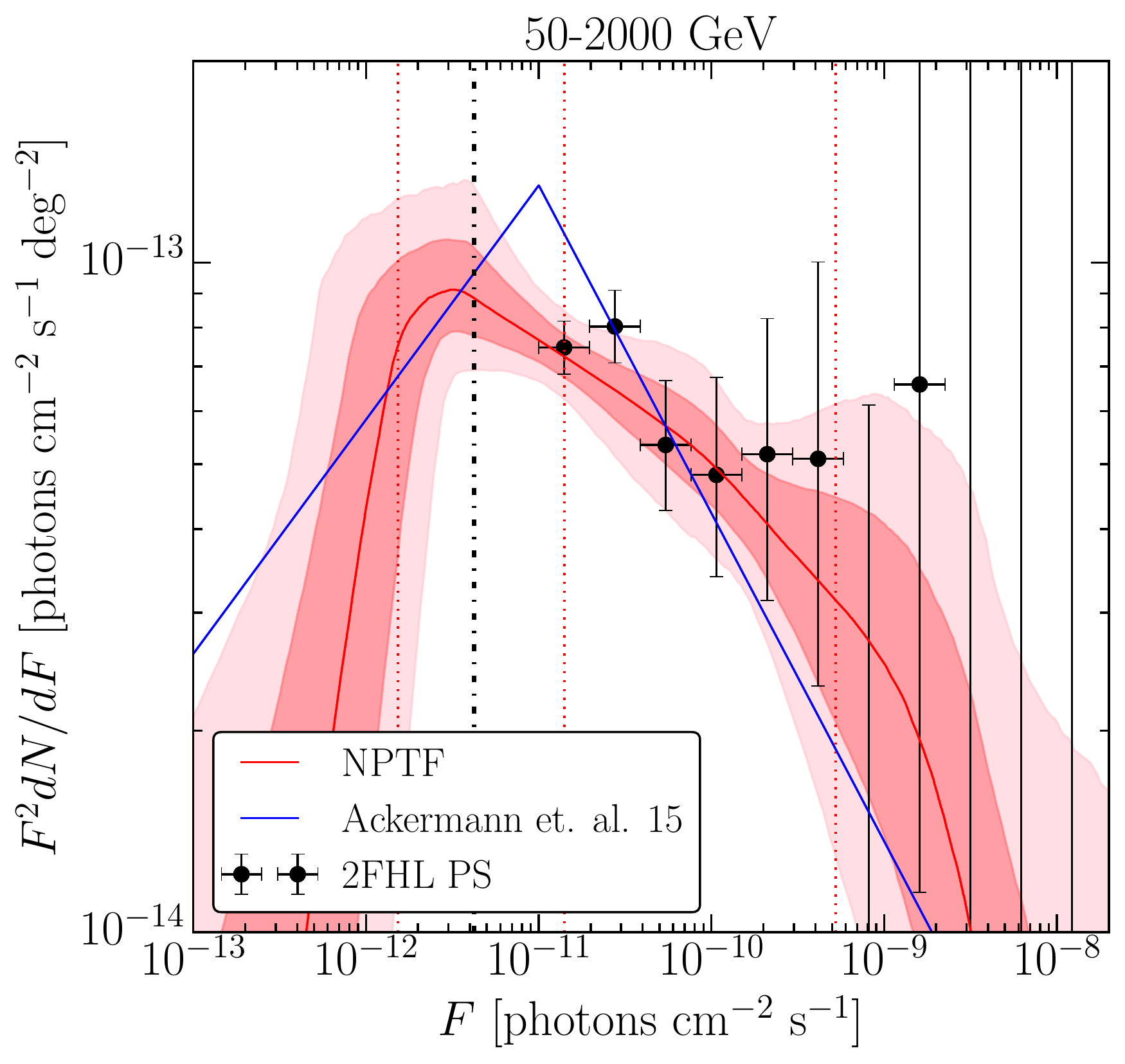}} &\scalebox{0.5}{\includegraphics{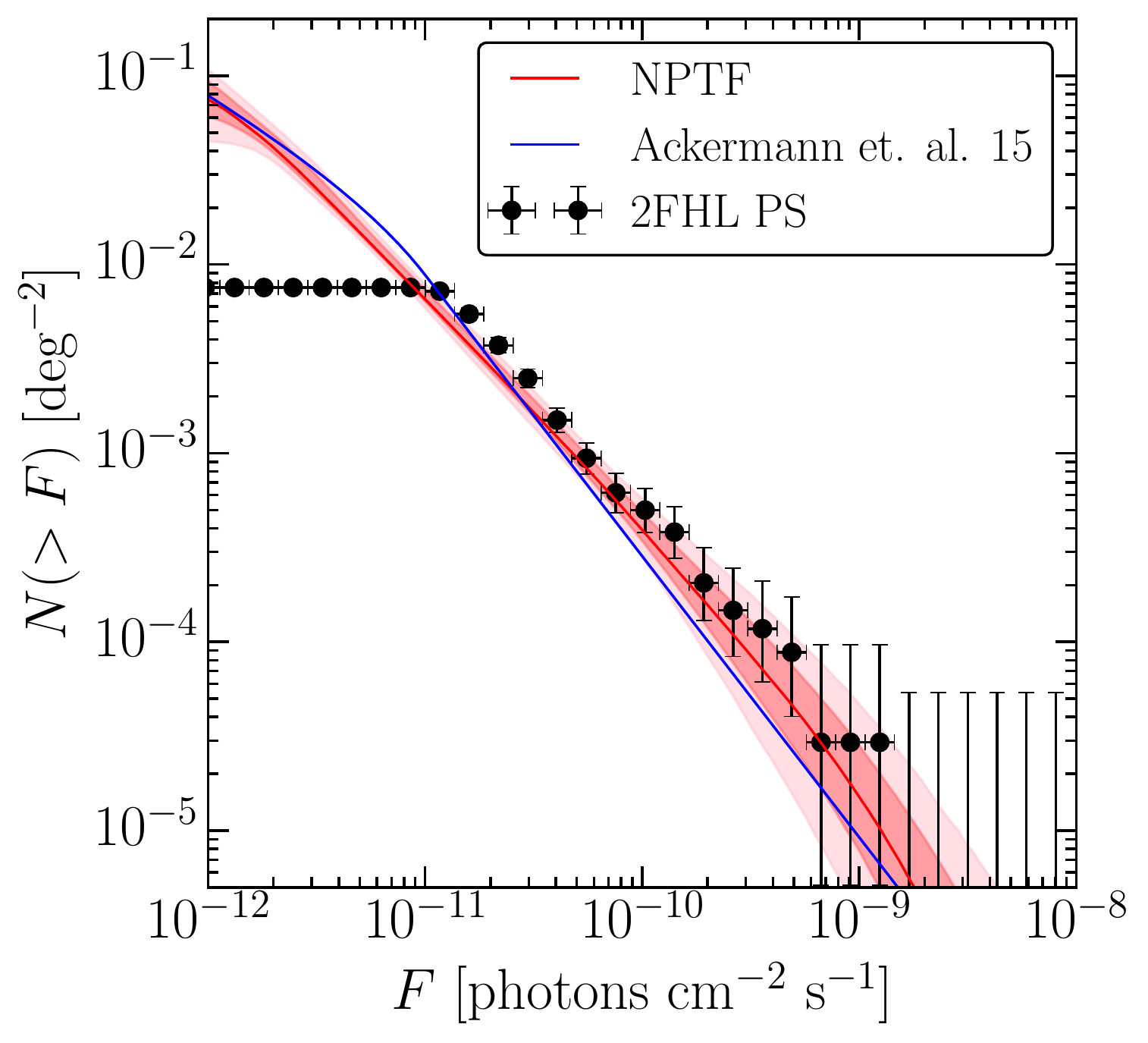}} 
	\end{array}$
	\end{center}
\caption{\emph{(left)}  Best-fit source-count distribution in the wide-energy bin from 50--2000~GeV using all quartiles of Pass~8 {\it ultracleanveto} data.  The black points indicate the 2FHL sources, and the blue line denotes the best-fit source-count from~\cite{TheFermi-LAT:2015ykq} that corresponds to the same energy bin. \emph{(right)}  A comparison of the cumulative source-count distribution for the same analysis.}
   \label{fig:dnds_0_10} 
\end{figure*}

It is possible to make stronger statements about the best-fit source-count distribution at high energies if we consider the wide-energy bin from 50--2000~GeV.  The results are shown in the left panel of Fig.~\ref{fig:dnds_0_10}.  Due to the improved statistics, the uncertainties on the source-count distribution are smaller than those for the three sub-bins.  Other than the low-flux sensitivity break, the NPTF finds no preference for  an additional break.  The intermediate-flux break, $F_{b,2}$, is essentially unconstrained as a result, and the power-law slope above (below) it are consistent within uncertainties: $n_2 = 2.28_{-0.22}^{+0.28}$ and $n_3 = 2.17_{-0.09}^{+0.12}$, respectively.  We compare this result to the best-fit source-count distribution (blue line) published by \emph{Fermi} for sources in this same energy range~\citep{TheFermi-LAT:2015ykq}.  There are important differences between the two analyses.  In the \emph{Fermi} study, simulated maps were created using several different source-count distributions, parametrized as singly broken power laws.  The histogram of the photon-count distribution for each of these maps, averaged over the full region of interest, was compared to the actual data, and a fit was done to select the simulated maps that most closely resembled the data.  
This method is related to but in many ways distinct from the NPTF.  The NPTF considers the difference between Poissonian and non-Poissonian photon probability distributions at the pixel-by-pixel level, instead of averaging the distributions over the full region.  Moreover, in our analysis we rely on semi-analytic techniques to calculate the photon-count probability distributions as we scan over the space of model parameters, instead of relying on Monte Carlo samples to numerically construct these distributions.  As a result, we are able to consider source-count distributions with additional degrees of freedom and also scan over the normalizations of all of the background templates, which tend to be well determined given the pixel-by-pixel nature of the fit.  In contrast, the intensity of all Poissonian models in~\citep{TheFermi-LAT:2015ykq}, including the smooth isotropic emission, was kept fixed while scanning over the source-count distribution degrees of freedom.

The cumulative source-count plot is provided in the right panel of Fig.~\ref{fig:dnds_0_10}.  Our result is in good agreement with the 2FHL sources above the catalog sensitivity threshold $\sim$$10^{-11}$~ph cm$^{-2}$~s$^{-1}$.  In the first few flux bins above this threshold, there appear to be more 2FHL sources than what is predicted by the NPTF, although the results are still consistent within uncertainties.  This may be due to the Eddington bias~\citep{Eddington} where extra sources are observed above threshold due to upward statistical fluctuations from sources immediately below.

Based on the results in Fig.~\ref{fig:dnds_0_10}, we can project the expected number of these sources that may be observed by the Cherenkov Telescope Array (CTA)~\citep{2011arXiv1111.2183C,Dubus:2012hm}.  For energies above 50~GeV, the CTA flux sensitivity is $\sim2.93\times10^{-12}$~cm$^{-2}$~s$^{-1}$ for 50 hours of observation per field-of-view (5$\sigma$ detection).\footnote{\url{https://portal.cta-observatory.org/Pages/CTA-Performance.aspx}}  For 250 hours total of observation time, this covers $\sim$190 deg$^2$ of sky, assuming a $7^\circ$ field-of-view.  As shown in Fig.~\ref{fig:dnds_0_10}, the NPTF predicts a density of $0.029^{+0.008}_{-0.005}$ deg$^{-2}$ for sources above this threshold.  This translates to $5.51_{-0.95}^{+1.52}$ detected sources, more than double what had previously been estimated for similar observing parameters~\citep{Dubus:2012hm}.  Relaxing the observing time per source and assuming, as in~\cite{TheFermi-LAT:2015ykq}, that a quarter of the sky is surveyed in 240 hours at 5mCrab sensitivity, then the NPTF predicts $161^{+30}_{-20}$ sources.  This is lower, and in slight tension, with the $200\pm45$ sources predicted by the \emph{Fermi} study using the blue source-count distribution illustrated in Fig.~\ref{fig:dnds_0_10}.

\section{Discussion and Conclusions}
\label{sec:conclusions}

The primary focus of this paper is to characterize the properties of the PSs contributing to the EGB in a data-driven manner.  To achieve this, we use a novel analysis method, referred to as Non-Poissonian Template Fitting (NPTF), which takes advantage of photon-count statistics to distinguish diffuse and PS contributions to gamma-ray maps with non-trivial spatial variations.  We presented the NPTF results on \emph{Fermi} Pass~8 data at low (1.89--94.9~GeV, $|b| > 30^\circ$) and high (50--2000~GeV, $|b| > 10^\circ$) energies.  For the first time, the intensity and source-count distributions for the isotropic PSs have been  obtained as a function of energy, up to 2~TeV.  The best-fit source-count distributions probe fluxes below the current detection threshold for the \emph{Fermi} 3FGL and 2FHL catalogs, providing information on the unresolved populations. 

Through extensive studies of how the NPTF responds to simulated populations, we have shown that the analysis procedure reproduces the properties of input source classes.  Therefore, the features of the best-fit source-count distributions obtained from the data provide a potential wealth of information about the source populations of the EGB.  While a detailed interpretation of the source-count distributions in terms of particular theoretical models is beyond the scope of this paper, several important trends were observed.  

In this work, the source-count distributions are parametrized as triply-broken power laws in the NPTF.  At all energies, a break is fit at low (high) fluxes, below (above) which the analysis method loses sensitivity.  Of particular interest is whether an additional break, $F_{b,2}$, is preferred at intermediate flux.  We find a break in the lowest energy bin (1.89--4.75~GeV) at $1.22_{-0.56}^{+2.00}\times10^{-10}$~cm$^{-2}$~s$^{-1}$ with slope $2.04_{-0.05}^{+0.05}$ above and $1.74_{-0.37}^{+0.19}$ below.  In the subsequent two energy bins, 4.75--11.9~GeV and 11.9--30.0~GeV, there is a mild indication that the source-count distribution hardens below the intermediate flux break, though the change in slope is not as robust and significant as in the lowest energy bin.  At higher energies, above $\sim$30 GeV, there is no indication that the source-count distribution changes slope at the intermediate break.  This trend is in line with the expectations from the blazar simulations in Sec.~\ref{sec:simulations}.  For example, in both Figs.~\ref{fig:bl1dnds} and~\ref{fig:bl2dnds} (see also Fig.~\ref{fig:dndsb2UCV}), which show the results of the NPTF run on simulated data with the Blazar--1 and Blazar--2 models, we find evidence for curvature in the source-count distribution at intermediate fluxes in the lowest energy bins, while at higher energies the recovered source-count distribution appears as a single power law at fluxes above the sensitivity threshold of the NPTF.  In the energy bin from 50--2000~GeV the best-fit value for $F_{b,2}$ is essentially unconstrained and the slopes above and below it are consistent within uncertainties:  $2.28_{-0.22}^{+0.28}$ and $2.17_{-0.09}^{+0.12}$.

\begin{figure}[b] 
   \centering
   \includegraphics[width=0.48\textwidth]{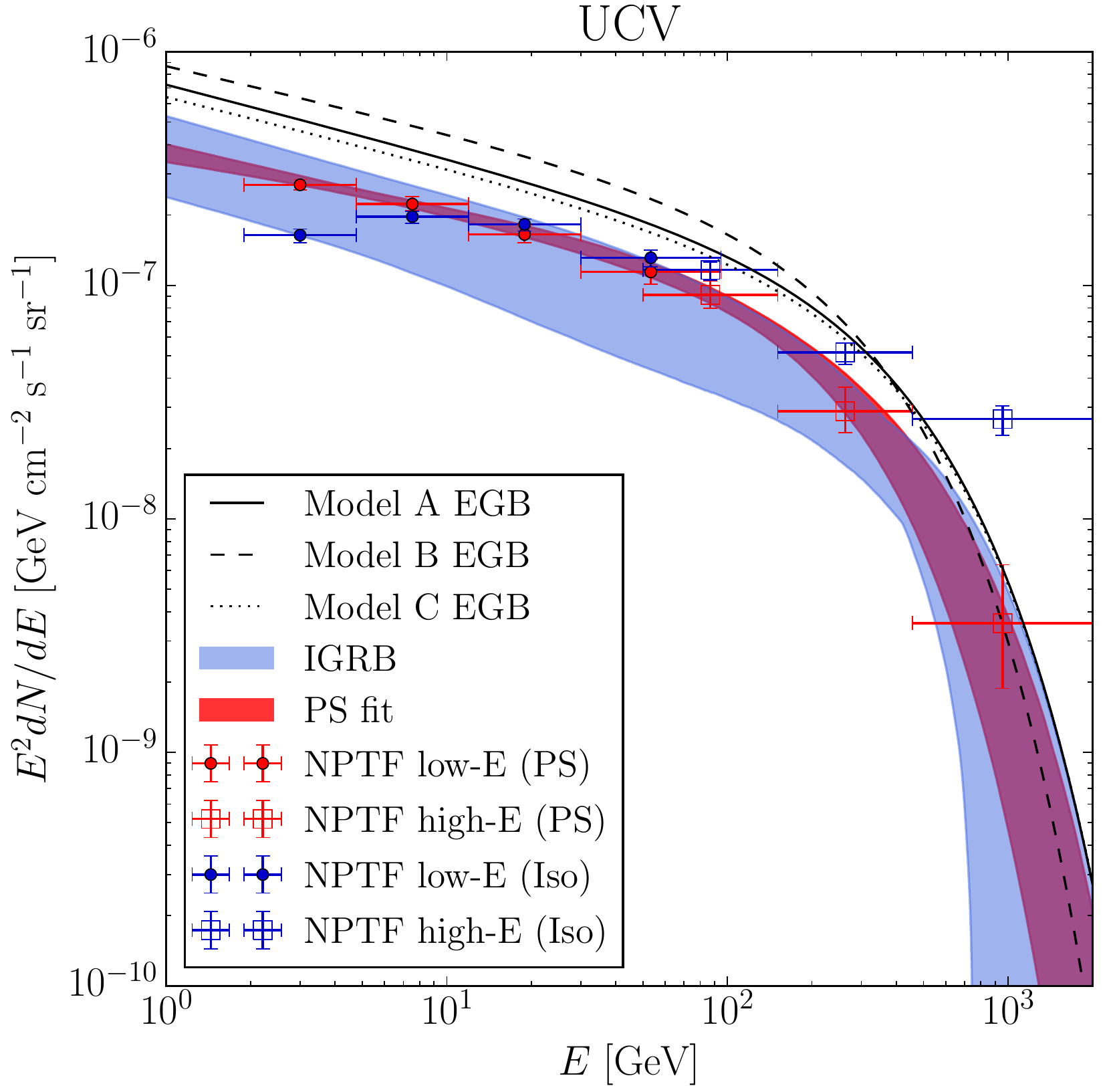} 
   \caption{Global fit to the PS intensity spectrum recovered by the NPTF. The results of the NPTF low-energy analysis on {\it ultracleanveto} PSF1--3 data and the high-energy analysis on {\it ultracleeanveto} PSF0--3 data are shown (filled red circles and open red boxes, respectively).  The red band indicates the best-fit (68\% credible interval) to a power law with exponential cutoff.  For comparison, the best-fit \emph{Fermi} EGB spectra from~\cite{Ackermann:2014usa} are shown for three different diffuse background models (Model A--C).  The blue band indicates the estimated IGRB spectrum, obtained by subtracting the PS spectrum from the \emph{Fermi} EGB; the spread includes the statistical uncertainty from the PS intensity as well as the systematic uncertainty on the EGB.  We also plot the best-fit smooth isotropic spectrum recovered by the NPTF (filled blue circles and open blue boxes).  The results are in good agreement with the estimated IGRB result (blue band) below $\sim$100 GeV, but overestimate the result at higher energies due to cosmic-ray contamination.  
    } 
   \label{fig:global}
\end{figure}

The NPTF also provides the best-fit intensities for the isotropic-PS populations as a function of energy.  Figure~\ref{fig:global} illustrates this spectrum for analyses done using the {\it ultracleanveto} event class.  The filled red circles (open red boxes) show the results for the dedicated low (high)-energy analysis, with PSF1--3 data used at low energies and PSF0--3 data at high energies.  For comparison, the \emph{Fermi} EGB spectrum is shown by the black line~\citep{Ackermann:2014usa}.  This corresponds to the best-fit intensity using the Model A diffuse background from that study.  To illustrate the systematic uncertainty on this curve, we also plot the spectra for diffuse models B and C (dashed and dotted, respectively).

\begin{table*}[t]
\renewcommand{\arraystretch}{1.5}
\setlength{\tabcolsep}{5.2pt}
\begin{center}
\begin{tabular}{  c | c   c  c  c || c c c c}
$I_\text{EGB}$ & \multicolumn{4}{c|| }{Low-Energy Analysis} & \multicolumn{4}{c }{High-Energy Analysis}  \Tstrut\Bstrut	\\   
& 1.89--4.75 & 4.75--11.9 & 11.9--30 & 30--94.9 & 50--151 & 151--457 & 457--2000 & 50--2000 \Tstrut\Bstrut \\
\hline
Scenario A& $0.62^{+0.04}_{-0.02}$ & $0.53^{+0.03}_{-0.03}$ & $0.48^{+0.03}_{-0.03}$ &  $0.47^{+0.05}_{-0.04}$ &$0.44^{+0.06}_{-0.05}$ & $0.36^{+0.08}_{-0.06}$ & $0.12^{+0.09}_{-0.06}$ & $0.43^{+0.05}_{-0.04}$\Tstrut\Bstrut \\
Scenario B& $0.54^{+0.03}_{-0.03}$ & $0.60^{+0.04}_{-0.03}$ &  $0.61^{+0.06}_{-0.05}$ &  $0.66^{+0.09}_{-0.07}$ & $0.67^{+0.10}_{-0.09}$ & $0.51^{+0.13}_{-0.09}$ &  $0.58^{+0.45}_{-0.27}$ & $0.68^{+0.09}_{-0.08}$ \Tstrut\Bstrut  \\ 
\end{tabular}
\end{center}
\caption{PS fractions ($I_\text{PS}/I_\text{EGB}$) for the low (PSF1--3) and high-energy (PSF0--3) analyses, using {\it ultracleanveto} data, with energy sub-bins in units of GeV.   The first row (`Scenario A') uses the EGB intensity obtained in this work using foreground model \texttt{p8r2}; however, this scenario likely overestimates the $I_\text{EGB}$ at energies above $\sim$100~GeV due to cosmic-ray contamination.  The second row shows the PS fractions calculated with respect to the \emph{Fermi} EGB intensity from~\cite{Ackermann:2014usa}, with foreground Model A (`Scenario B').  Although the \emph{Fermi} analysis uses a different foreground model, it takes advantage of a dedicated event selection above $\sim$100~GeV that mitigates effects of additional contamination.  }
\label{tab:fractions}
\end{table*}

The PS fraction, defined as $I_\text{PS}/I_\text{EGB}$, is provided in Tab.~\ref{tab:fractions} for each energy bin.  While using the EGB intensity derived in this work (`Scenario A') is the most self-consistent comparison, this may underestimate the PS contribution above $\sim$100~GeV, where the NPTF appears to recover too much smooth isotropic emission due to increased cosmic-ray contamination in the data sets used, as already discussed.  Therefore, we also show the PS fractions calculated relative to the \emph{Fermi} EGB intensity from~\cite{Ackermann:2014usa} for diffuse model A (`Scenario B').  The comparison to the EGB as measured in~\cite{Ackermann:2014usa} is not fully self consistent, since, for example, the foreground modeling and data sets in~\cite{Ackermann:2014usa} differ from those used in this work to measure $I_\text{PS}$.  However, the advantage of this comparison is that the \emph{Fermi} analysis uses special event-quality cuts to mitigate contamination, and thus their measure of $I_\text{EGB}$ is likely more faithful than that presented in this work.  These results are shown in the second row of Tab.~\ref{tab:fractions}.  For the low-energy analysis, the PS fractions are consistent, within uncertainties, when $I_\text{EGB}$ is taken from our work or \emph{Fermi}'s.\footnote{For `Scenario B', the quoted uncertainties only include those measured in this work for $I_\text{PS}$.  For $I_\text{EGB}$, we use the best-fit value given in~\citep{Ackermann:2014usa}.}  The substantial differences occur at high-energies, where our result is systematically lower than the fractions based on \emph{Fermi}'s EGB intensity.

In general, we find that approximately 50--70\% of the EGB consists of PSs in the energy ranges considered.  To interpret these results, we use the ratios $I_\text{iso}^\text{PS} / I_\text{blazar-sim}$ obtained in the simulation studies of Sec.~\ref{sec:simulations}.  In that section, we showed that the efficiency for the NPTF to recover the flux for the Blazar--2 model (with PSF1--3) is 
$\sim$100\% in the first energy bin and drops to $\sim$60\% in the fourth energy bin.  For the Blazar--1 model, the efficiencies are consistently higher than the Blazar--2 scenario.  These two blazar models are meant to illustrate extreme scenarios, with the Blazar--1 model having a significant fraction of the total flux arising from high-flux sources, while low-flux sources dominate instead in the Blazar--2 case. 
The high efficiency of the NPTF to recover the blazar component at low energies, combined with the PS fractions observed in the data (Tab.~\ref{tab:fractions}), clearly suggests that there is a substantial non-blazar component of the EGB up to energies $\sim$30 GeV.   The interpretation of the results in the energy bin from 30.0--94.9 GeV is less clear.  A proper interpretation of the results at higher energies in terms of evidence for or against a non-blazar component of the EGB requires dedicated blazar simulations, which we leave to future work.  
 
Our results tend to predict fewer PSs (and photons from PSs) where we do overlap with previous studies.  For example, a similar photon-count analysis was used by~\citep{Zechlin:2015wdz} to study 1--10~GeV energies in the Pass~7 Reprocessed data.  They found an $\sim$80\% PS fraction at these energies.  At the lowest energies that we probe---which admittedly do not extend down as low as $\sim$1~GeV---we only find a $\sim$54\% PS fraction (relative to Model A).  
  Systematic uncertainties, as shown in Fig.~\ref{fig:systematicsplot}, can affect the recovered PS intensities at the ${\cal O}(10\%)$ level, which can partially alleviate the tension between our results.  

Above 50~GeV, the NPTF procedure predicts that $0.68_{-0.08}^{+0.09}$ of the EGB consists of PSs, with systematic uncertainties estimated at approximately $\pm 10\%$.  This fraction is smaller, and in slight tension, with the predicted value $0.86_{-0.14}^{+0.16}$ obtained in previous work~\citep{TheFermi-LAT:2015ykq}.  The fact that our results suggest that there is more diffuse isotropic emission at high energies may help alleviate the tension between \cite{TheFermi-LAT:2015ykq} and the hadronuclear ($pp$) interpretation of IceCube's PeV neutrinos \citep{Murase:2013rfa}.  Some models suggest, for example, that these very-high-energy neutrinos are produced in hadronuclear interactions, along with high-energy gamma-rays that would contribute to the IGRB~\citep{Murase:2013rfa, Tamborra:2014xia,Ando:2015bva,Hooper:2016gjy}.  If the smooth isotropic gamma-ray spectrum (\emph{i.e.}, the non-blazar spectrum) is suppressed above 50~GeV in the \emph{Fermi} data, it could put such scenarios in tension with the data ~\citep{Bechtol:2015uqb,Murase:2015xka}; however, that does not necessarily appear to be the case  given the results of our analysis~\citep{Murase:2016gly}.  With that said, and as already mentioned, dedicated blazar simulations at high energies are needed to properly interpret our results at these energies.

The PS spectrum in Fig.~\ref{fig:global} is well-modeled (\mbox{$\chi^2 = 1.18$}) as a power law with an exponential cut-off:
\es{powerexp}{
{d N \over dE} = C \left( {E \over 0.1 \, \, \text{GeV} } \right)^{-\gamma} \exp{ \left( - {E \over E_\text{cut}} \right) } \,,
}
where $C=6.91_{-1.29}^{+1.44} \times 10^{-5}$ GeV$^{-1}$cm$^{-2}$s$^{-1}$sr$^{-1}$,  $\gamma= 2.26_{-0.05}^{+0.05}$, and $E_\text{cut} = 289_{-86.3}^{+127}$~GeV are the best-fit parameters.\footnote{Repeating the fit using the results from the NPTF analyses with {\it source} data returns similar results, though the PS spectrum is slightly enhanced relative to the {\it ultracleanveto} result.  In particular, with {\it source} data, we find $C=7.98_{-1.40}^{+1.58} \times 10^{-5}$ GeV$^{-1}$cm$^{-2}$s$^{-1}$sr$^{-1}$,  $\gamma= 2.29_{-0.05}^{+0.04}$, and $E_\text{cut} = 325_{-78.1}^{+117}$~GeV, with $\chi^2 = 0.93$. }  Note that the fit is done taking into account the uncertainties on the PS intensities in the energy sub-bins.  The global fit for the PS spectrum is shown in Fig.~\ref{fig:global} by the red band, which denotes the 68\% credible interval.  Interestingly, the index $\gamma$ and cut-off $E_\text{cut}$ that we extract from the fit are very similar to the values found in~\cite{Ackermann:2014usa}, which used the same functional form to fit the EGB spectrum.  Subtracting our PS spectrum from the EGB spectral fits gives the blue band in Fig.~\ref{fig:global}.  The band includes statistical uncertainties from our global fit as well as systematic uncertainties associated with varying between Models A-C.  The blue band is an estimate of the IGRB spectrum and we compare it to the smooth isotropic spectrum recovered by the NPTF (blue points).  Note that the two are consistent, within the large uncertainties, below $\sim$100~GeV; above this energy, our IGRB value is expectedly high. 

\begin{figure*}[phtb] 
   	\begin{center}
	$\begin{array}{c}
\includegraphics[scale=0.8]{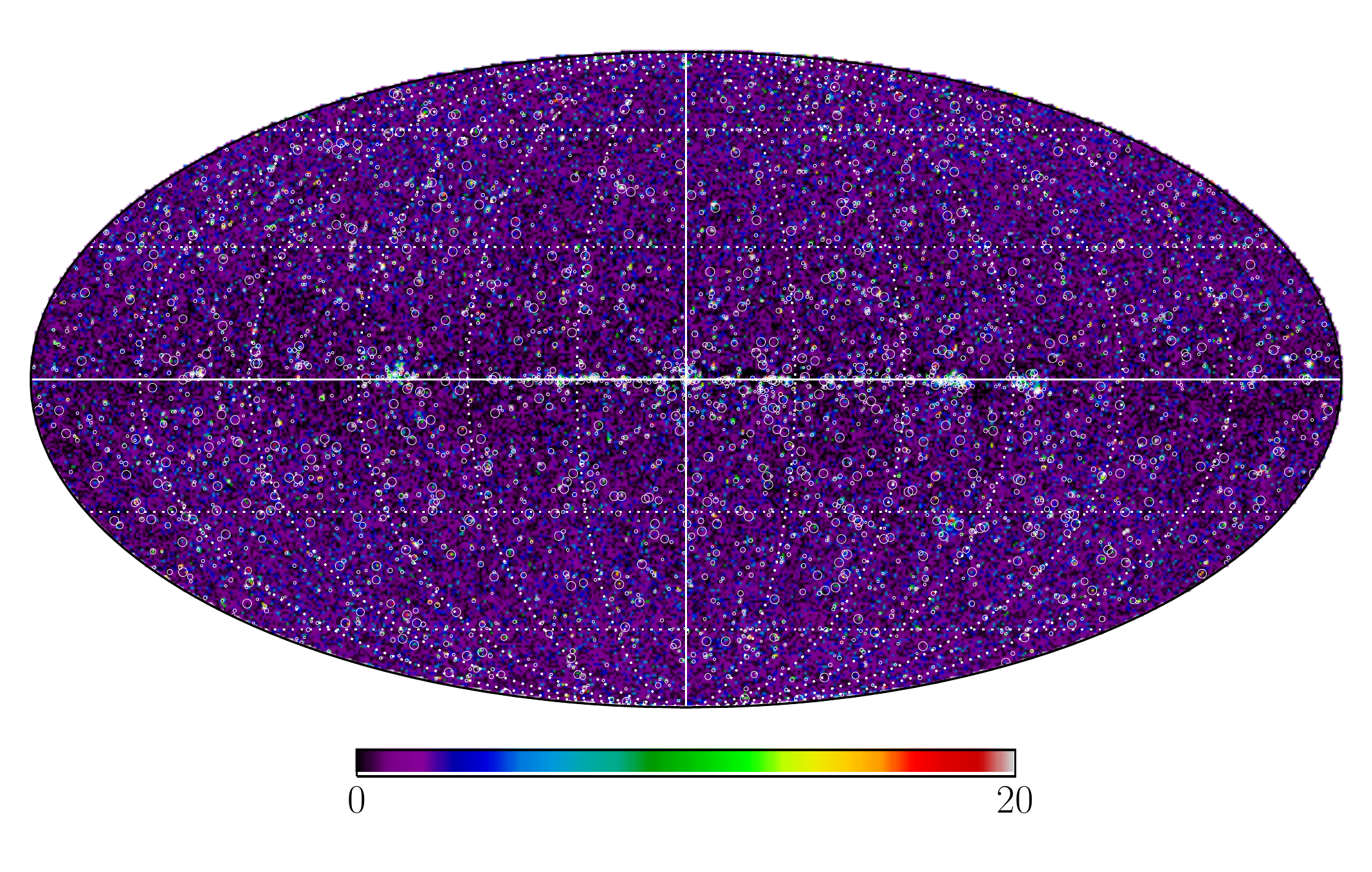} \\
\includegraphics[scale=0.8]{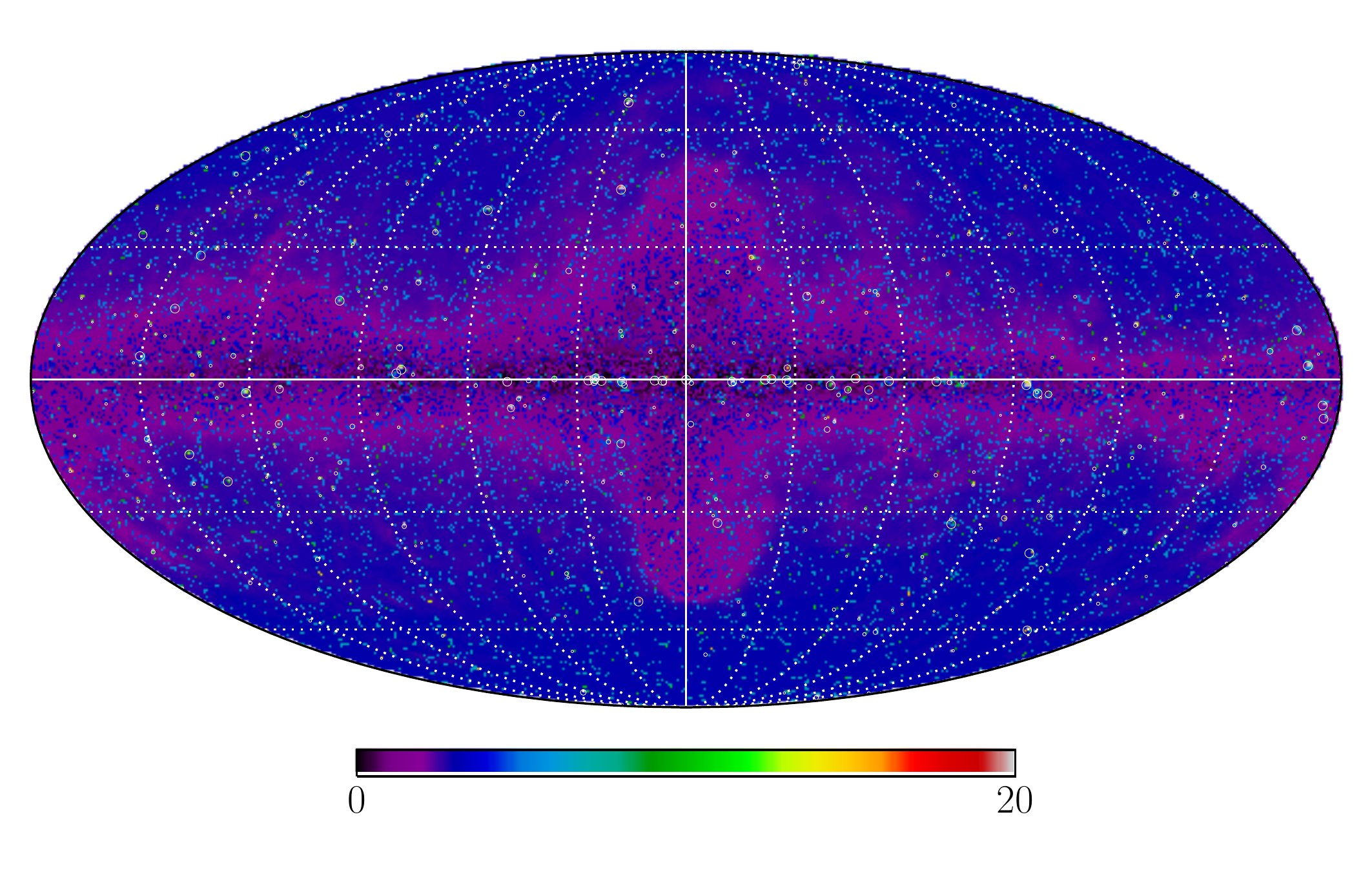}
	\end{array}$
	\end{center}
\caption{Full-sky maps showing the value (clipped at 20) of $-\log \epsilon_p$ in each pixel $p$.  The larger the value of $-\log \epsilon_p$, the more likely the pixel contains a point source.  \emph{(top)}  Results using {\it ultracleanveto} data (PSF3) for energies 1.89--94.9~GeV.  \emph{Fermi} 3FGL sources are indicated by the white circles, with radii weighted by the predicted number of photon counts for a given source.  \emph{(bottom)} Results using all quartiles of {\it ultracleanveto} data for 50--2000~GeV.  Circles now represent \emph{Fermi} 2FHL sources.  The data for this figure is available upon request. }
   \label{fig:hotspot} 
\end{figure*}

The NPTF allows us to make statistical statements about the properties of source populations contributing to the EGB, but at the expense of identifying the precise locations of these sources.  However, it is still possible to make probabilistic statements about these locations.  To do so, we compare the observed photon count in a given pixel, $n_p$, to the mean expected value, $\mu_p$, without accounting for PSs.  To determine $\mu_p$ we include the diffuse background, smooth isotropic emission, and the {\it Fermi} bubbles templates, with normalizations as determined from the NPTF.
The pixel-dependent survival function is defined as
\begin{equation}
\epsilon_p \equiv 1- \text{CDF}\left[ \mu_p, n_p \right] \, ,
\end{equation}
where CDF is the Poisson cumulative distribution function.  The smaller the value of $\epsilon_p$ (or, conversely, the larger the value of $-\log \epsilon_p$), the more probable it is that the pixel contains a PS.  Figure~\ref{fig:hotspot} shows full-sky maps of $-\log \epsilon_p$ for both low (1.89--94.9~GeV) and high (50--2000~GeV) energies.\footnote{ Digital versions of these maps are available upon request.}  The white circles indicate the presence of a 3FGL (2FHL) source for the low- \mbox{(high-)}energy map, with the radii proportional to the predicted photon counts for the sources.  There is good correspondence between the hottest pixels, as determined by $-\log \epsilon_p$,  and the brightest resolved sources.  Pixels that are correspondingly less ``hot" tend to be associated with less-bright 3FGL (or 2FHL) sources.  Of particular interest are the hot pixels not already identified by the published catalogs.  In the region $|b| \gtrsim 30^\circ$ ($|b| \gtrsim 10^\circ$) in the low- (high-)energy analysis, these are likely the sources lending the most weight to the NPTF below the catalog sensitivity thresholds.  While more sophisticated algorithms are needed to further refine the candidate source locations, Fig.~\ref{fig:hotspot} provides a starting point for identifying the spatial locations of potential new sources to help guide, for example, future TeV gamma-ray observations and cross-correlations with other data sets, such as the IceCube ultra-high-energy neutrinos.

Deciphering the constituents of the EGB remains an important goal in the study of high-energy gamma-ray astrophysics, with broad implications extending from the production of PeV neutrinos to signals of dark-matter annihilation or decay.     The \emph{Fermi} LAT has already played an important role in the discovery of many new sources in the GeV sky.  By taking advantage of the statistical properties of unresolved populations, our results provide a glimpse at the aggregate properties of the sources that lie below the detection threshold of these published catalogs and suggest a wealth of detections for future observatories.

\section*{Note Added}

As this work was being completed, we became aware of a complementary analysis that also takes advantage of photon-count statistics to obtain the source-count distributions in the energy range from 1--171~GeV~\citep{Zechlin:2016pme}.  That study used the {\it clean} event class of the Pass 7 Reprocessed data to study the source-count distributions in five energy bins from 1--171~GeV.  A direct comparison between the two results is challenging given the different data sets, foreground models, priors, energy sub-bins, pixelation, and regions of interest that were used between the two studies.  In the energy range where the two studies overlap, we tend to find smaller PS fractions.   (For a discussion of why we do not consider energies below 1.89~GeV, see Appendix~\ref{app:verylowenergies}.)  Specifically,~\cite{Zechlin:2016pme} finds PS fractions of  $0.79_{-0.16}^{+0.04}$, $0.66_{-0.07}^{+0.20}$, $0.66_{-0.05}^{+0.28}$, and $0.81_{-0.19}^{+0.52}$ in the energy ranges [1.99, 5.0], [5.0, 10.4], [10.4, 50.0], and [50.0, 171]~GeV.  (PS fractions are quoted relative to the \emph{Fermi} EGB intensity for foreground Model A~\citep{Ackermann:2014usa}.)  These numbers can be compared to our results, summarized in Tab.~\ref{tab:fractions}.  The best-fit source-count distributions recovered by~\cite{Zechlin:2016pme} closely resemble the Blazar--2 scenario that we studied in simulations (see Sec.~\ref{sec:simulations}).  Our studies on simulated data show that the NPTF can successfully recover the energy spectrum and source-count distributions for this blazar population, both for the case where it singularly contributes to the EGB, as well as the case where it contributes along with SFGs, modeled according to~\cite{Tamborra:2014xia}.  However, we observe different features in the best-fit source-count distributions when running the NPTF on the actual data.  

\section*{Acknowledgements}

We thank the \emph{Fermi} Collaboration for the use of the \emph{Fermi} public data and the Fermi Science Tools.  We also thank J. Balkind, K.~Bechtol, K. Blum, M. Di Mauro, N.~Rodd, and T.~Slatyer for helpful conversations.  LN is supported by the U.S. Department of Energy under grant Contract
Number DE-SC00012567. ML is supported by the DoE under grant Contract Number DE-SC0007968.  BRS is supported by a Pappalardo Fellowship
in Physics at MIT.

\onecolumngrid

\newpage
\appendix
\setcounter{equation}{0}
\setcounter{figure}{0}
\setcounter{table}{0}
\setcounter{section}{0}
\makeatletter
\renewcommand{\theequation}{S\arabic{equation}}
\renewcommand{\thefigure}{S\arabic{figure}}
\renewcommand{\thetable}{S\arabic{table}}

\section{Simulations of Extragalactic Gamma-ray Point Sources}
In this Appendix, we provide further details about the simulations and analyses of extragalactic point sources.

\subsection{Energy-Binned Source-Count Distributions}
\label{app:sims}

We generate simulated maps directly from the source-count distribution $dN/dF_{\gamma}$. To obtain this, we need two inputs: the gamma-ray luminosity function, $\Phi(L_{\gamma},z,\Gamma)$, and the source energy spectrum, $dF/dE$~\citep{DiMauro:2014wha}.  Typically, the luminosity function (LF) is given by
\begin{equation}
\Phi(L_{\gamma},z,\Gamma)=\frac{d^3N}{dL_\gamma\,dV\,d\Gamma} \,  ,
\end{equation}
where $V$ is the comoving volume, $\Gamma$ is the photon spectral index, $z$ is the redshift, $N$ is the number of sources, and $L_\gamma$ is the rest-frame luminosity for energies from 0.1--100~GeV in units of GeV\,s$^{-1}$.     

The photon flux in this energy range, $F_\gamma$, 
is defined in terms of the source energy spectrum,
\begin{equation}
F_\gamma(\Gamma) = \int_{E_\text{min}}^{E_\text{max}}\frac{dF}{dE} \, dE \, ,
\label{eq: Fgamma}
\end{equation}
where the units are cm$^{-2}$\,s$^{-1}$, and $E_\text{min(max)} = 0.1(100)$~GeV.

The source-count distribution is then given by
\begin{equation}
\frac{dN}{dF_\gamma} = \frac{1}{4\pi} \int_{\Gamma_\text{min}}^{\Gamma_\text{max}} d\Gamma \int_{z_\text{min}}^{z_\text{max}} dz \, \Phi(L_\gamma,z,\Gamma) \, \frac{dV}{dz} \, \frac{dL_\gamma}{dF_{\gamma}} \, ,
\label{eq: dNdFexact}
\end{equation}
which can be accurately estimated as
\begin{equation}
\frac{dN}{dF_\gamma} \approx \frac{1}{\Delta F_\gamma} \, \frac{1}{4\pi} \int_{\Gamma_\text{min}}^{\Gamma_\text{max}} d\Gamma \int_{z_\text{min}}^{z_\text{max}} dz \, \int_{L_\gamma(F_\gamma, \Gamma,z)}^{L_\gamma(F_\gamma+\Delta F_\gamma, \Gamma,z)} dL_\gamma \, \Phi(L_\gamma,z,\Gamma) \, \frac{dV}{dz} \, ,
\label{eq: dNdF}
\end{equation}
where $4\pi$ is the full-sky solid angle, $dV/dz$ is the comoving volume slice for a given redshift and $\Delta F_\gamma$ is sufficiently small.  To calculate $dN/dF_\gamma$, we need the following expression, which relates the luminosity to the energy flux:
\begin{equation} 
L_\gamma(F_\gamma, \Gamma,z) = \frac{4\pi d_L^2}{(1+z)^{2-\Gamma}}  \, \int_{E_\text{min}}^{E_\text{max}} \, E\,\frac{dF}{dE}\, dE \, ,
\label{eq:lumi}
\end{equation}
where $d_L$ is the luminosity distance. For a given $F_\gamma$ and $\Gamma$, one can use~\eqref{eq: Fgamma} to solve for the normalization of $dF/dE$, which can be substituted into~\eqref{eq:lumi}, along with $z$ and $\Gamma$, to obtain the associated value of the luminosity. The photon flux, $F_\gamma$, is related to the photon count, $S_\gamma$, via the mean exposure $\langle \bar{\mathcal{E}} \rangle$, which is averaged over 0.1--100~GeV and the ROI.  This allows us to finally obtain $dN/dS_\gamma$ from \eqref{eq: dNdF}.

The procedure outlined above allows one to obtain the source-count distributions based on models of luminosity functions and spectral energy distributions provided in the literature. For the AGN and SFG examples we consider in detail in this work, the luminosity functions correspond to photon energies from 0.1--100~GeV.  However, we also need the source-count distributions in subset energy ranges corresponding to our energy bins of interest, with $E'_\text{min, max}\in [0.1, 100]$~GeV.  We rescale the fluxes for these individual energy bins of interest to those in the provided 0.1--100~GeV range using a procedure similar to~\cite{DiMauro:2014wha}. Denoting quantities associated with this energy bin with a prime, we can write the new source-count distribution as
\begin{equation}
\frac{dN}{dF'_\gamma}\approx \frac{1}{\Delta F'_\gamma} \, \frac{1}{4\pi} \int_{\Gamma_\text{min}}^{\Gamma_\text{max}} d\Gamma \int_{z_\text{min}}^{z_\text{max}} dz \int_{L_\gamma(F_\gamma(F'_\gamma,\Gamma),\Gamma,z)}^{L_\gamma(F_\gamma(F'_\gamma+\Delta F'_\gamma,\Gamma),\Gamma,z)} dL_\gamma \, \Phi(L_\gamma,z,\Gamma) \, \frac{dV}{dz} \, ,
\label{eq: dNdFprime}
\end{equation}
where $\Delta F'_\gamma$ is again sufficiently small---we set $\Delta F'_\gamma \equiv 10^{-3} F'_\gamma$, and verify that the answer is robust to this choice. Note that the integral must still be done over $L_\gamma$ (unprimed) because the luminosity function is explicitly defined in terms of it.  So, we must solve for the photon flux over the full energy, $F_\gamma$, in terms of the value in the sub-bin, $F_\gamma'$.  The two are related via a proportionality relation
\begin{equation}
F_\gamma(F'_\gamma,\Gamma) = F'_\gamma \, \frac{\int_{E_\text{min}}^{E_\text{max}}  dE \int_{L_{\gamma,\text{min}}}^{L_{\gamma,\text{max}}} dL_{\gamma}\int_{z_\text{min}}^{z_\text{max}} dz \,  \Phi(L_\gamma,z,\Gamma) \, \frac{dV}{dz} \, \frac{dF}{dE}\,e^{-\tau_\text{\tiny{EBL}}(E,z)}}{\int_{E'_\text{min}}^{E'_\text{max}}  dE \int_{L_{\gamma,\text{min}}}^{L_{\gamma,\text{max}}} dL_{\gamma}\int_{z_\text{min}}^{z_\text{max}} dz \, \Phi(L_\gamma,z,\Gamma) \, \frac{dV}{dz} \, \frac{dF}{dE}\,e^{-\tau_\text{\tiny{EBL}}(E,z)}} \, ,
\end{equation}
where the exponential factor accounts for the attenuation due to extragalactic background light (EBL)~\citep{Gould:1966pza, Fazio:1970pr, 1992ApJ...390L..49S, Franceschini:2008tp, 2012Sci...338.1190A,Abramowski:2012ry,Dominguez:2013lfa}.  It arises from pair annihilation of high-energy gamma-ray photons with other background photons in infrared, optical, and/or ultraviolet, and is described by the optical depth, $\tau_\text{\tiny{EBL}}$.  We use the EBL attenuation model from~\cite{2010ApJ...712..238F}.  

Additionally, the expected gamma-ray spectrum can be calculated from the luminosity function as
\begin{equation}
\frac{dN}{dE} = \frac{1}{4\pi} \int_{\Gamma_\text{min}}^{\Gamma_\text{max}} d\Gamma \int_{z_\text{min}}^{z_\text{max}} dz \, \int_{L_{\gamma,\text{min}}}^{L_{\gamma,\text{max}}} dL_\gamma \, \Phi(L_\gamma,z,\Gamma) \, \frac{dV}{dz}\,\frac{dF}{dE} \, e^{-\tau_\text{\tiny{EBL}}(E,z)} \, .
\label{eq: dIdE}
\end{equation}
We use this equation to appropriately weight the number of photons per energy sub-bin for the individual sources when creating simulated maps. This ensures that the variations in PSF and exposure within the larger energy bins used in the NPTF analyses are properly accounted for in the simulation procedure.

\subsection{Simulations at Energies $\lesssim$ 1.89~GeV}
\label{app:verylowenergies}

The main analyses presented in the text do not consider energies below 1.89~GeV because we find that systematic effects may become more important at these low energies.  In particular, simulations done with the set of priors presented in Sec.~\ref{sec:methodology} show that the NPTF can over-estimate the PS intensity at very low energies.   
That is, in simulations with both blazars and SFGs, the isotropic-PS template tends to absorb more emission than is simulated in blazars, while the smooth isotropic template absorbs less emission than is simulated in SFGs.
 
\begin{figure*}[htbp] 
   	\begin{center}$
	\begin{array}{c}
	\scalebox{0.5}{\includegraphics{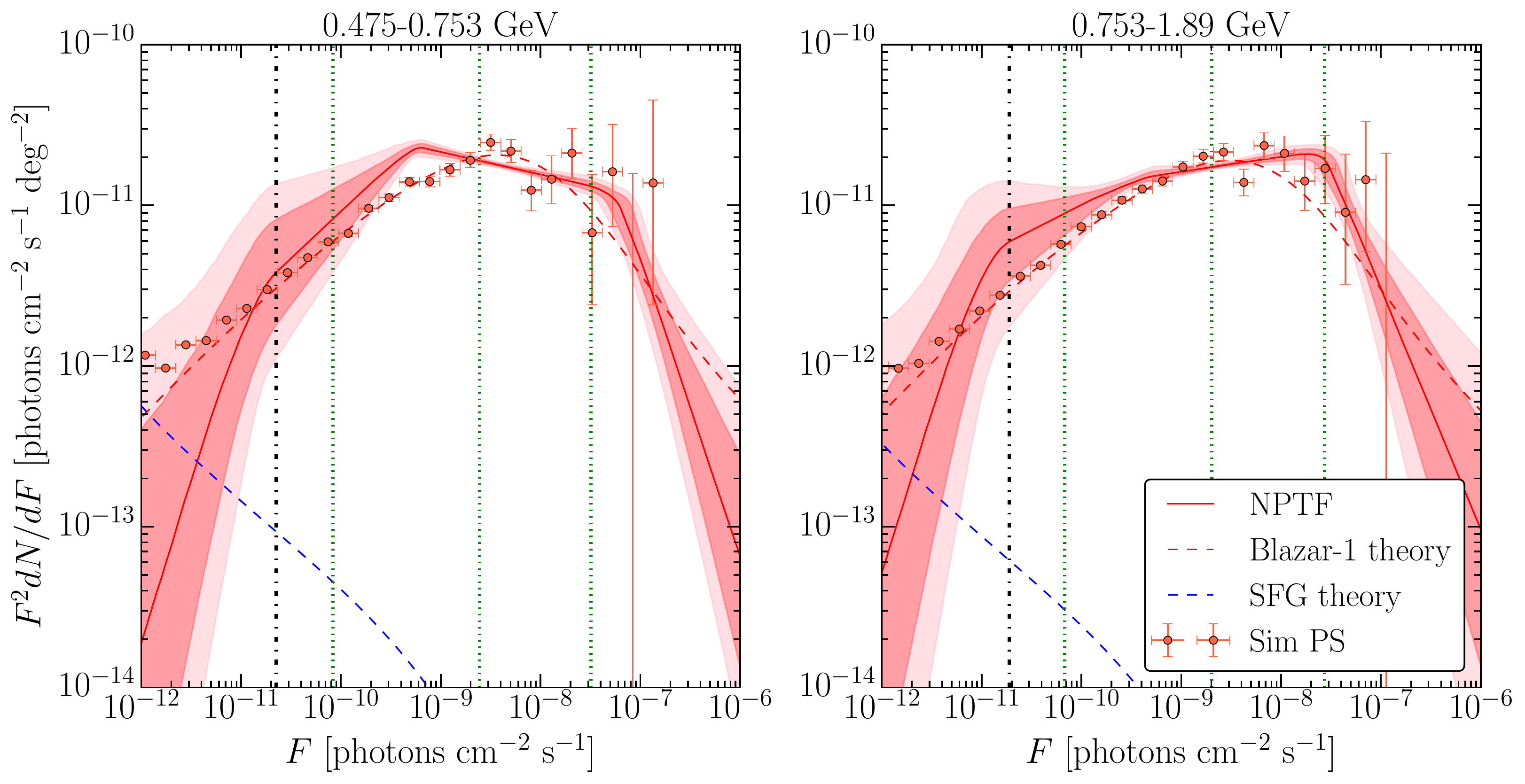}} 
	\end{array}$
	\end{center}
\caption{Best-fit source-count distribution for a simulated map containing both Blazar--1 and SFG sources in the 0.475--0.753 \emph{(left)} and 0.753--1.89 \emph{(right)}~GeV energy bins.  (Formatted as in Fig.~\ref{fig:bl1dnds}.) }
   \label{fig:dnds_verylow} 
\end{figure*}

As an illustration, we show the results when the NPTF is run on a simulated map of Blazar--1 and SFG sources.  Figure~\ref{fig:dnds_verylow} shows the best-fit source-count distributions for the energy ranges 0.475--0.753 and 0.753--1.89~GeV.  The PS fractions in these bins are $I_\text{iso}^\text{PS} / I_\text{blazar-sim} = [1.17_{-0.09}^{+0.13}, 1.26_{-0.08}^{+0.10}]$, illustrating that the PS template is absorbing more PS emission than is simulated in blazars.  From Fig.~\ref{fig:dnds_verylow}, we see that this is likely the result of the fit typically predicting more sources than it should at intermediate and low fluxes.  This could be due to a variety of factors.  For example, 
at very low energies the angular resolution of the detector quickly degrades, and the PS flux also becomes more and more sub-dominant compared to foreground emission.  For these reasons---and out of an abundance of caution---we do not present results of the NPTF on data below 1.89 GeV.  It is certainly possible that analyses at low energies could provide a wealth of interesting observations.  However, having confidence in the NPTF results at such low-energies requires a more careful consideration of systematics, which goes beyond the scope of the present work.  

\clearpage
\subsection{{\it Ultracleanveto} PSF1-3 Simulation Analysis}
\label{app:ucvsims}

\begin{figure*}[b] 
   \centering
   \includegraphics[width=\textwidth]{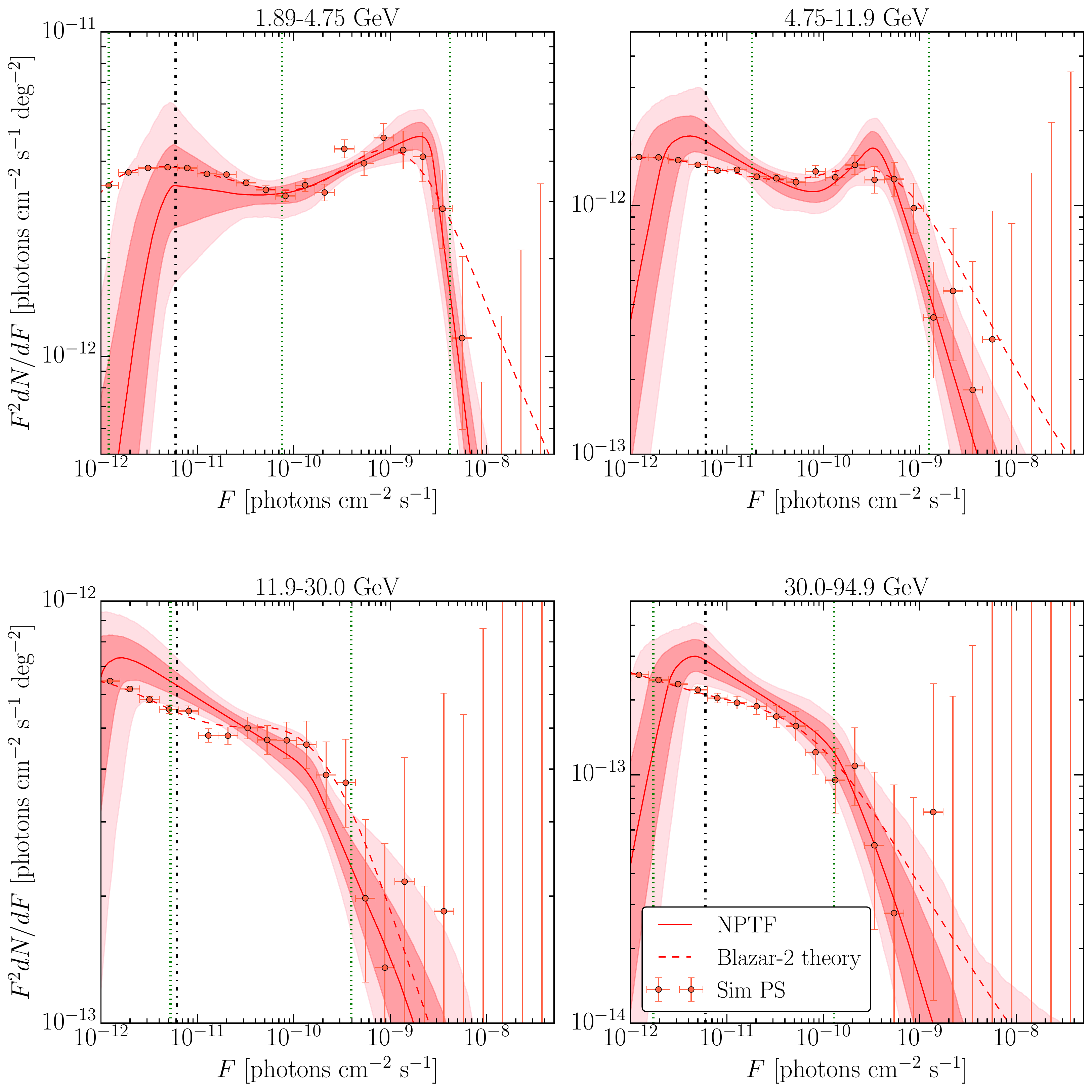} 
   \caption{Best-fit source-count distribution, as a function of energy, obtained from simulating the Blazar--2 model with top three quartiles of \emph{ultracleanveto} data.  (Formatted as in Fig.~\ref{fig:bl1dnds}.)}
   \label{fig:dndsb2UCV}
\end{figure*}

The simulated-data studies in the main text used the PSF3 instrument response function for the Pass~8 {\it ultracleanveto} data set.  Here, we show what happens when using the PSF1--3 instrument response function instead.  Including the top three quartiles of data increases the total exposure, though at the cost of decreased angular resolution.  Figure~\ref{fig:dndsb2UCV} illustrates the result for the Blazar--2 simulated data set.  The best-fit source-count distribution extends to lower fluxes than the corresponding distribution for top-quartile data in Fig.~\ref{fig:bl2dnds}. 

\subsection{Simulations of SFGs}
\label{app:sfgsims}

\begin{figure}[b] 
   	\begin{center}$
	\begin{array}{c}
	\scalebox{0.5}{\includegraphics{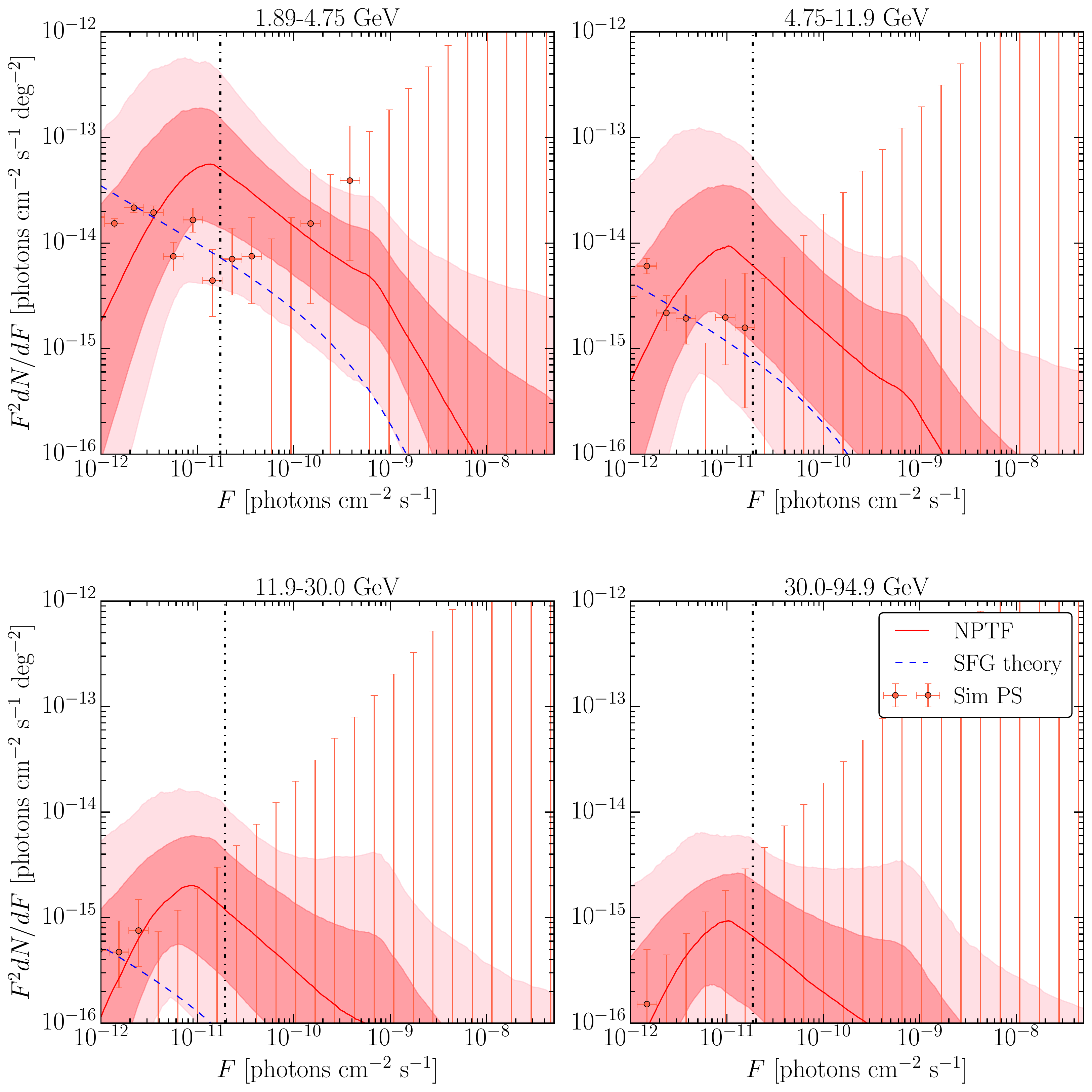}} 
		\end{array}$
	\end{center}
   \caption{Best-fit source-count distribution for a simulated map containing SFGs~\citep{Tamborra:2014xia}.  (Formatted as in Fig.~\ref{fig:bl1dnds}.) 
 }
   \label{fig:simdata_sfg} 
\end{figure}

\begin{figure}[htbp] 
   	\begin{center}$
	\begin{array}{c}
	\scalebox{0.5}{\includegraphics{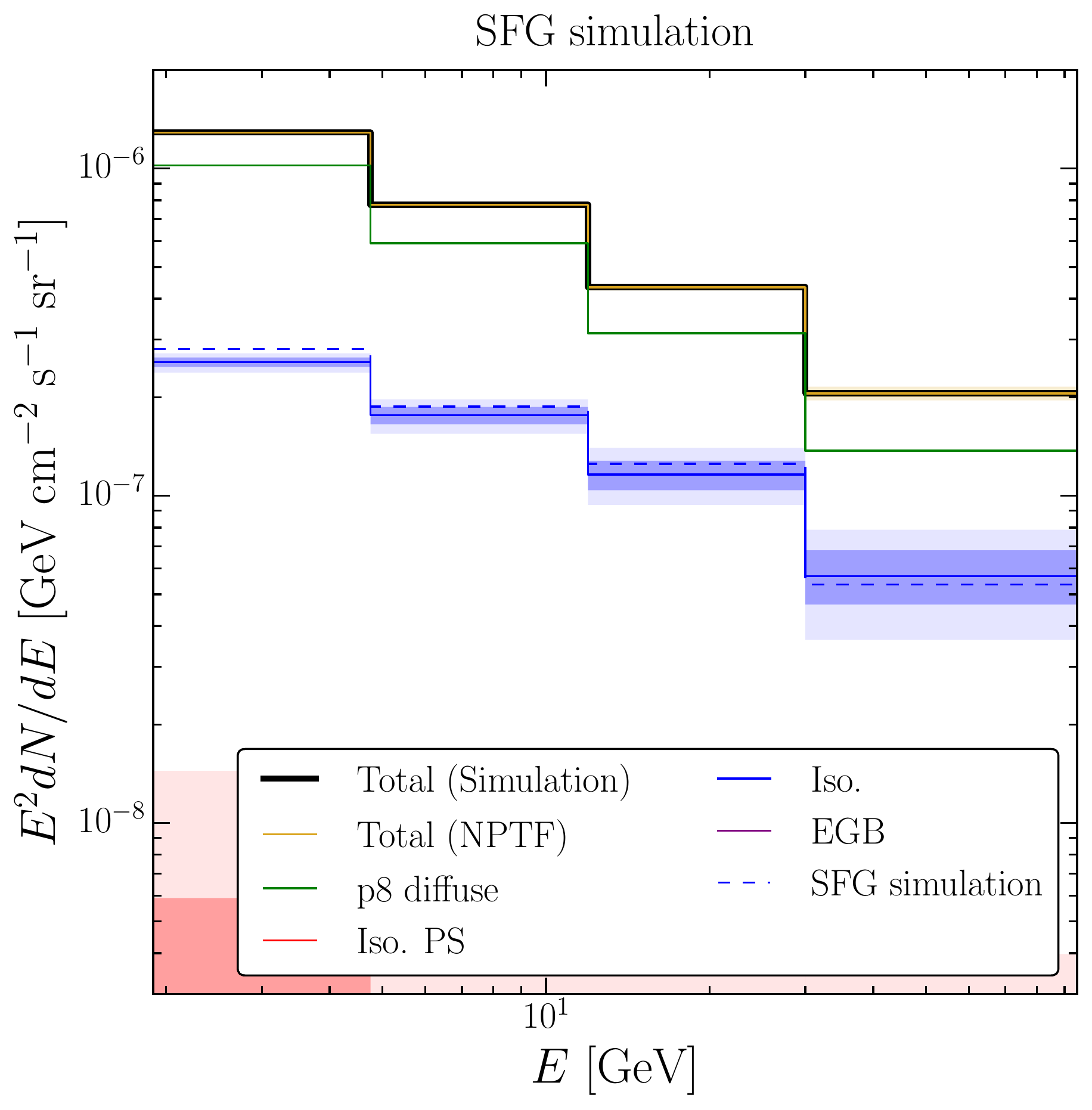}} 
		\end{array}$
	\end{center}
   \caption{Energy spectra for a simulated map containing SFGs~\citep{Tamborra:2014xia}.  (Formatted as in Fig.~\ref{fig:blESpec}.) }
   \label{fig:spec_sfg} 
\end{figure}

Here, we show the results of running the NPTF analysis on simulated data where the EGB arises solely from SFGs.  The resulting best-fit source-count distributions are shown in Fig.~\ref{fig:simdata_sfg}.  In the first energy bin, the brightest SFGs, which contribute little more than $\sim$1 photon, are detected as PSs by the NPTF.  At higher energies, the best-fit source-count distributions are consistent with zero and no significant evidence for a PS population is found.  The energy spectrum plot in Fig.~\ref{fig:spec_sfg} shows that the SFG flux is absorbed by the smooth isotropic template.  In comparison, the intensity absorbed by the isotropic-PS template is completely subdominant and is several orders of magnitude lower than its Poissonian counterpart at all energies.

%
%
\clearpage
\pagebreak

\section{Supplementary Results: Low-Energy Analysis}

\subsection{Best-Fit Intensities and Posterior Distributions}
\label{app:suppanalysis_low}

\begin{figure}[p] 
   \centering
   \includegraphics[trim={2cm 0 6cm 0},clip,width=\textwidth]{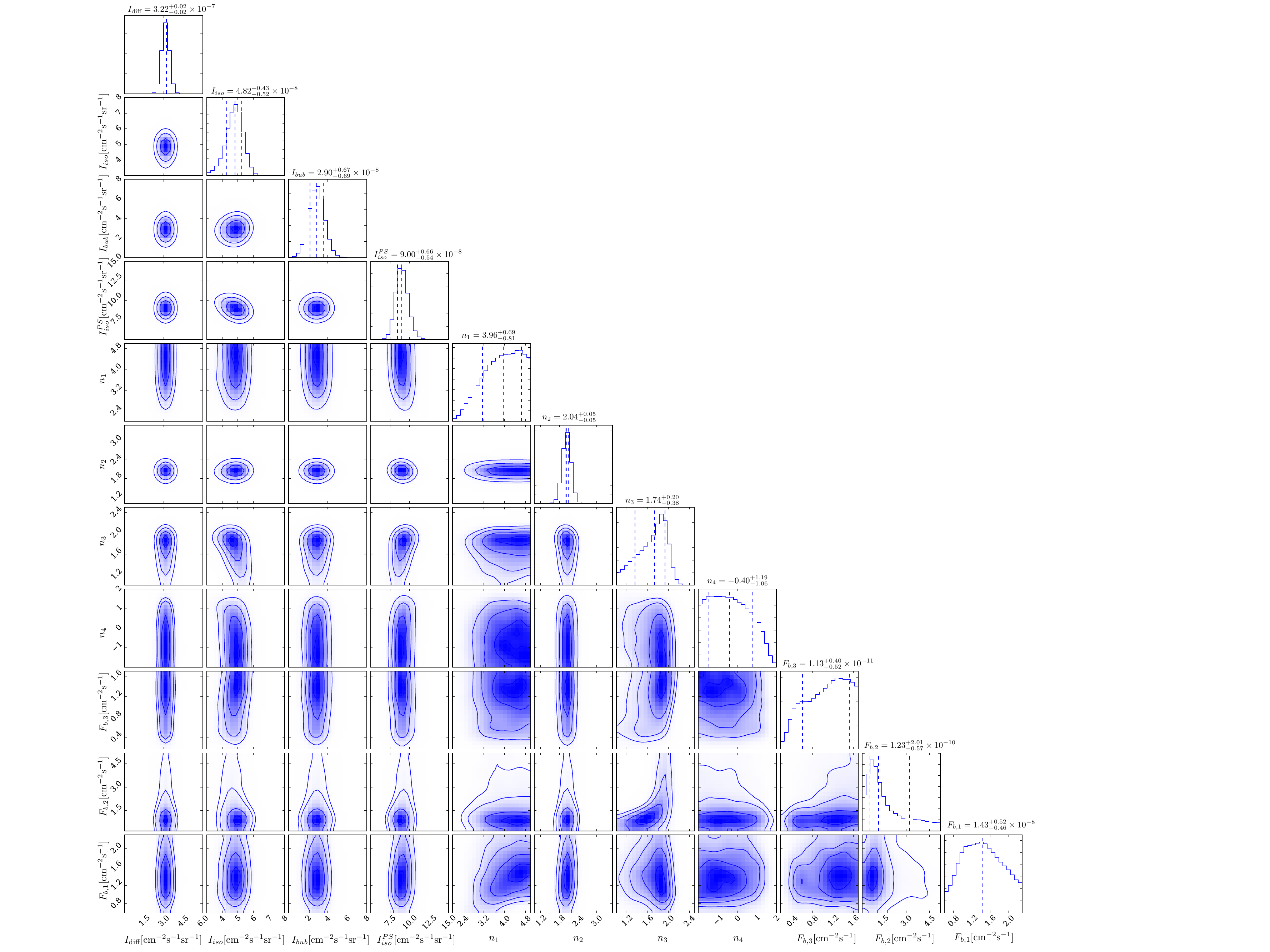} 
   \caption{Triangle plot for the 1.89--4.75~GeV bin.  The posterior distributions correspond to the NPTF analysis for Pass~8 {\it ultracleanveto} PSF3 data using the \texttt{p8r2} foreground model.  The \emph{Fermi} bubbles intensity is defined relative to the interior of the bubbles, while the intensities of the other templates are computed with respect to the region $\abs{b} \geq 30^\circ$. }  
   \label{fig:p8triangle1} 
\end{figure}

This section includes supplementary information pertaining to the low-energy analysis presented in Sec.~\ref{sec:benchmark}.  In particular, Tabs.~\ref{tab:bestfit_lowQ3} and~\ref{tab:bestfit_dndf_lowQ3} present the best-fit intensities and source-count parameters for the NPTF analysis of the top-three quartiles of {\it ultracleanveto} data.  Figures~\ref{fig:p8triangle1}--\ref{fig:p8triangle4} show the posterior distributions, in each energy bin, for the benchmark analysis.  
\vspace{-1.5in}
\begin{table}[phtb]
\renewcommand{\arraystretch}{1.4}
\setlength{\tabcolsep}{5pt}
\begin{center}
\begin{tabular}{ c  | c  c  c c  c   }
 Energy & $I_\text{EGB}$&$I_\text{iso}^\text{PS}$ & $I_\text{iso}$ & $I_\text{diff}$ & $I_\text{bub}$   \\
$[\text{GeV}]$ &  \multicolumn{5}{c}{$\left[\text{cm}^{-2}\text{ s}^{-1}\text{ sr}^{-1}\right]$}    \\
\hline
1.89--4.75  
&  $1.47_{-0.04}^{+0.05} \times 10^{-7}$ & $8.99_{-0.45}^{+0.53} \times 10^{-8}$ & $5.72_{-0.36}^{+0.30} \times 10^{-8}$ & $3.28_{-0.01}^{+0.01} \times 10^{-7}$ & $3.34_{-0.41}^{+0.39} \times 10^{-8}$\\
4.75--11.9 &    
$5.60_{-0.17}^{+0.21} \times 10^{-8}$ & $2.85_{-0.18}^{+0.23} \times 10^{-8}$ & $2.74_{-0.16}^{+0.12} \times 10^{-8}$ & $7.69_{-0.08}^{+0.09} \times 10^{-8}$ & $1.52_{-0.23}^{+0.22} \times 10^{-8}$   \\
11.9--30.0  & 
$1.85_{-0.07}^{+0.08} \times 10^{-8}$ & $8.54_{-0.71}^{+0.85} \times 10^{-9}$ & $9.92_{-0.63}^{+0.53} \times 10^{-9}$ & $1.68_{-0.04}^{+0.04} \times 10^{-8}$ & $5.59_{-1.31}^{+1.35} \times 10^{-9}$  \\  
30.0-94.9  &  
$6.19_{-0.29}^{+0.42} \times 10^{-9}$ & $2.85_{-0.33}^{+0.39} \times 10^{-9}$ & $3.36_{-0.30}^{+0.25} \times 10^{-9}$ & $3.77_{-0.19}^{+0.17} \times 10^{-9}$ & $1.10_{-0.61}^{+0.72} \times 10^{-9}$  \\
\end{tabular}
\end{center}
\caption{Same as Tab.~\ref{tab:bestfit}, except using the top three quartiles (PSF1--3) of the Pass~8 {\it ultracleanveto} data. }
\label{tab:bestfit_lowQ3}
\end{table}

\vspace{-4in}
\begin{table}[phtb]
\renewcommand{\arraystretch}{1.4}
\setlength{\tabcolsep}{3pt}
\begin{center}
\begin{tabular}{ c  | c  c  c c |  c c c   }
 Energy & $n_1$ & $n_2$ & $n_3$ & $n_4$ & $F_{b,1}$ & $F_{b,2}$ & $F_{b,3}$   \\
$[\text{GeV}]$ &  & & & & \multicolumn{3}{c}{$\left[\text{cm}^{-2}\text{ s}^{-1}\right]$}    \\
\hline
1.89--4.75 &  
$3.97_{-0.83}^{+0.69}$ & $2.10_{-0.03}^{+0.04}$ & $1.75_{-0.14}^{+0.09}$ & $-0.45_{-1.01}^{+1.10}$ & $8.16_{-3.00}^{+6.26} \times 10^{-9}$& $6.63_{-2.31}^{+3.45} \times 10^{-11}$ &$3.85_{-1.73}^{+1.38} \times 10^{-12}$     \\
4.75--11.9 &    
$3.94_{-0.85}^{+0.68}$ & $2.08_{-0.05}^{+0.08}$ & $1.94_{-0.19}^{+0.07}$ & $-0.46_{-1.01}^{+1.07}$ & $3.85_{-1.65}^{+2.69} \times 10^{-9}$& $8.05_{-5.66}^{+7.34} \times 10^{-11}$ & $4.00_{-1.64}^{+1.36} \times 10^{-12}$    \\
11.9--30.0  & 
$3.70_{-0.90}^{+0.82}$ & $2.20_{-0.14}^{+0.17}$ & $2.02_{-0.08}^{+0.07}$ & $-0.37_{-1.04}^{+1.05}$ & $2.69_{-1.34}^{+1.37} \times 10^{-9}$& $1.19_{-0.63}^{+0.45} \times 10^{-10}$ & $3.78_{-1.50}^{+1.33} \times 10^{-12}$ 
  \\  
30.0-94.9  &  
$3.54_{-0.90}^{+0.94}$ & $2.15_{-0.28}^{+0.24}$ & $2.21_{-0.11}^{+0.13}$ & $-0.09_{-1.20}^{+1.06}$ &  $1.59_{-0.73}^{+0.75} \times 10^{-9}$& $1.07_{-0.59}^{+0.49} \times 10^{-10}$ &  $3.37_{-1.48}^{+1.51} \times 10^{-12}$ \\
\end{tabular}
\end{center}
\caption{Same as Tab.~\ref{tab:bestfit_dndf}, except using the top three quartiles (PSF1--3) of the Pass~8 {\it ultracleanveto} data.  }
\label{tab:bestfit_dndf_lowQ3}
\end{table}

\afterpage{%
\begin{figure}[p] 
   \centering
   \includegraphics[trim={2cm 0 6cm 0},clip,width=\textwidth]{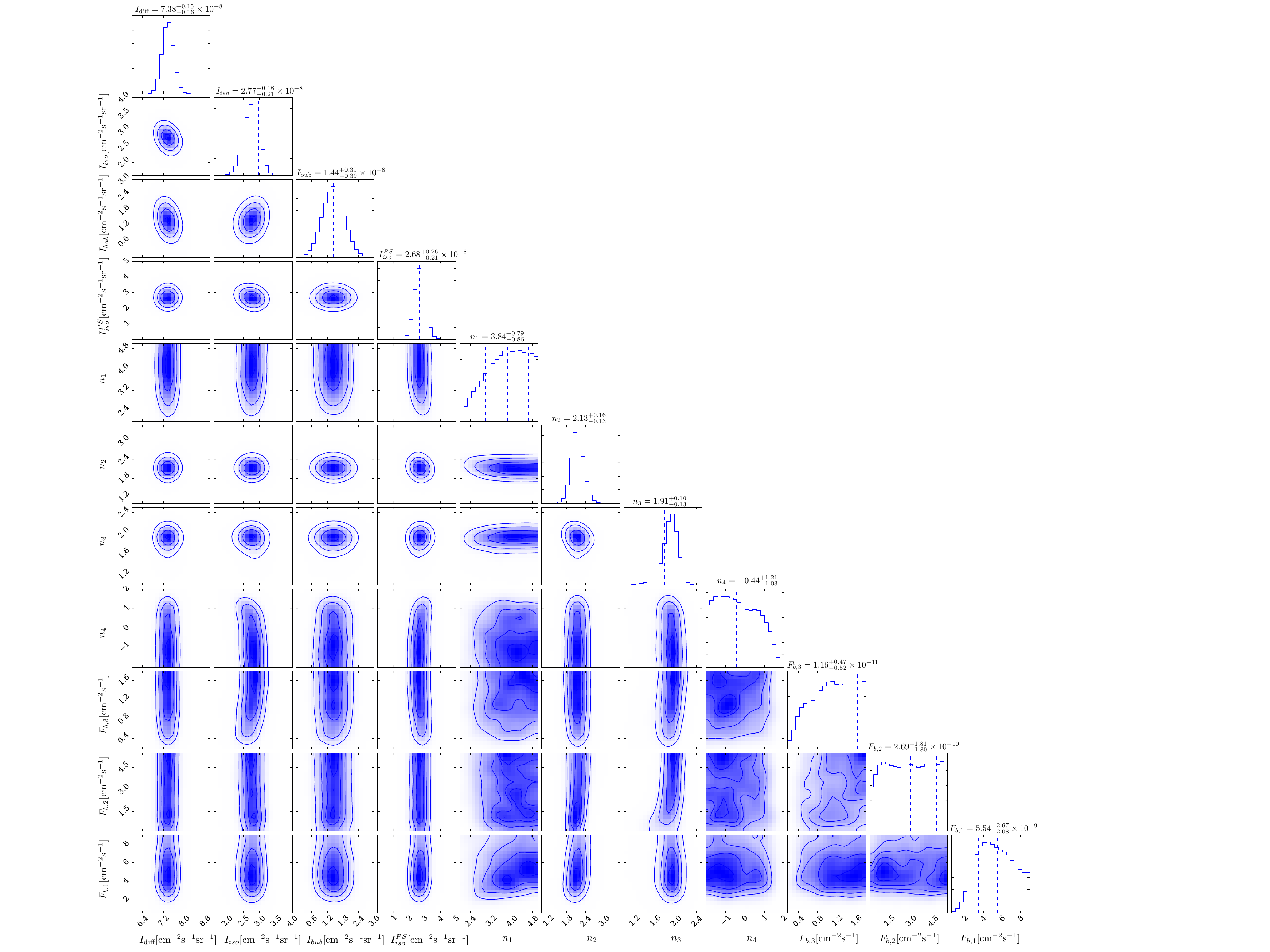} 
   \caption{Same as Fig.~\ref{fig:p8triangle1}, except for 4.75--11.9~GeV. }  
   \label{fig:p8triangle2} 
\end{figure}
\clearpage
}

\afterpage{%
\begin{figure}[tb] 
   \centering
   \includegraphics[trim={2cm 0 6cm 0},clip,width=\textwidth]{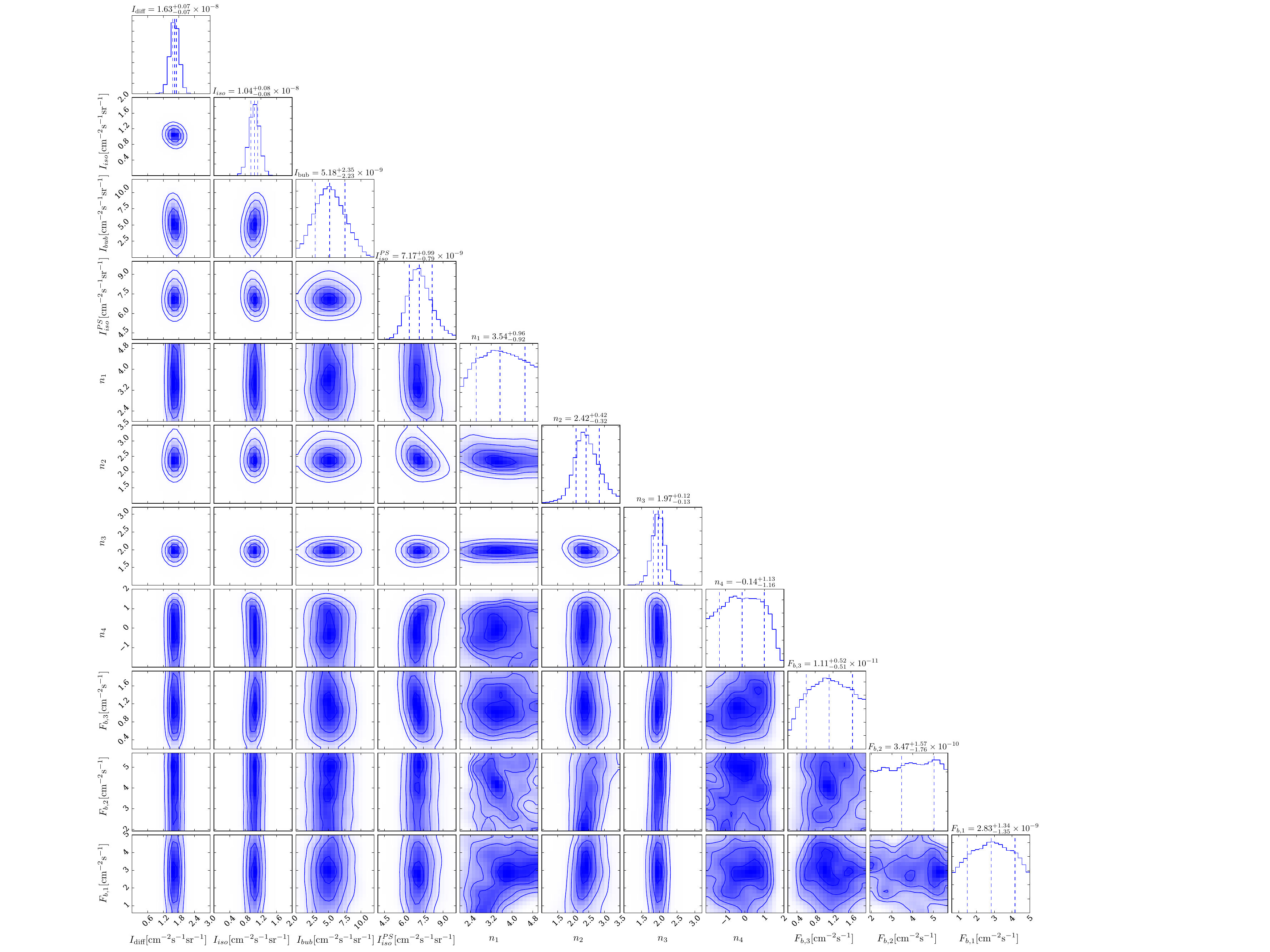} 
   \caption{Same as Fig.~\ref{fig:p8triangle1}, except for 11.9--30.0~GeV.}  
   \label{fig:p8triangle3} 
\end{figure}
\clearpage
}

\afterpage{%
\begin{figure}[tb] 
   \centering
   \includegraphics[trim={4cm 0 4cm 0},clip,width=\textwidth]{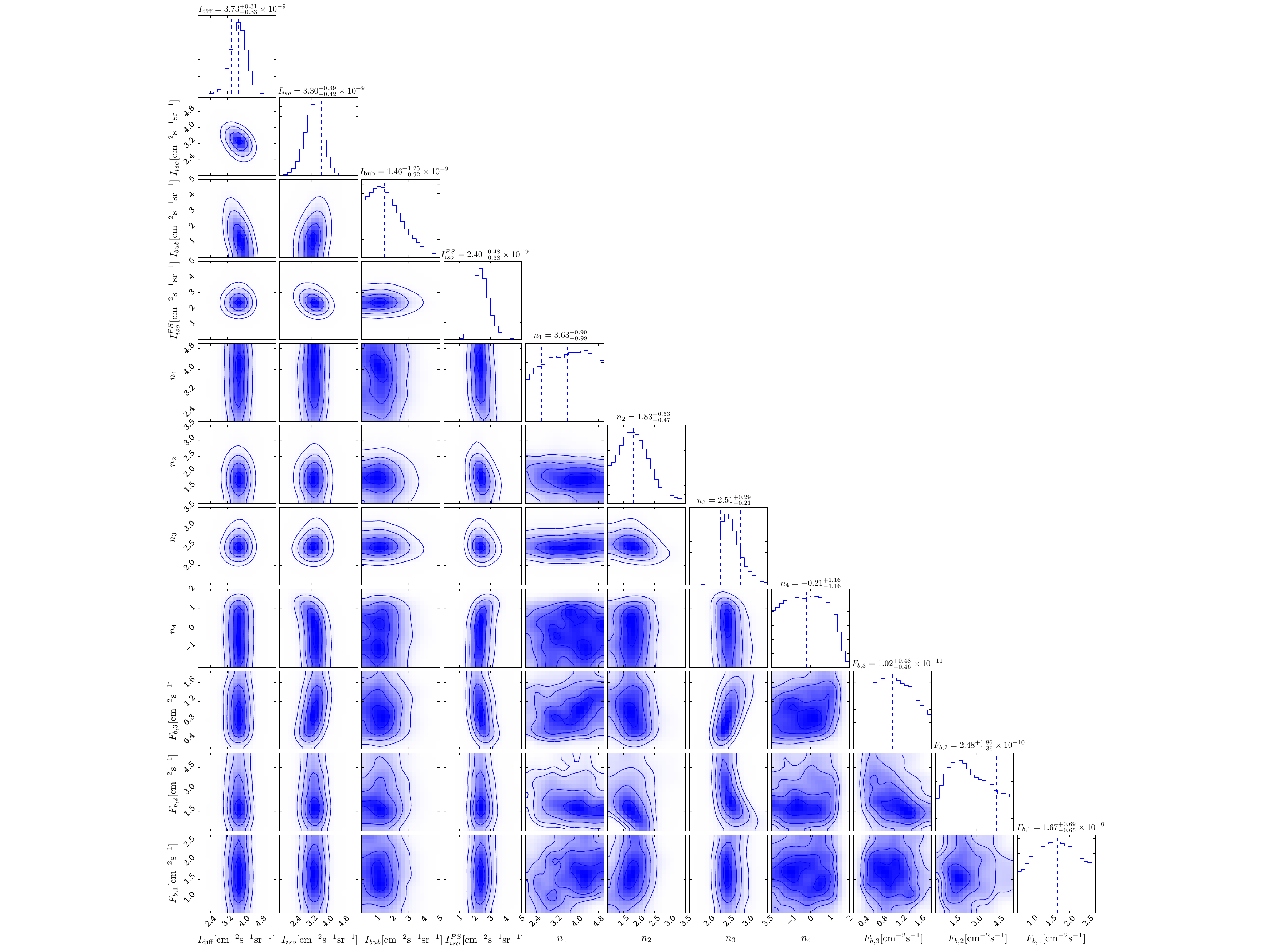} 
   \caption{Same as Fig.~\ref{fig:p8triangle1}, except for 30.0-94.9~GeV.}   
   \label{fig:p8triangle4} 
\end{figure}
\clearpage
}
\pagebreak

\subsection{Systematic Tests}
\label{app:systematics}

We now describe in detail the systematic tests that were conducted for the low-energy analysis, the results of which are summarized in Fig. \ref{fig:systematicsplot}.  The primary conclusion that we draw is that the the PS fraction is stable under the variety of tests that we have explored.  

\subsubsection{Region of Interest}

As a first cross-check on the stability of the results presented in Sec.~\ref{sec:benchmark}, we explore the effects of altering the region of interest.  While we previously defined the region of interest with $|b| \geq 30^\circ$, we now loosen this constraint and consider the case $|b| \geq 10^\circ$.  Extending the region of interest closer to the Galactic disk increases the amount of data being analyzed, but at the cost of potentially more contamination from diffuse foreground emission and local PSs.  As shown in Fig.~\ref{fig:systematicsplot}, the best-fit intensities for the isotropic and isotropic-PS components are equivalent, within errors, to their counterparts in the benchmark analysis. The best-fit source-counts are shown in Fig.~\ref{fig:dndsdata_m10}.

We also ran the NPTF on the Northern ($b > 30^\circ$) and Southern ($b < -30^\circ$) hemispheres separately.  The intensities for the EGB, IGRB, and PS components are systematically lower (higher) for the Northern (Southern) analysis than for the benchmark case.  Similar behavior is apparent in the source-count plots, shown in Figs.~\ref{fig:dndsdata_north} and \ref{fig:dndsdata_south}.

\subsubsection{Event class}

We explored the implications of broadening the {\it ultracleanveto} data set to include the top three quartiles in Sec. \ref{sec:benchmark_top3}.   Now, we consider the implications of repeating the NPTF analysis on the {\it source} data with PSF1--3.  This event class has looser photon-quality cuts, which leads to larger overall exposure, but significantly more cosmic-ray contamination.  In general, it is not recommended to use {\it source} data for IGRB studies; for our purposes, however, it will be intriguing to see how the increased photon statistics affect the recovered source-count distribution for the PS component.  As shown in Fig.~\ref{fig:systematicsplot}, the EGB intensity is far larger than that recovered by the benchmark analysis and overpredicts \emph{Fermi}'s EGB result in most energy bins.  The sharp rise in the EGB intensity can be traced to a substantial fraction of smooth isotropic emission, which is expected for this event class at most energies.  Most importantly, the intensity of the isotropic-PS component is consistent, within uncertainties, with that found in the benchmark analysis.\footnote{The recovered PS intensity is slightly larger with {\it source} PSF1--3 data as compared to {\it ultracleanveto} PSF3 data, which is likely due to the increased exposure in the {\it source} PSF1--3 data set. 
}   This is a  confirmation that the NPTF is able to successfully constrain the source-count distribution even in a data set with significantly more smooth isotropic flux.  

The source-count distributions are provided in Fig.~\ref{fig:dndsdata_source}.  In general, they exhibit similar behavior to the {\it ultracleanveto} PSF1--3 functions, extending to lower fluxes due to the increased exposure for this event class.  One potential new feature of interest in the {\it source}-data source-count distributions is that, in the second energy bin from $4.75$--$11.9$ GeV, there is a more pronounced hardening of the source-count distribution below the second break $F_{b,2}$, as compared to the {\it ultracleanveto} PSF1--3 analyses.

\subsubsection{Foreground Model}
\label{sec:foreground}

A potentially significant source of systematic uncertainty in the NPTF analysis is due to mis-modeling of high-energy gamma-rays produced in cosmic-ray propagation in the Milky Way~\citep{Ackermann:2012pya}.  These high-energy photons arise from  bremsstrahlung of electrons off the interstellar medium, boosted pion decay, and inverse Compton (IC)  emission off the interstellar radiation field.  Our benchmark analysis uses 
the associated foreground model for the Pass~8 data set (\emph{gll\_iem\_v06.fits}), denoted here as \texttt{p8r2}.  The total diffuse emission in \texttt{p8r2} is modeled as a linear combination of several sources, some of which are traced by maps of gas column densities, which serve as templates for the pion and bremsstrahlung emission.  The IC component is modeled using the \texttt{GALPROP} package~\citep{Strong:2007nh}.\footnote{\url{http://galprop.stanford.edu/}}  These individual templates are fit to the data, and used to identify `extended emission excesses' that are identified directly and then added back into the model~\citep{Acero-2016}.

To better assess the uncertainties due to the foreground modeling, we repeat the NPTF analysis using several other foreground models made available by \emph{Fermi}.  In particular, we use the \emph{gll\_iem\_v02\_P6\_V11\_DIFFUSE.fits} diffuse emission model, denoted as \texttt{p6v11}, which was initially developed for the Pass~6 data set.\footnote{\url{http://fermi.gsfc.nasa.gov/ssc/data/access/lat/ring_for_FSSC_final4.pdf}}  \texttt{p6v11} is distinct from \texttt{p8r2} in that it uses older gas and IC maps and does not include templates for large-scale structure or extended emission excesses.  The Pass~7 model \emph{gal\_2yearp7v6\_v0.fits}, denoted as \texttt{p7v6},\footnote{\url{http://fermi.gsfc.nasa.gov/ssc/data/access/lat/Model_details/Pass7_galactic.html}} is a compromise as it uses updated gas and IC maps and includes some large-scale extended structures, such as Loop~1 and the \emph{Fermi} bubbles.

The NPTF results using the \texttt{p6v11} and \texttt{p7v6} foreground models are summarized in Fig.~\ref{fig:systematicsplot}, with source-count distributions  shown in Figs.~\ref{fig:dndsdata_p6} and~\ref{fig:dndsdata_p7}, respectively. In general, we observe that the intensity of the PS components is consistent with that for the benchmark analysis in all energy bins.  However, variations occur in the smooth isotropic intensity.  Typically, more IGRB intensity is recovered with \texttt{p6v11} and \texttt{p7v6}, versus \texttt{p8r2}.  The differences are particularly dramatic in the first two energy bins and are more severe for \texttt{p6v11}.  The net consequence is that the EGB intensity is higher than the expected range from \emph{Fermi}.  The enhancement in the isotropic component may arise from the fact that each foreground model incorporates large-scale diffuse structures differently---with \texttt{p6v11} being the least inclusive and \texttt{p8r2} being the most inclusive.  We note, however, that the fit to data with the \texttt{p8r2} foreground model, from the point of view of the Bayesian evidence, is much better than the analogous fit with the  \texttt{p6v11} model; the fit with the \texttt{p7v6} model is intermediate. 

\subsubsection{The Bubbles Template}

To better understand how dependent the analysis is on the details of the \emph{Fermi} bubbles template, we simply removed the template from the analysis.  This has indiscernible effects on the final results.  We see in Fig.~\ref{fig:systematicsplot} that the EGB, IGRB, and PS intensities are consistent, within uncertainties, to the corresponding values in the benchmark study.  The source-count distributions, shown in Fig.~\ref{fig:dndsdata_nobub}, are also degenerate with those found including the \emph{Fermi} bubbles template.   

\subsubsection{Point Spread Function}

The PSF can affect the photon-count distribution because it can redistribute photons between pixels, and must therefore be properly accounted for in the calculation of the photon-count probability distributions.  For the primary analyses presented in this work, the PSF is modeled using a King function.  However, to test the sensitivity of the results to mis-modeling of the PSF, we have also repeated the NPTF analysis using a two-dimensional Gaussian in the calculation of the photon-count probability distributions, with a width set to give the correct 68\% containment radius.  As shown in Fig.~\ref{fig:systematicsplot}, the NPTF results remain unchanged with this substitution.  The best-fit source-count distribution for 1.89--4.75~GeV shows some variation at the lowest fluxes, but within uncertainties (see Fig.~\ref{fig:dndsdata_gauss}).    

\subsubsection{Priors}

Our choice of priors, given in Tab.~\ref{tab:priors}, is carefully chosen to both avoid biasing the posterior for the source-count distribution while at the same time allowing breaks at both high and low flux.  This is meant to properly account for the fact that the source-count distribution is not well constrained by the data at very high fluxes, where the mean expected number of sources over the full region is much less than unity, and at very low fluxes, where the mean photon-count per source is much less than unity.  Our choice of priors is further justified by the simulated data studies, presented in Sec.~\ref{sec:simulations}, which show that the NPTF can successfully constrain the emission from blazar models.  However, one may still be concerned that these particular choice of priors might bias the recovered source-count distribution in a particular way.  For that reason, we have tried many variations to the priors shown in Tab.~\ref{tab:priors}, three of which (labeled `alternate priors 1--3')  are described below and shown in Fig.~\ref{fig:systematicsplot}:
\begin{itemize}
\item {\it Alternate prior 1}:  All priors are the same as in Tab.~\ref{tab:priors}, except for those on the breaks, which are changed to $[0.1,10]$, $[10,40]$, and $[40, 2 \times S_\text{b,max}]$ ph for $S_{b,1}$, $S_{b,2}$, and $S_{b,3}$, respectively. 
\item {\it Alternate prior 2}:  As above, except changing the priors for the breaks to $[1,20]$, $[20,S_\text{b,max}/2]$, and $[S_\text{b,max}/2, 2 \times S_\text{b,max}]$ ph, respectively.
\item {\it Alternate prior 3}:  All priors are the same as in Tab.~\ref{tab:priors}, except for that of $n_4$, which is changed to $[1,1.99]$.
\end{itemize}

The first two examples address the possibility that the break priors might artificially sculpt the source-count distribution and the recovered PS intensity, while the third example addresses how the source-count distribution is dealt with at fluxes below the lowest break, where the distribution is not well constrained by the data.  In many classes of blazar models, such as those considered in Sec.~\ref{sec:simulations}, the index below the lowest break ($n_4$) is greater than unity, so that the total number of PSs $\sim$$\int_{F_\text{min}} dF \, dN / dF$ diverges as the minimum flux cut-off $F_\text{min}$ is taken to zero.

It is useful to know if the recovered PS intensity, $I_\text{iso}^\text{PS}$, tends to under or overshoot the simulated blazar intensity, $I_\text{blazar-sim}$, when using the alternate priors.  With that in mind, we run the NPTF on simulated maps, as in Sec.~\ref{sec:simulations}, constructed from both the SFG + Blazar--1 model as well as the SFG + Blazar--2 model.   For \emph{Alternate prior 1}, we find that
\begin{equation}
{I_\text{iso}^\text{PS} \over I_\text{blazar-sim}} = [0.87_{-0.04}^{+0.05},0.93_{-0.08}^{+0.17},0.92_{-0.15}^{+0.23},0.61_{-0.07}^{+0.11}]  \nonumber
\end{equation}
and
\begin{equation}
{I_\text{iso}^\text{PS} \over I_\text{blazar-sim}} = [0.68_{-0.05}^{+0.06},0.59_{-0.09}^{+0.15},0.52_{-0.05}^{+0.07},0.37_{-0.03}^{+0.05}]  \nonumber
\end{equation} 
for the SFG + Blazar--1 and SFG + Blazar--2 models, respectively, with {\it ultracleanveto} PSF3 instrument response function.
With {\it Alternate prior 1}, we see larger uncertainties, with the PS template capable of absorbing more flux in particular.  With \emph{Alternate prior 2}, on the other hand, we find more noticeable differences in the medians as well as in the uncertainties.  In particular, for the SFG + Blazar--1 and SFG + Blazar--2 models, we find
\begin{equation}
{I_\text{iso}^\text{PS} \over I_\text{blazar-sim}} = [1.01_{-0.10}^{+0.12}, 1.27_{-0.31}^{+0.16}, 1.25_{-0.15}^{+0.12}, 0.73_{-0.12}^{+0.21}]  \nonumber
\end{equation}
and
\begin{equation}
{I_\text{iso}^\text{PS} \over I_\text{blazar-sim} }= [0.74_{-0.06}^{+0.19}, 0.94_{-0.19}^{+0.20}, 0.61_{-0.10}^{+0.17}, 0.41_{-0.05}^{+0.09}] \,, \nonumber
\end{equation} 
respectively.  In the Blazar--1 model case, it is important to notice that at intermediate energies the NPTF tends to over-predict $I_\text{blazar-sim}$ at the $\sim$20\% level.  With \emph{Alternate prior 3}, the results are 
\begin{equation}
{I_\text{iso}^\text{PS} \over I_\text{blazar-sim}} = [1.06_{-0.09}^{+0.15}, 1.10_{-0.09}^{+0.14}, 1.00_{-0.10}^{+0.14}, 0.85_{-0.11}^{+0.15}]  \nonumber
\end{equation}
and
\begin{equation}
{I_\text{iso}^\text{PS} \over I_\text{blazar-sim}} = [0.92_{-0.09}^{+0.16}, 0.77_{-0.14}^{+0.39}, 0.69_{-0.08}^{+0.12}, 0.53_{-0.06}^{+0.10}] \,, \nonumber
\end{equation} 
for the Blazar--1 and Blazar--2 models.  The \emph{Alternate prior 3} results are consistently closer to unity than the first two alternate prior results. 

As may be seen in Fig.~\ref{fig:systematicsplot}, the median values for the PS intensities recovered from the NPTF analyses with alternate priors are generally consistent with those found in the baseline study.  The {\it Alternate prior 3} PS intensities are slightly enhanced in all energy bins compared to the baseline results---following our expectations from the simulation results presented above---though the two results are consistent within statistical uncertainties.  The recovered source-count distributions, shown in Fig.~\ref{fig:dndsdata_altpriors3}, illustrate that the {\it Alternate prior 3} results are consistent with our baseline results at fluxes above the $\sim$1 photon threshold.  At lower fluxes, the source-count distributions are slightly softer than in our baseline analysis, as is almost guaranteed by that fact that $n_4 > 1$ with {\it Alternate prior 3} while $n_4 > -2$ in our baseline analysis.   

The {\it Alternate prior 1} and {\it Alternate prior 2} results have mean PS intensities similar to those in the baseline analysis, though in the second and third energy bin the upper limits of the credible intervals extend to higher values.  This is a reflection of the fact that the recovered source-count distributions, shown in Figs.~\ref{fig:dndsdata_altpriors1}, \ref{fig:dndsdata_altpriors2}, are softer at low fluxes compared to those in the baseline analysis.  This is perhaps due to the fact that the lowest break tends to be at higher flux, and thus the index $n_4$ is influenced by the data in the vicinity of the break.  

\afterpage{
\begin{figure*}[phtb] 
   \centering
   \includegraphics[width=\textwidth]{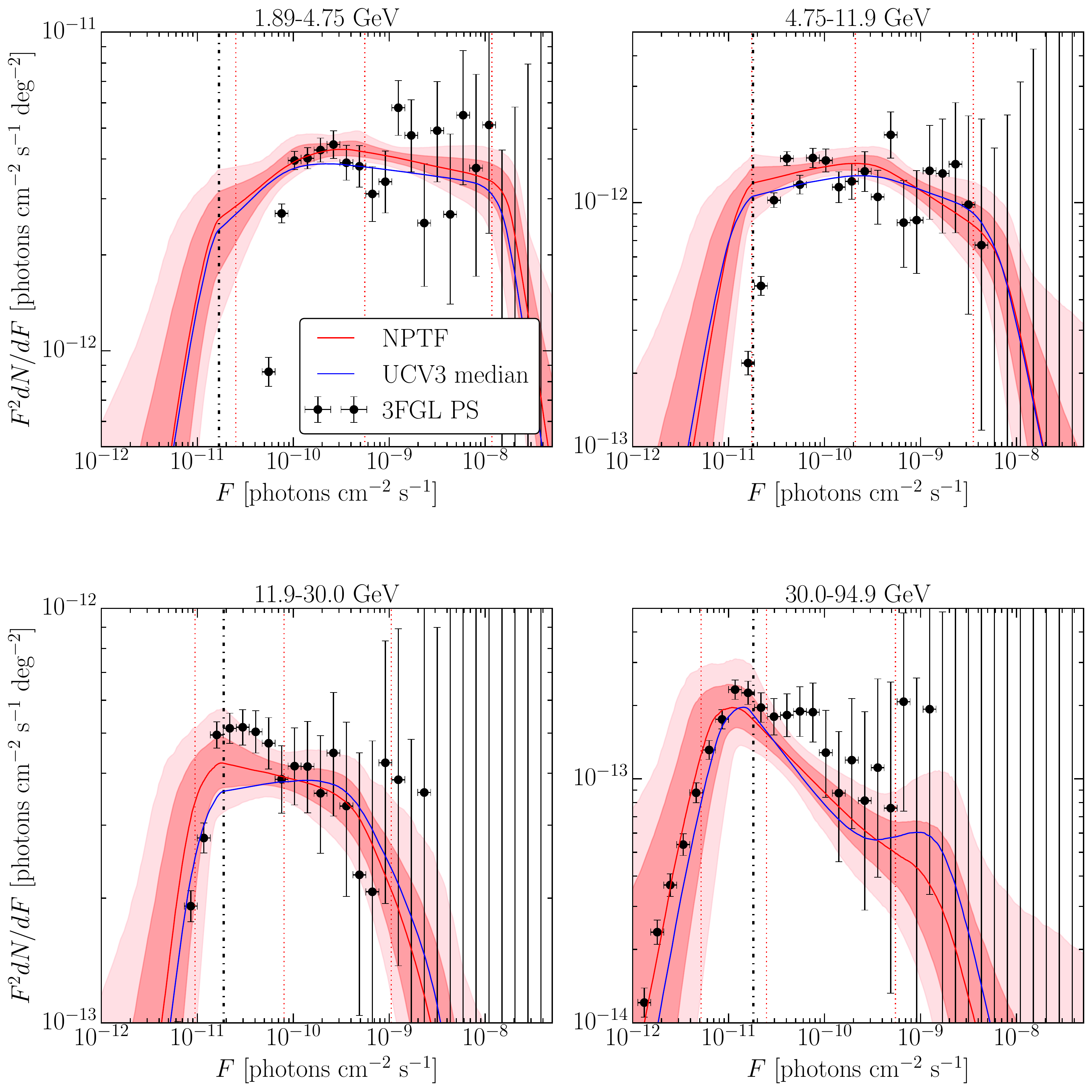} 
   \caption{Best-fit source-count distribution using the Pass 8 {\it ultracleanveto} PSF3 data set and \texttt{p8r2} foreground model, but with $|b| > 10^\circ$.  The median source-count distribution for the benchmark analysis is shown in blue. (Formatted as in Fig.~\ref{fig:dndsdata}.)}
   \label{fig:dndsdata_m10}
\end{figure*}
\clearpage}

\afterpage{
\begin{figure*}[phtb] 
   \centering
   \includegraphics[width=\textwidth]{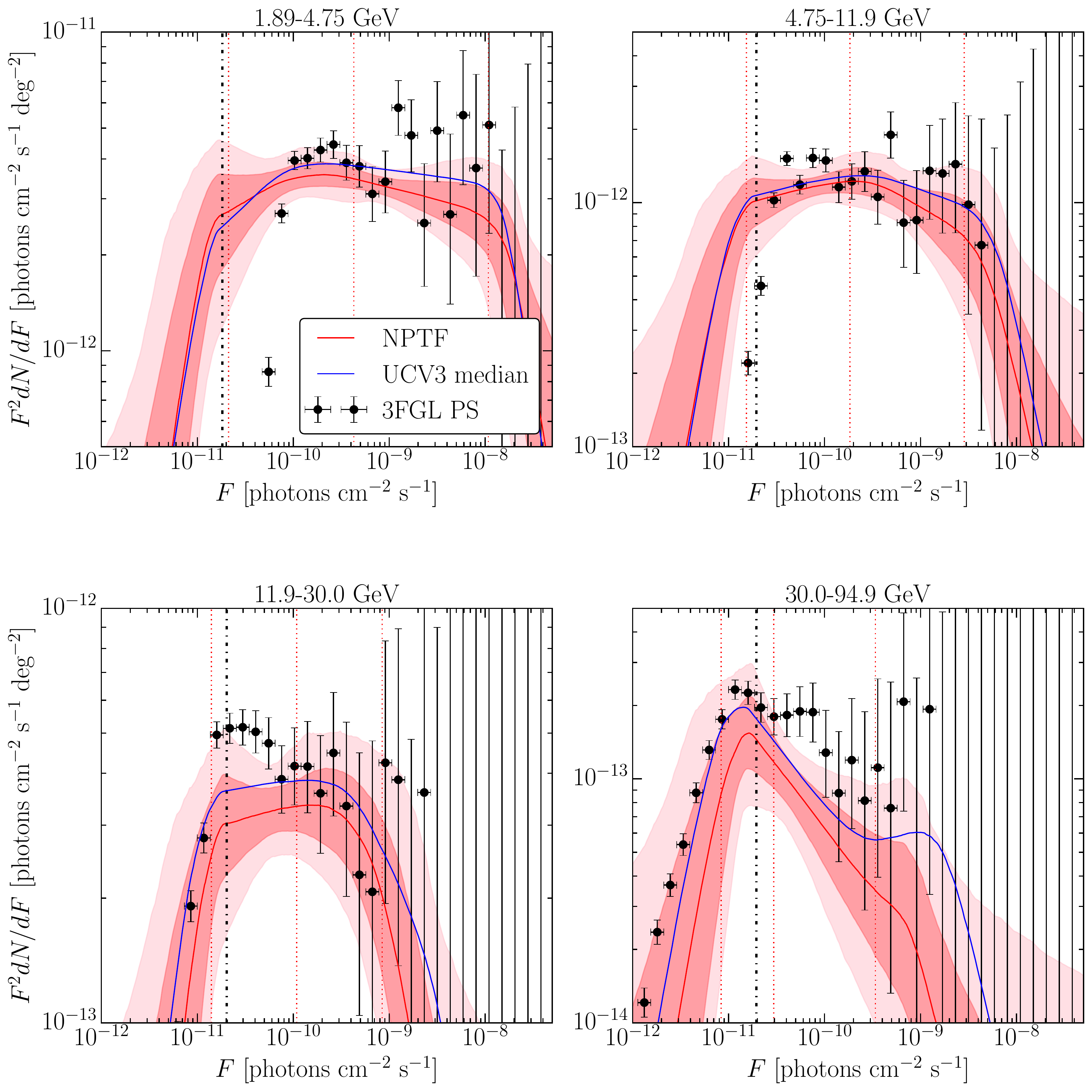} 
   \caption{Best-fit source-count distribution using the Pass 8 {\it ultracleanveto} PSF3 data set and \texttt{p8r2} foreground model, but with $b > 30^\circ$.  The median source-count distribution for the benchmark analysis is shown in blue.  (Formatted as in Fig.~\ref{fig:dndsdata}.)}
   \label{fig:dndsdata_north}
\end{figure*}
\clearpage}

\afterpage{
\begin{figure*}[phtb] 
   \centering
   \includegraphics[width=\textwidth]{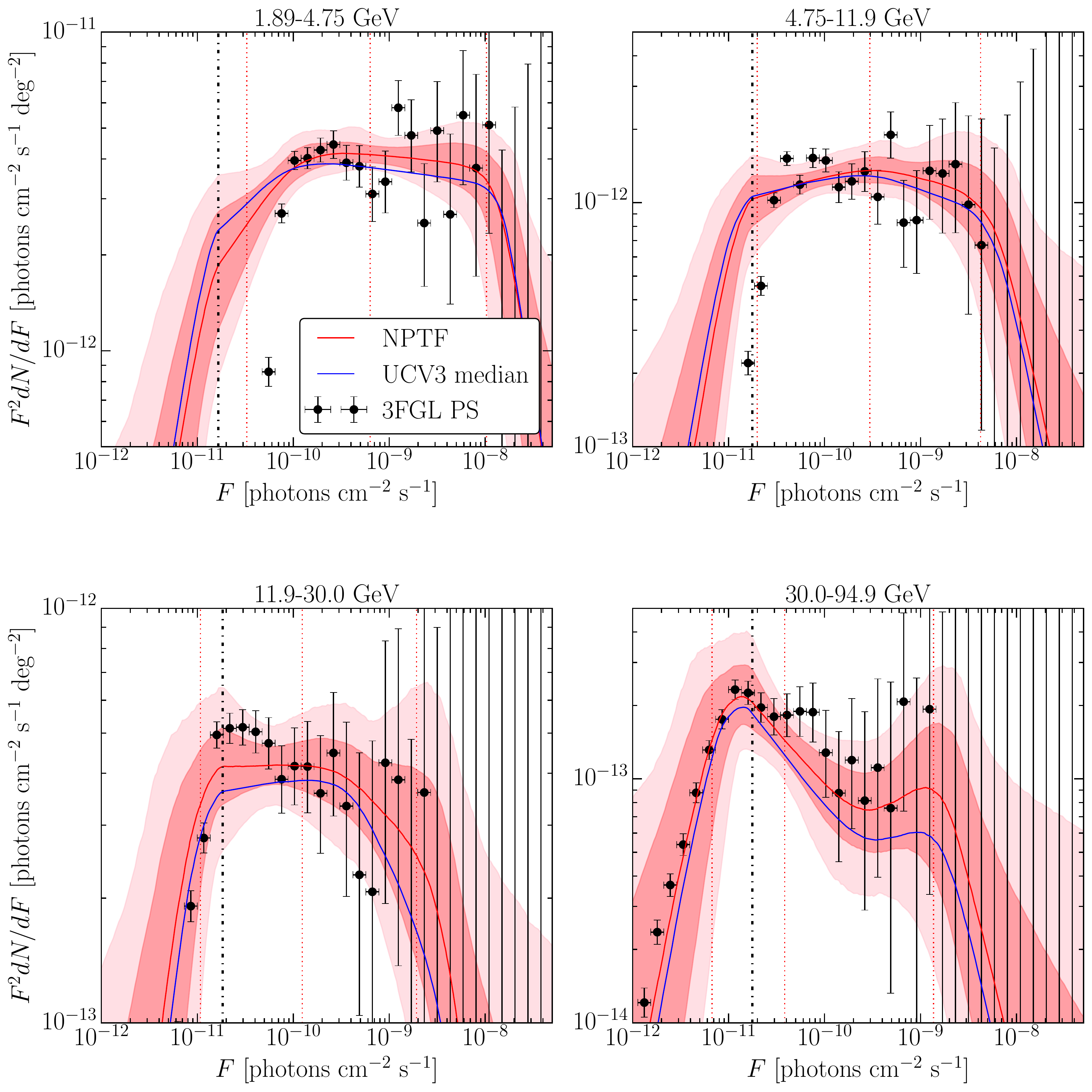} 
   \caption{Best-fit source-count distribution using the Pass 8 {\it ultracleanveto} PSF3 data set and \texttt{p8r2} foreground model, but with $b < -30^\circ$.  The median source-count distribution for the benchmark analysis is shown in blue. (Formatted as in Fig.~\ref{fig:dndsdata}.)}
   \label{fig:dndsdata_south}
\end{figure*}
\clearpage}

\afterpage{
\begin{figure*}[phtb] 
   \centering
   \includegraphics[width=\textwidth]{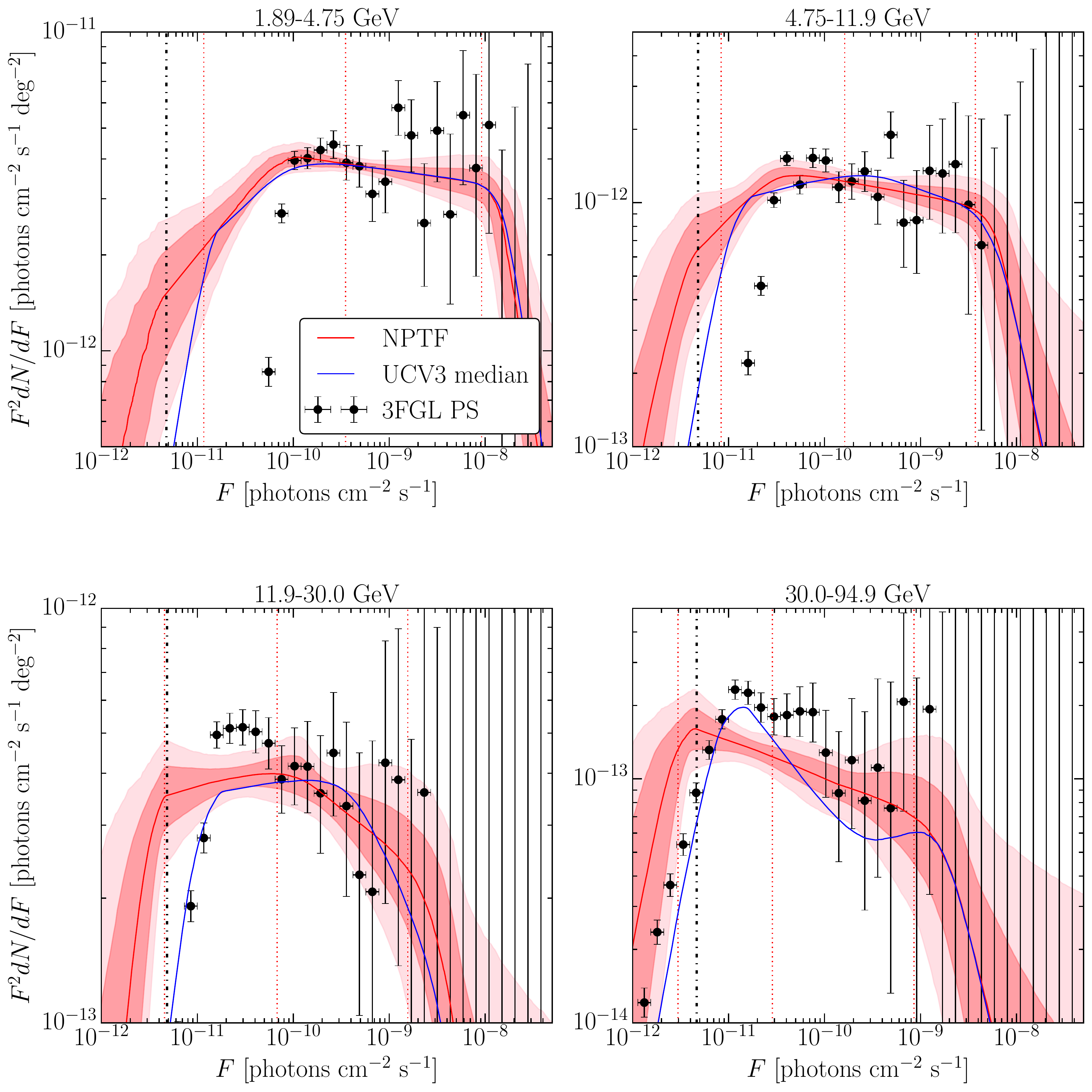} 
   \caption{Best-fit source-count distribution using the top three quartiles of Pass 8 {\it source} data and the \texttt{p8r2} foreground model.  The median source-count distribution for the benchmark analysis is shown in  blue.  (Formatted as in Fig.~\ref{fig:dndsdata}.)}
   \label{fig:dndsdata_source}
\end{figure*}
\clearpage}

\afterpage{
\begin{figure*}[phtb] 
   \centering
   \includegraphics[width=\textwidth]{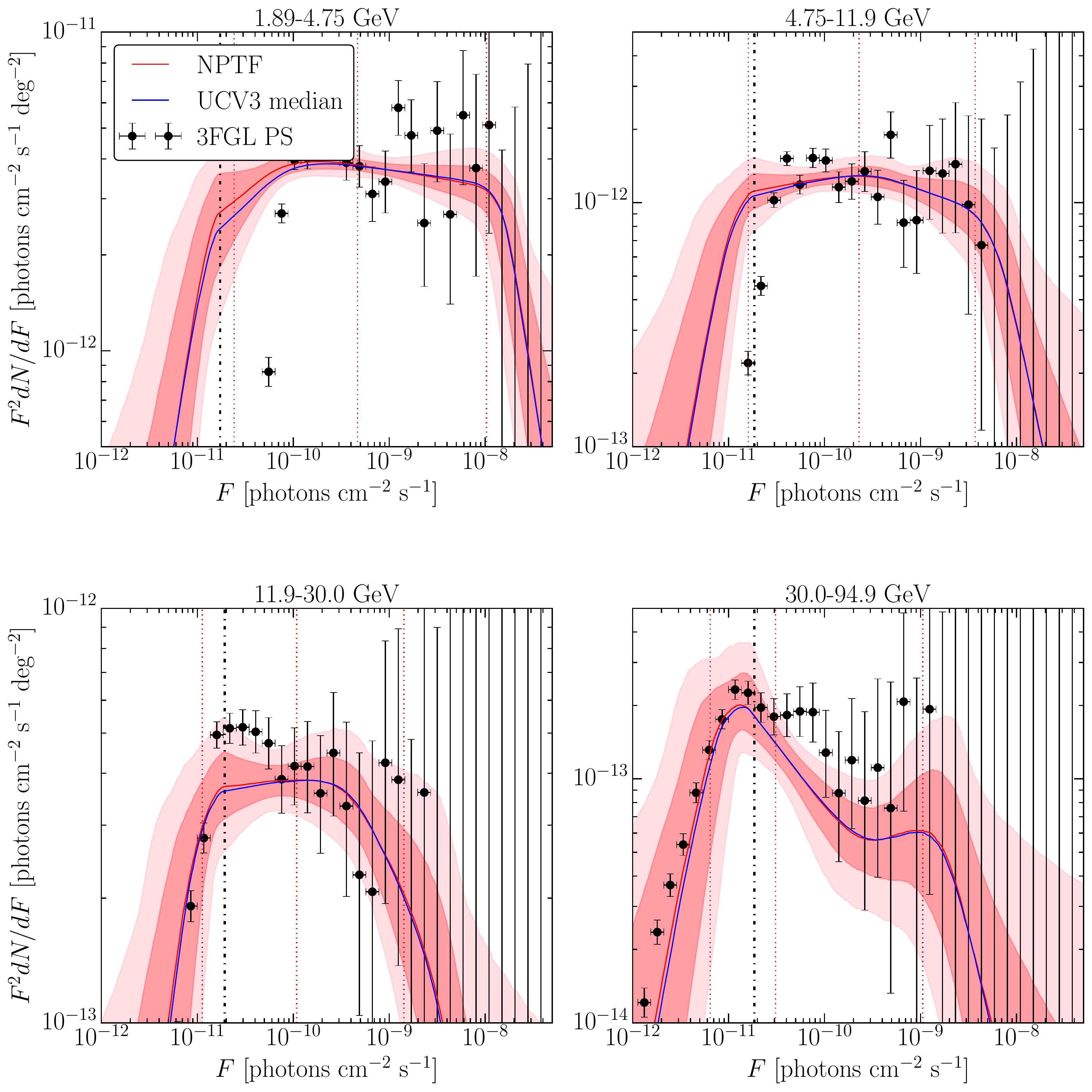} 
   \caption{Best-fit source-count distribution using the Pass 8 {\it ultracleanveto} PSF3 data set and \texttt{p6v11} foreground model.  The median source-count distribution for the benchmark analysis is shown in blue. (Formatted as in Fig.~\ref{fig:dndsdata}.)}
   \label{fig:dndsdata_p6}
\end{figure*}
\clearpage}

\afterpage{
\begin{figure*}[phtb] 
   \centering
   \includegraphics[width=\textwidth]{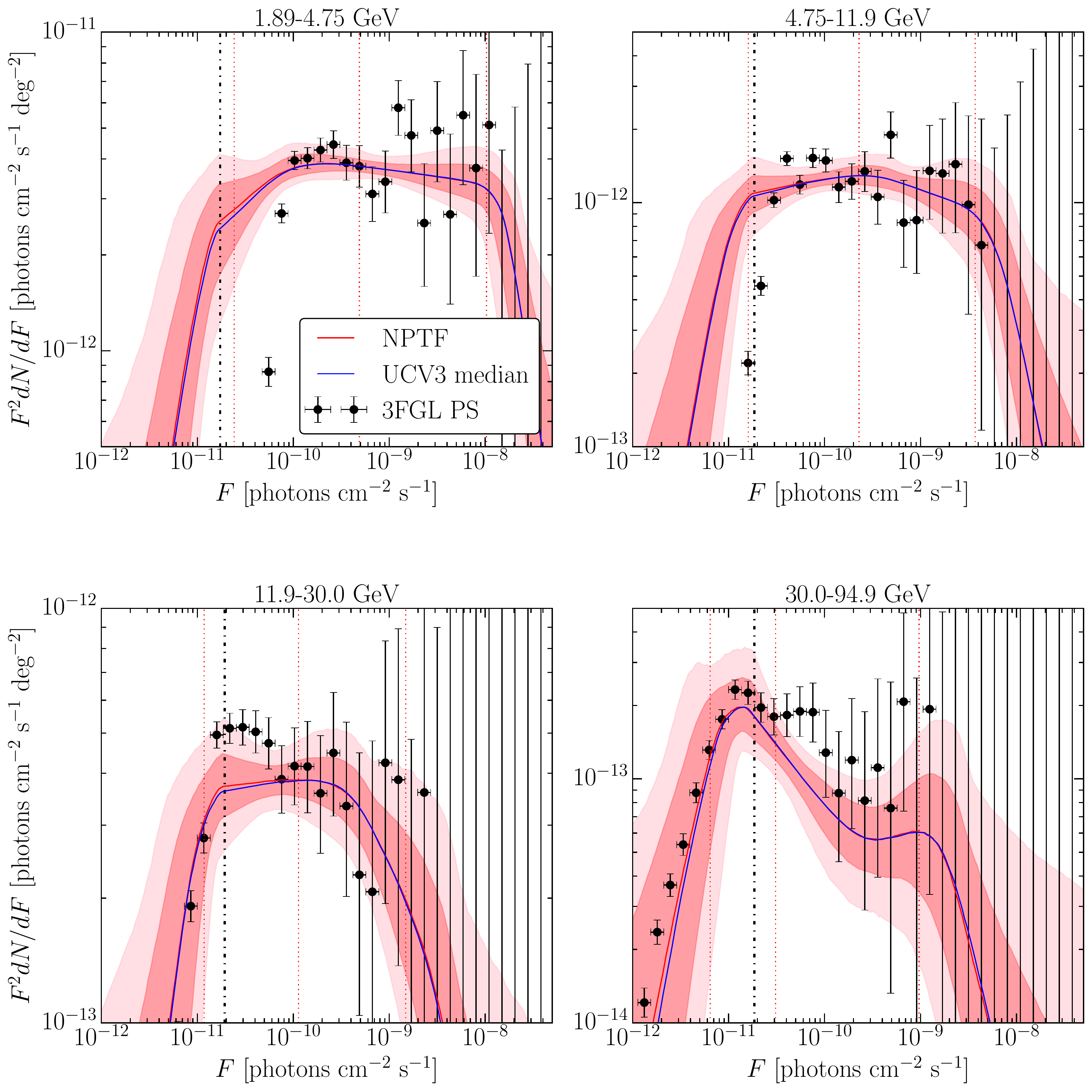} 
   \caption{Best-fit source-count distribution using the Pass 8 {\it ultracleanveto} PSF3 data set and \texttt{p7v6} foreground model.  The median source-count distribution for the benchmark analysis is shown in blue. (Formatted as in Fig.~\ref{fig:dndsdata}.)}
   \label{fig:dndsdata_p7}
\end{figure*}
\clearpage}

\afterpage{
\begin{figure*}[phtb] 
   \centering
   \includegraphics[width=\textwidth]{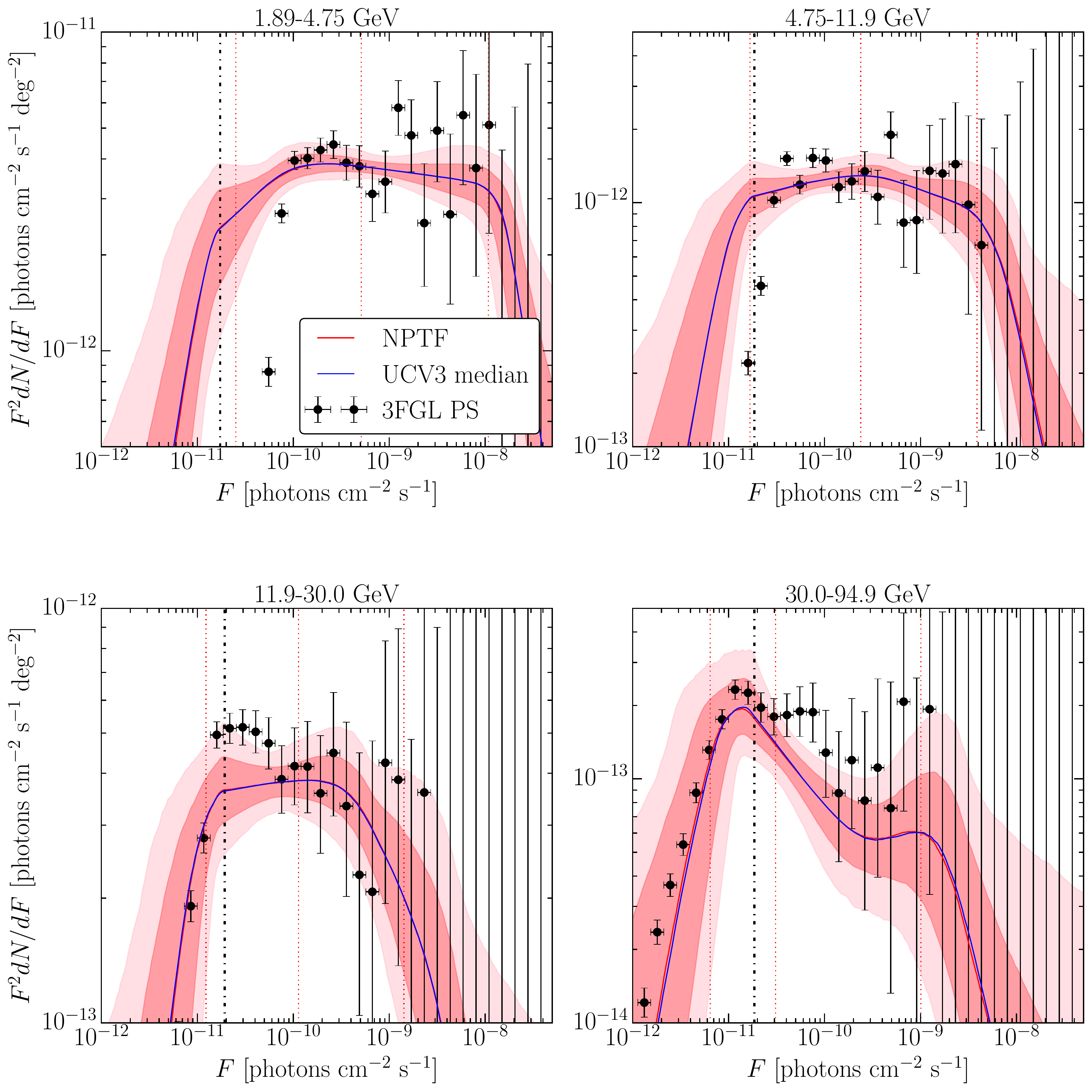} 
   \caption{Best-fit source-count distribution using the Pass 8 {\it ultracleanveto} PSF3 data set and \texttt{p8r2} foreground model, but removing the \emph{Fermi} bubbles template from the analysis..  The median source-count distribution for the benchmark analysis is shown in blue. (Formatted as in Fig.~\ref{fig:dndsdata}.)}
   \label{fig:dndsdata_nobub}
\end{figure*}
\clearpage}

\afterpage{
\begin{figure*}[phtb] 
   \centering
   \includegraphics[width=\textwidth]{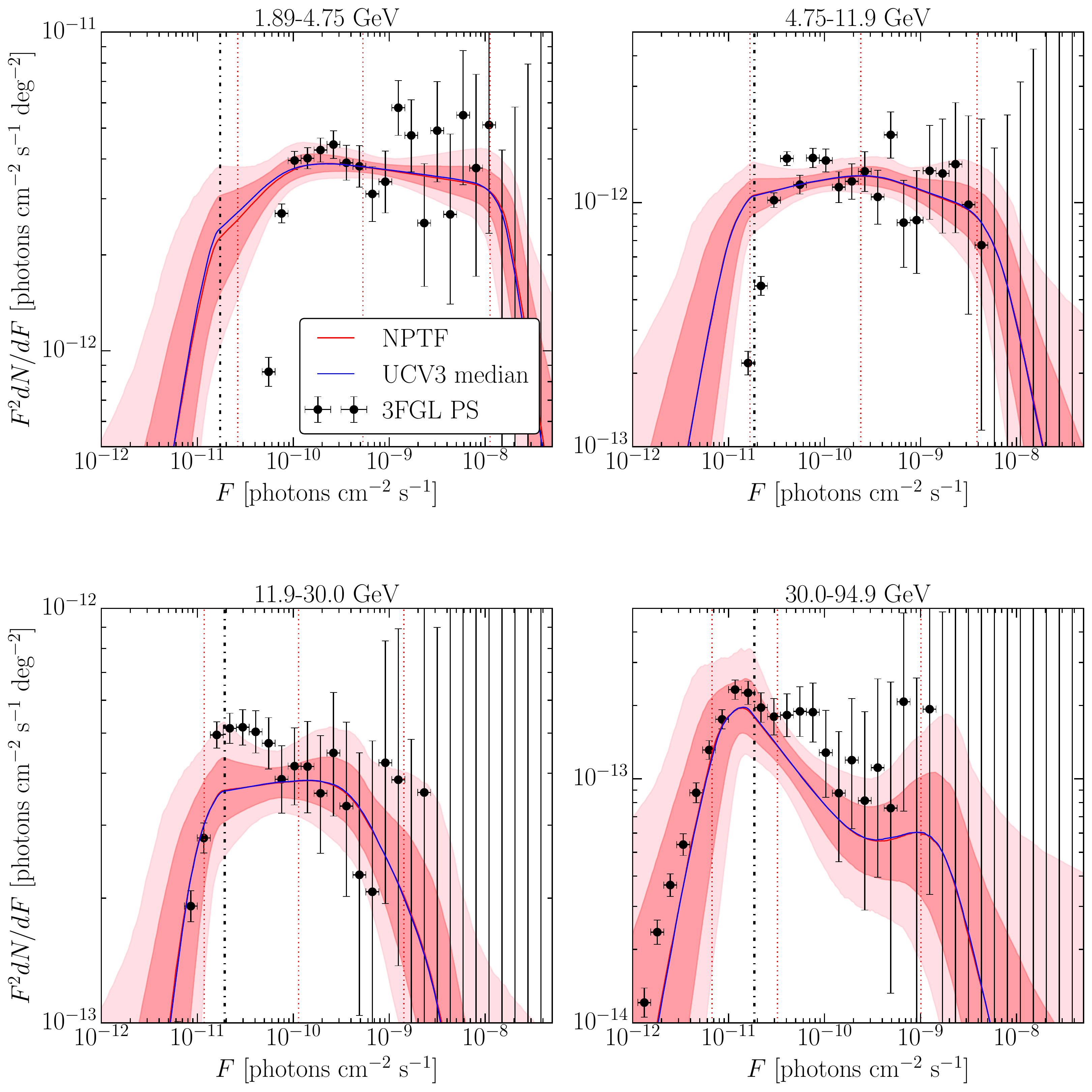} 
   \caption{Best-fit source-count distribution using the Pass 8 {\it ultracleanveto} PSF3 data set and \texttt{p8r2} foreground model, but with a Gaussian PSF.  The median source-count distribution for the benchmark analysis is shown in blue. (Formatted as in Fig.~\ref{fig:dndsdata}.)}
   \label{fig:dndsdata_gauss}
\end{figure*}
\clearpage}
\clearpage

\afterpage{
\begin{figure*}[phtb] 
   \centering
   \includegraphics[width=\textwidth]{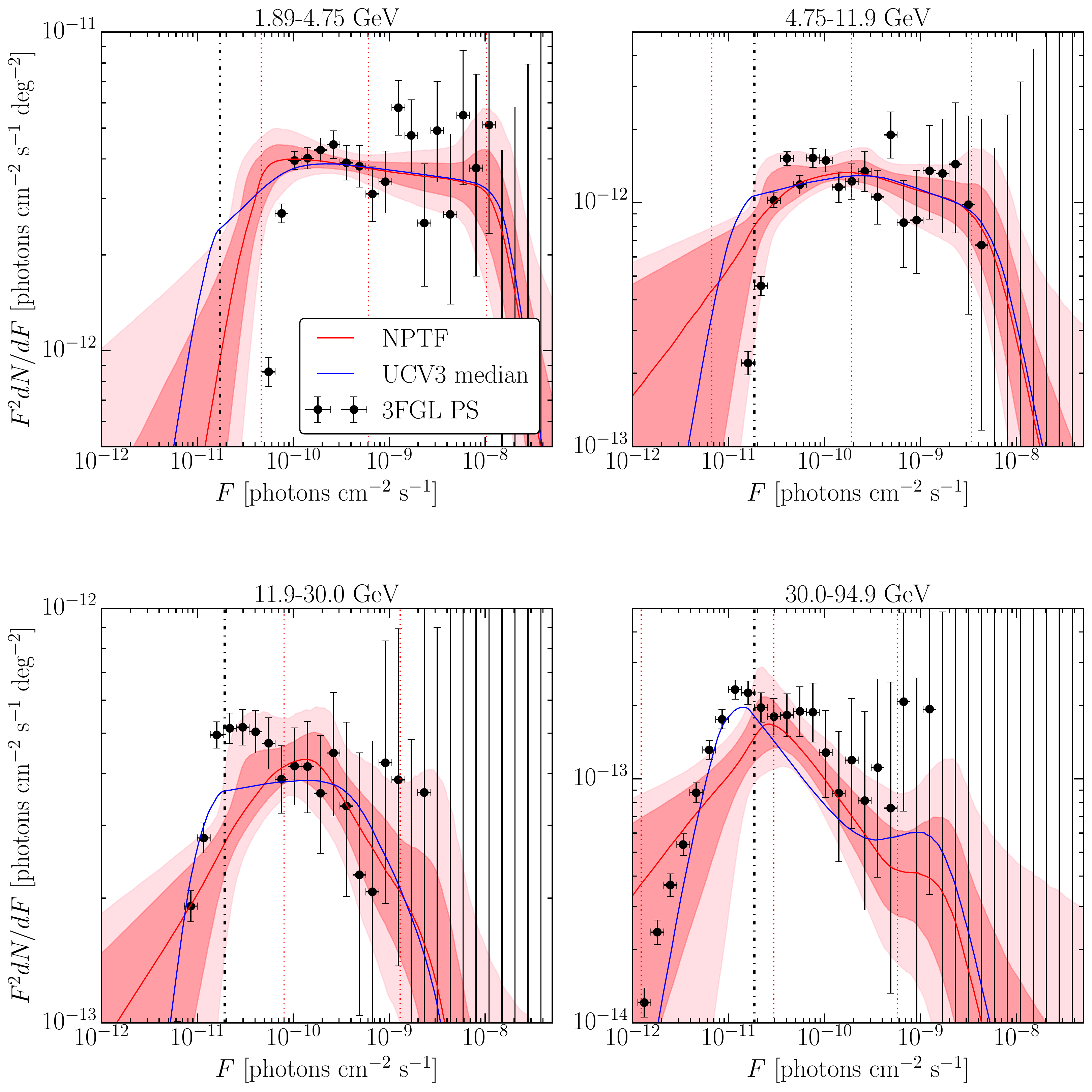} 
   \caption{Best-fit source-count distribution using the Pass 8 {\it ultracleanveto} PSF3 data set and \texttt{p8r2} foreground model, but with the break priors set to $[0.1, 10], [10,40], [40, 2\times S_\text{b,max}]$~ph, where $S_\text{b,max}$ is the maximum number of photons in the 3FGL catalog in the energy bin of interest.  The median source-count distribution for the benchmark analysis is shown in  blue. (Formatted as in Fig.~\ref{fig:dndsdata}.)}
   \label{fig:dndsdata_altpriors1}
\end{figure*}
\clearpage}
\clearpage

\afterpage{
\begin{figure*}[phtb] 
   \centering
   \includegraphics[width=\textwidth]{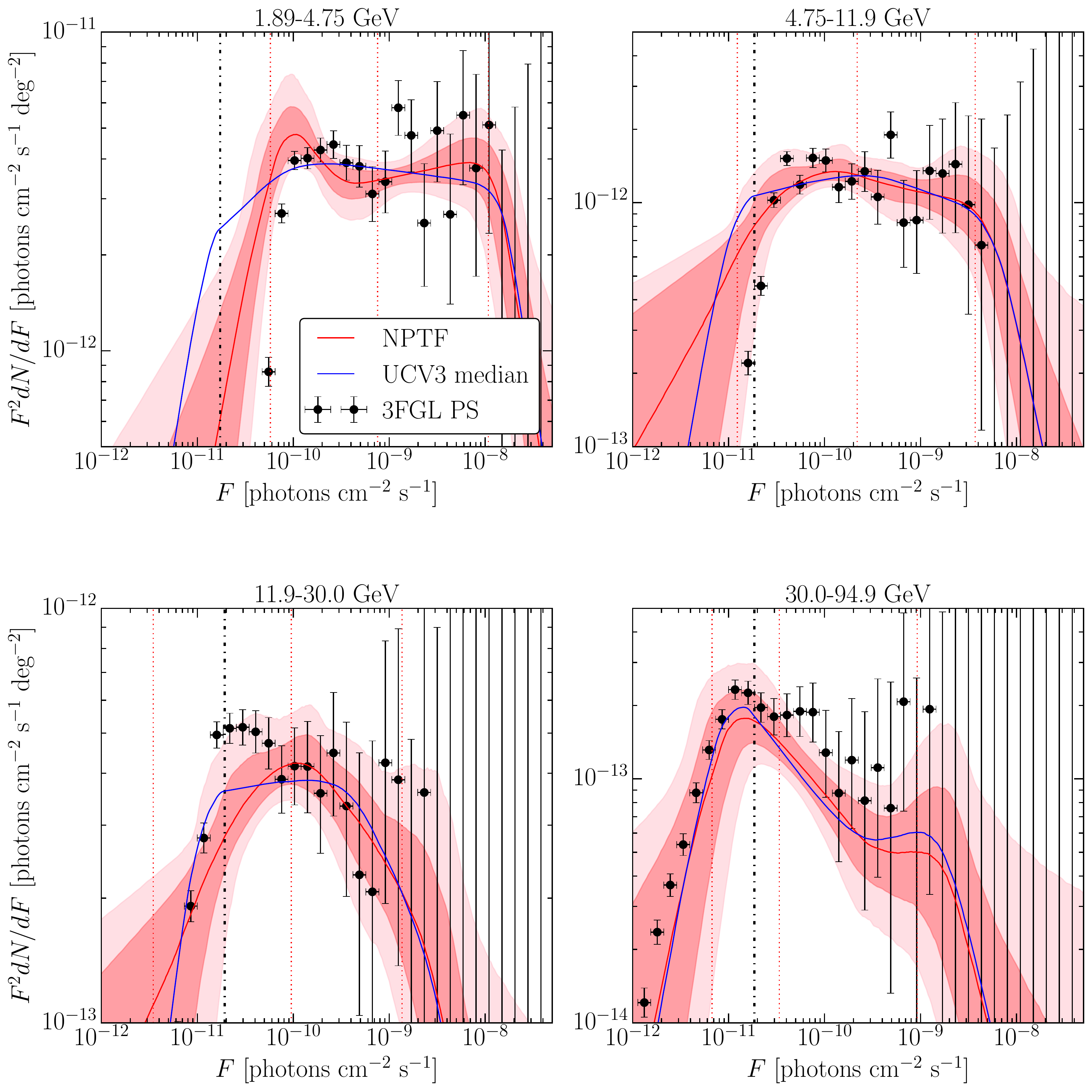} 
   \caption{Best-fit source-count distribution using the Pass 8 {\it ultracleanveto} PSF3 data set and \texttt{p8r2} foreground model, but with the break priors set to $[1, 20], [20, S_\text{b,max}/2], [S_\text{b,max}/2, 2\times S_\text{b,max}]$~ph, where $S_\text{b,max}$ is the maximum number of photons in the 3FGL catalog in the energy bin of interest.  The median source-count distribution for the benchmark analysis is shown in blue.  (Formatted as in Fig.~\ref{fig:dndsdata}.)}
   \label{fig:dndsdata_altpriors2}
\end{figure*}
\clearpage}
\clearpage

\afterpage{
\begin{figure*}[phtb] 
   \centering
   \includegraphics[width=\textwidth]{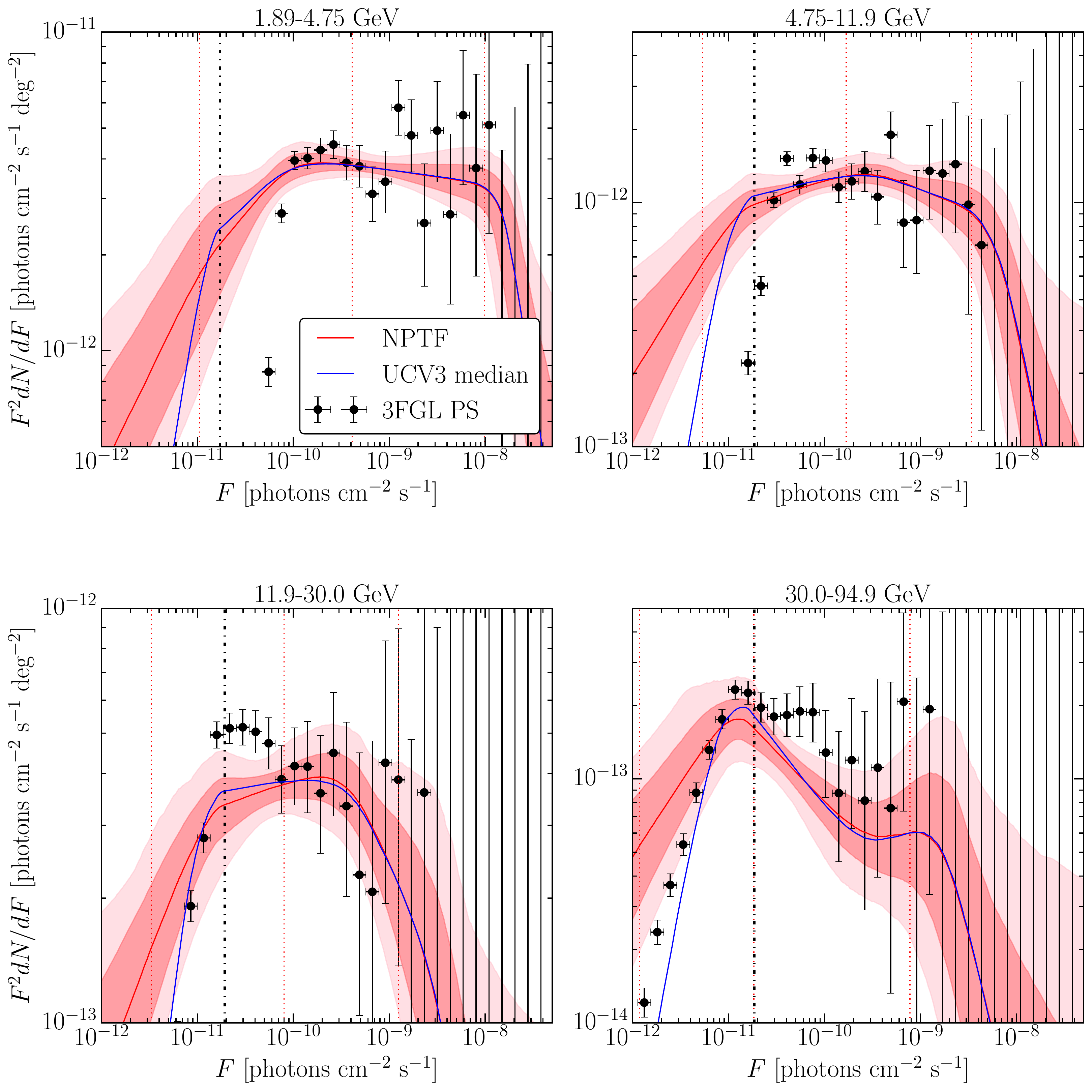} 
   \caption{Best-fit source-count distribution using the Pass 8 {\it ultracleanveto} PSF3 data set and \texttt{p8r2} foreground model, but with the prior for the lowest slope restricted to $n_{4}\in[1,2]$.  The median source-count distribution for the benchmark analysis is shown in blue. (Formatted as in Fig.~\ref{fig:dndsdata}.)}
   \label{fig:dndsdata_altpriors3}
\end{figure*}
\clearpage}
\clearpage

\section{Supplementary Results:  High-Energy Analysis}
\label{app:suppanalysis_high}

This section includes supplementary information pertaining to the high-energy analysis presented in Sec.~\ref{sec:highenergy}.  In particular, Tabs.~\ref{tab:bestfit_highE_intensity} and~\ref{tab:bestfit_highE_dnds} present the best-fit intensities and source-count parameters for the NPTF analysis of all quartiles of {\it ultracleanveto} data.  Figures~\ref{fig:dndsdata_HE_m30}--\ref{fig:dndsdata_HE_S_m30} show the best-fit source-count distributions for the various systematic studies described in the text.

\vspace{0.5in}

\begin{table}[hpbt]
\renewcommand{\arraystretch}{1.4}
\setlength{\tabcolsep}{5pt}
\begin{center}
\begin{tabular}{ c  | c  c  c c  c   }
 Energy & $I_\text{EGB}$&$I_\text{iso}^\text{PS}$ & $I_\text{iso}$ & $I_\text{diff}$ & $I_\text{bub}$   \\
$[\text{GeV}]$ &  \multicolumn{5}{c}{$\left[\text{cm}^{-2}\text{ s}^{-1}\text{ sr}^{-1}\right]$}    \\
\hline
50.0--151 
&  $3.10_{-0.11}^{+0.13} \times 10^{-9}$ & $1.36_{-0.16}^{+0.19} \times 10^{-9}$ & $1.74_{-0.16}^{+0.13} \times 10^{-9}$ & $2.69_{-0.06}^{+0.06} \times 10^{-9}$ & $5.26_{-2.51}^{+2.60} \times 10^{-10}$\\
151--457 &    
$4.38_{-0.32}^{+0.42} \times 10^{-10}$ & $1.56_{-0.29}^{+0.42} \times 10^{-10}$ & $2.80_{-0.31}^{+0.27} \times 10^{-10}$ & $4.12_{-0.23}^{+0.21} \times 10^{-10}$ & $5.40_{-3.81}^{+7.15} \times 10^{-11}$   \\
457--2000  & 
$1.10_{-0.13}^{+0.15} \times 10^{-10}$ & $1.29_{-0.61}^{+0.99} \times 10^{-11}$ & $9.61_{-1.40}^{+1.32} \times 10^{-11}$ & $6.29_{-1.37}^{+1.22} \times 10^{-11}$ & $7.18_{-4.11}^{+5.02} \times 10^{-11}$  \\  
50.0-2000  &  
$3.74_{-0.12}^{+0.16} \times 10^{-9}$ & $1.61_{-0.18}^{+0.22} \times 10^{-9}$ & $2.13_{-0.19}^{+0.15} \times 10^{-9}$ & $3.29_{-0.07}^{+0.07} \times 10^{-9}$ & $5.26_{-2.58}^{+3.01} \times 10^{-10}$ \\
\end{tabular}
\end{center}
\caption{Same as Tab.~\ref{tab:bestfit}, except using all quartiles (PSF0--3) of the Pass~8 {\it ultracleanveto} data for the high-energy analysis.  Note that the \emph{Fermi} bubbles template intensity is defined relative to the interior of the bubbles, while the intensities of the other templates are computed with respect to the region $|b| > 10^\circ$.  }
\label{tab:bestfit_highE_intensity}
\end{table}

\vspace{0.5in}

\begin{table}[hpbt]
\renewcommand{\arraystretch}{1.4}
\setlength{\tabcolsep}{3pt}
\begin{center}
\begin{tabular}{ c  | c  c  c c |  c c c   }
 Energy & $n_1$ & $n_2$ & $n_3$ & $n_4$ & $F_{b,3}$ & $F_{b,2}$ & $F_{b,1}$   \\
$[\text{GeV}]$ &  & & & & \multicolumn{3}{c}{$\left[\text{cm}^{-2}\text{ s}^{-1}\right]$}    \\
\hline
50.0--151 &  
$3.56_{-0.90}^{+0.92}$ & $2.34_{-0.27}^{+0.36}$ & $2.18_{-0.11}^{+0.12}$ & $-0.05_{-1.21}^{+1.12}$ & $1.58_{-0.76}^{+1.21} \times 10^{-12}$ & $7.93_{-4.10}^{+2.78} \times 10^{-11}$ & $1.55_{-0.73}^{+0.77} \times 10^{-9}$
   \\
151--457 &    
$3.56_{-0.97}^{+0.95}$ & $1.87_{-0.53}^{+0.65}$ & $2.42_{-0.28}^{+0.39}$ & $-0.06_{-1.26}^{+1.08}$ & $2.53_{-1.15}^{+1.16} \times 10^{-12}$ & $6.41_{-3.86}^{+4.34} \times 10^{-11}$ & $4.80_{-2.29}^{+2.63} \times 10^{-10}$
   \\
457--2000  & 
$3.57_{-0.96}^{+0.91}$ & $2.26_{-0.78}^{+0.78}$ & $2.16_{-0.67}^{+0.68}$ & $-0.01_{-1.25}^{+1.25}$ & $3.16_{-1.68}^{+1.74} \times 10^{-12}$ & $8.90_{-5.73}^{+5.94} \times 10^{-11}$ & $2.35_{-0.37}^{+0.38} \times 10^{-10}$
  \\  
50.0-2000  &  
$3.63_{-0.99}^{+0.90}$ & $2.28_{-0.22}^{+0.28}$ & $2.17_{-0.09}^{+0.12}$ & $-0.05_{-1.24}^{+1.10}$ & $1.72_{-0.79}^{+1.29} \times 10^{-12}$ & $7.87_{-4.37}^{+3.16} \times 10^{-11}$ & $2.15_{-1.06}^{+1.18} \times 10^{-9}$
  \\
\end{tabular}
\end{center}
\caption{Same as Tab.~\ref{tab:bestfit_dndf}, except using all quartiles (PSF0--3) of the Pass~8 {\it ultracleanveto} data for the high-energy analysis.  }
\label{tab:bestfit_highE_dnds}
\end{table}

\afterpage{
\begin{figure*}[phtb] 
   \centering
   \includegraphics[width=\textwidth]{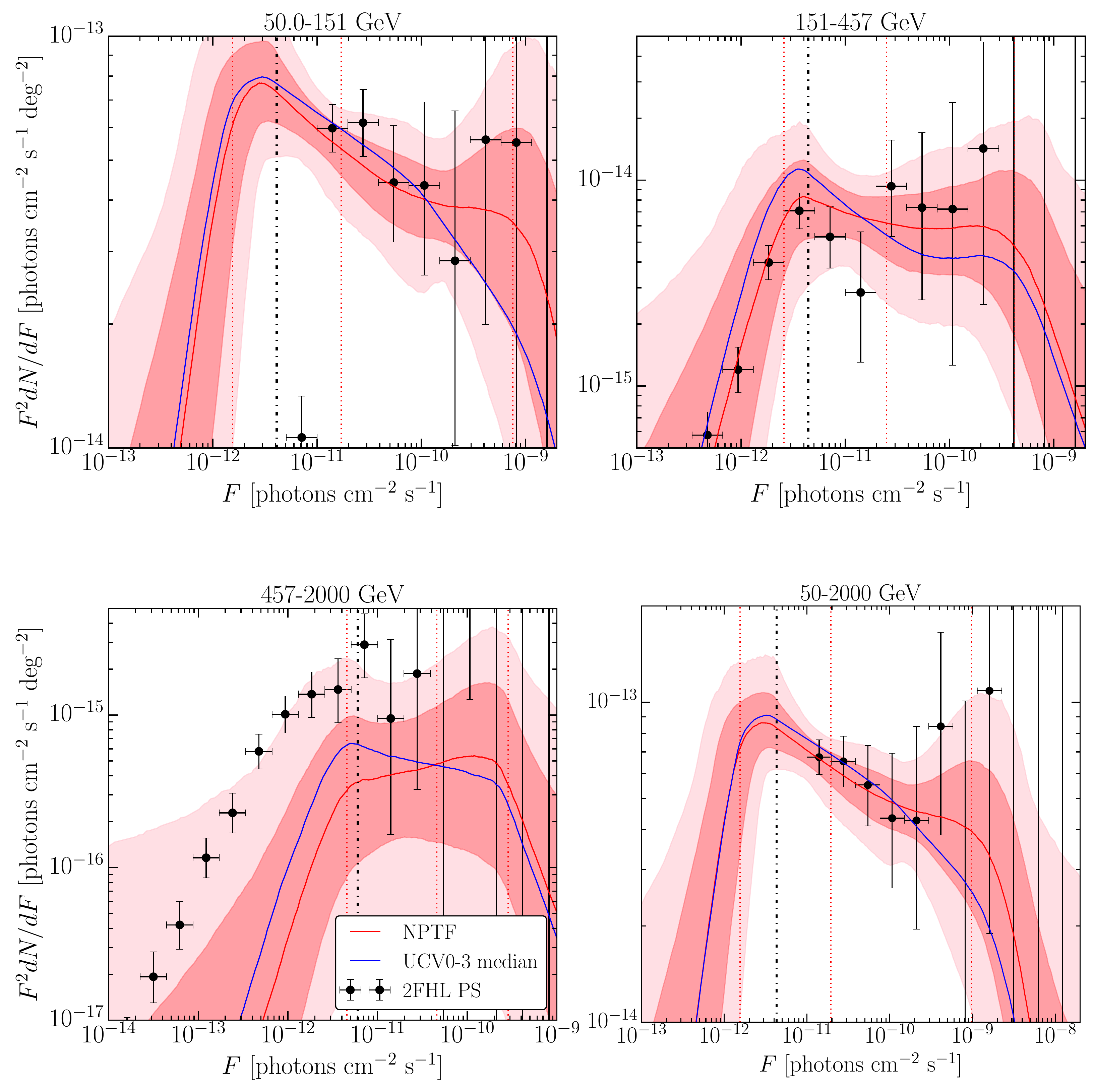} 
   \caption{   Best-fit source-count distribution using the Pass 8 {\it ultracleanveto} PSF0--3 data set and \texttt{p8r2} foreground model, but with $|b| > 30^\circ$.  The median source-count distribution for the benchmark analysis is shown in blue.  (Formatted as in Fig.~\ref{fig:dndsdata_HE}.)}
   \label{fig:dndsdata_HE_m30}
\end{figure*}
\clearpage}

\afterpage{
\begin{figure*}[phtb] 
   \centering
   \includegraphics[width=\textwidth]{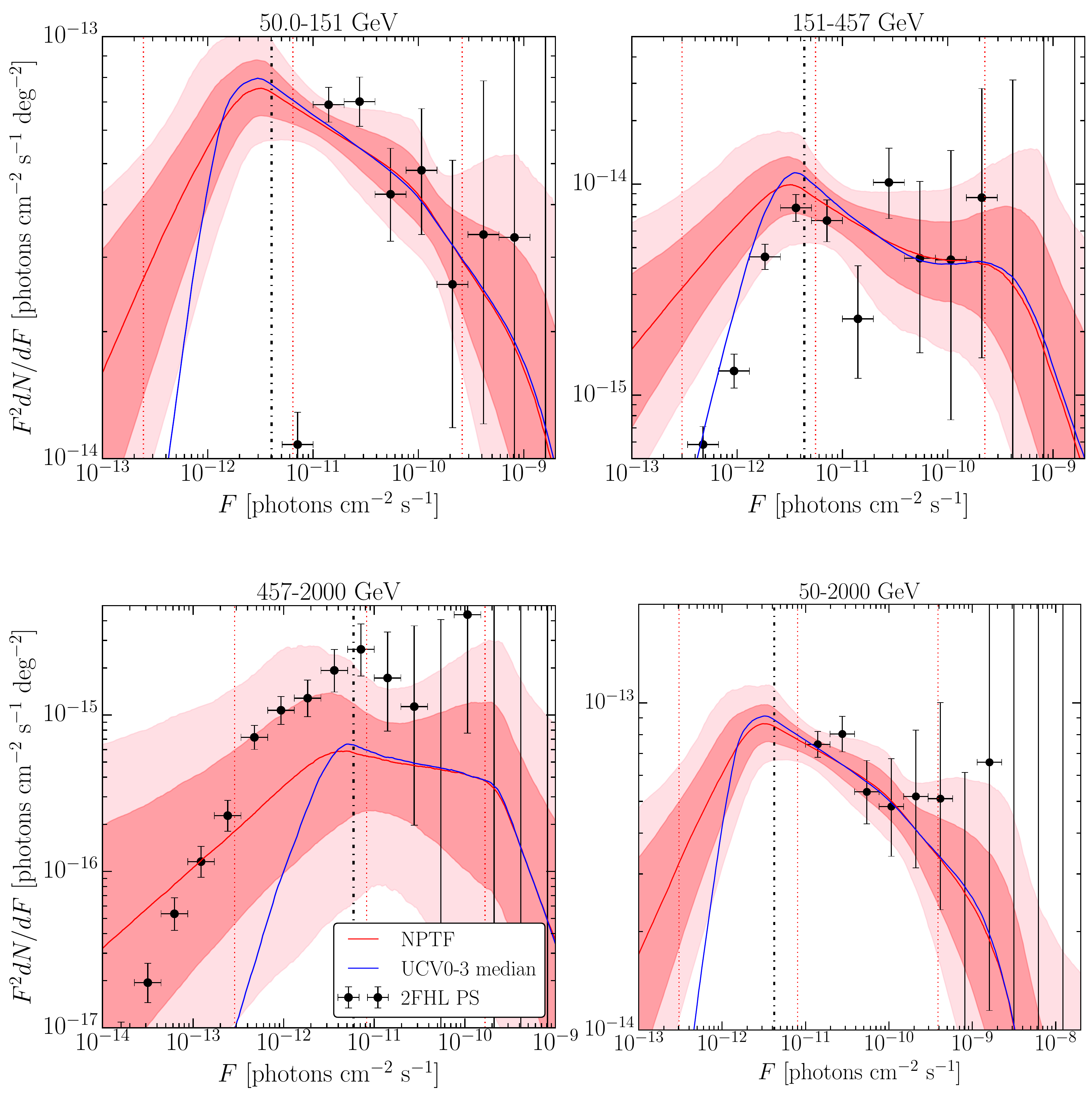} 
   \caption{Best-fit source-count distribution using the Pass 8 {\it ultracleanveto} PSF0--3 data set and \texttt{p8r2} foreground model, but with the prior on the lowest slope restricted to $n_{4}\in[1,2]$.  The median source-count distribution for the benchmark analysis is shown in blue.  (Formatted as in Fig.~\ref{fig:dndsdata_HE}.)}
   \label{fig:dndsdata_HE_m30}
\end{figure*}
\clearpage}

\afterpage{
\begin{figure*}[phtb] 
   \centering
   \includegraphics[width=\textwidth]{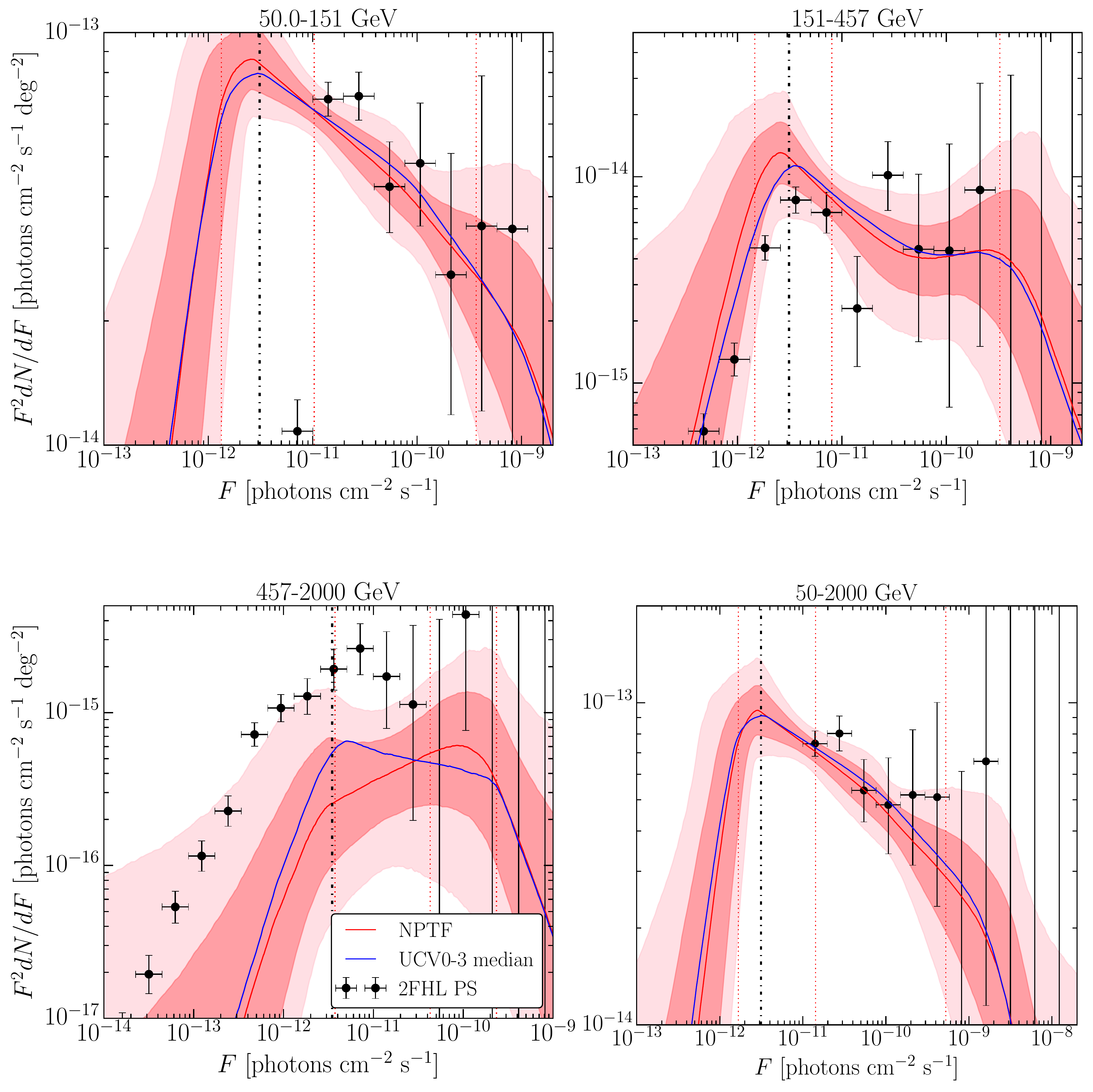} 
   \caption{Best-fit source-count distribution using the Pass 8 {\it source} PSF0--3 data set and \texttt{p8r2} foreground model, with $|b| > 10^\circ$.  The median source-count distribution for the benchmark analysis is shown in blue.  (Formatted as in Fig.~\ref{fig:dndsdata_HE}.)}
   \label{fig:dndsdata_HE_S_m10}
\end{figure*}
\clearpage}

\afterpage{
\begin{figure*}[phtb] 
   \centering
   \includegraphics[width=\textwidth]{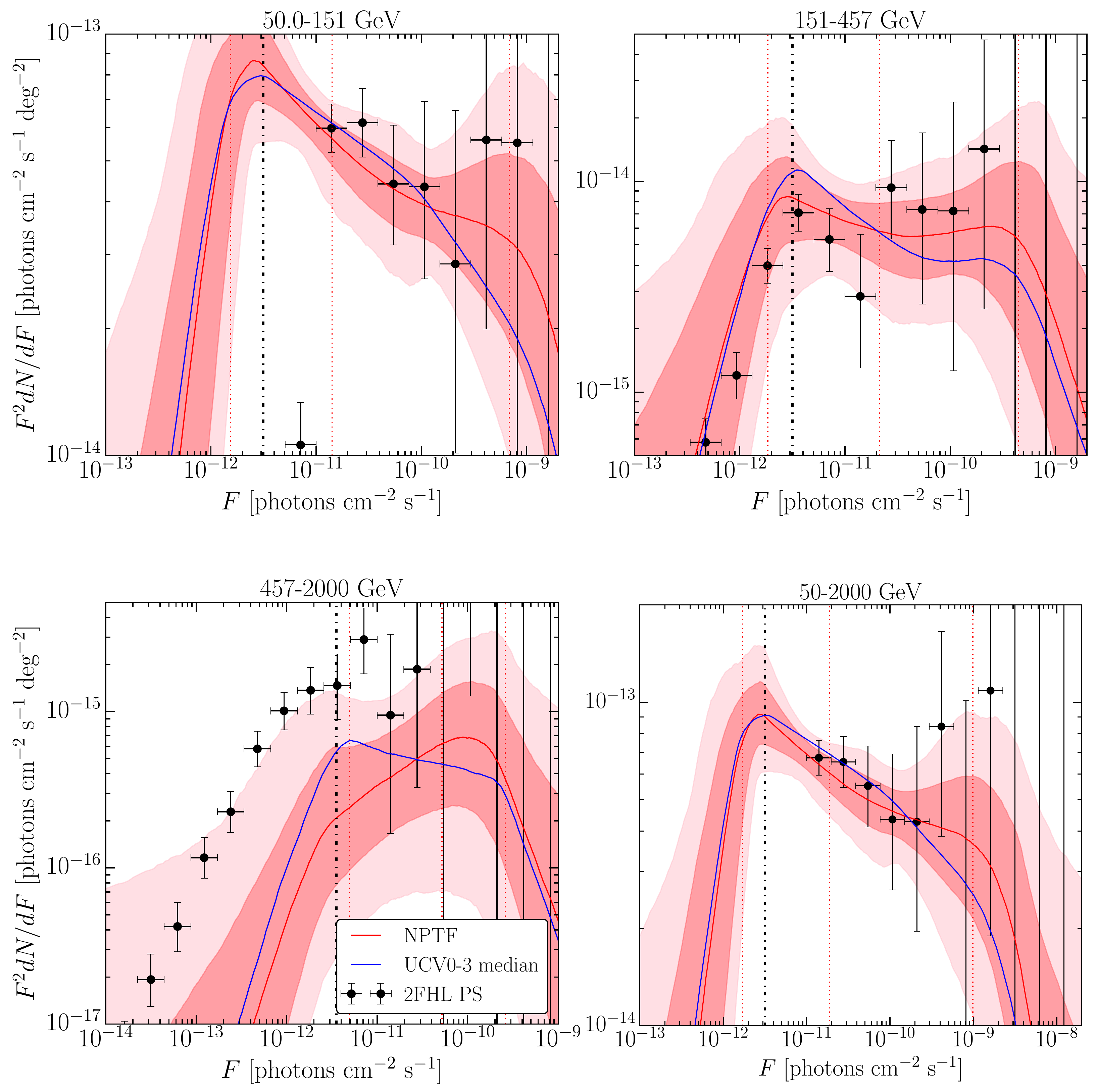} 
   \caption{Best-fit source-count distribution using the Pass 8 {\it source} PSF0--3 data set and \texttt{p8r2} foreground model, but with $|b| > 30^\circ$.  The median source-count distribution for the benchmark analysis is shown in blue. (Formatted as in Fig.~\ref{fig:dndsdata_HE}.)}
   \label{fig:dndsdata_HE_S_m30}
\end{figure*}
\clearpage}

\clearpage
\def\bibsection{} 
\bibliographystyle{aasjournal}
\bibliography{fermi_igrb}

\end{document}